\documentclass[a4paper]{article}
\usepackage{amssymb}
\usepackage{amsfonts}
\usepackage{amsmath}
\usepackage[margin=2cm]{geometry}
\usepackage{graphicx}
\usepackage{caption}
\usepackage{subfig}
\usepackage{multirow}

\input{tcilatex}
\begin{document}

\title{Fractional Burgers models in \\
creep and stress relaxation tests}
\author{Aleksandar S. Okuka\thanks{%
Department of Mechanics, Faculty of Technical Sciences, University of Novi
Sad, Trg D. Obradovi\'{c}a 6, 21000 Novi Sad, Serbia, aokuka@uns.ac.rs}, Du%
\v{s}an Zorica\thanks{%
Mathematical Institute, Serbian Academy of Arts and Sciences, Kneza Mihaila
36, 11000 Belgrade, Serbia, dusan\textunderscore zorica@mi.sanu.ac.rs and
Department of Physics, Faculty of Sciences, University of Novi Sad, Trg D.
Obradovi\'{c}a 4, 21000 Novi Sad, Serbia}}
\maketitle

\begin{abstract}
\noindent Classical and thermodynamically consistent fractional Burgers
models are examined in creep and stress relaxation tests. Using the Laplace
transform method, the creep compliance and relaxation modulus are obtained
in integral form, that yielded, when compared to the thermodynamical
requirements, the narrower range of model parameters in which the creep
compliance is a Bernstein function while the relaxation modulus is
completely monotonic. Moreover, the relaxation modulus may even be
oscillatory function with decreasing amplitude. The asymptotic analysis of
the creep compliance and relaxation modulus is performed near the initial
time-instant and for large time as well.

\end{abstract}

\section{Introduction}

\begin{figure}[htbp]
\centering
\includegraphics[width=0.5\columnwidth]{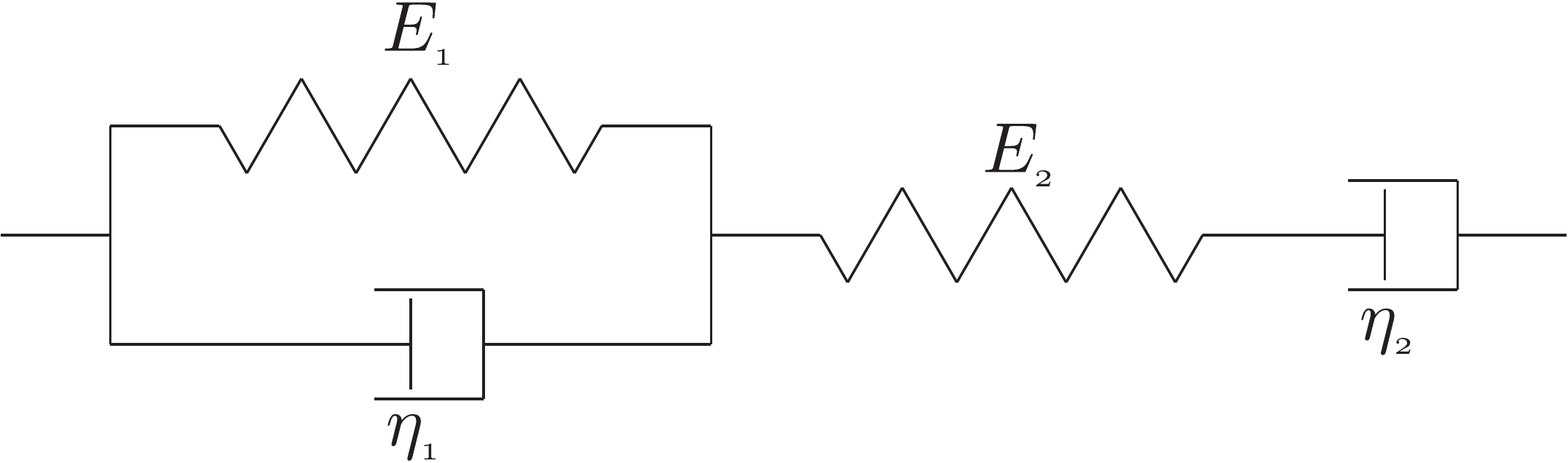}
\caption{Rheological representation of the classical Burgers model.}
\label{rrbm}
\end{figure}

The classical Burgers model 
\begin{equation}
\left( 1+a_{1}\frac{\mathrm{d}}{\mathrm{d}t}+a_{2}\frac{\mathrm{d}^{2}}{%
\mathrm{d}t^{2}}\right) \sigma \left( t\right) =\left( b_{1}\frac{\mathrm{d}%
}{\mathrm{d}t}+b_{2}\frac{\mathrm{d}^{2}}{\mathrm{d}t^{2}}\right)
\varepsilon \left( t\right) ,  \label{cbm}
\end{equation}%
where $\sigma $ and $\varepsilon $ denote stress and strain, that are
functions of time $t>0,$ while $a_{1},a_{2},b_{1},b_{2}>0$ are model
parameters, see \cite{FindleyLaiOnaran,Mai-10}, is obtained using the
rheological representation shown in Figure \ref{rrbm}. Thermodynamical
constraints 
\begin{equation}
\frac{a_{2}}{a_{1}}\leq \frac{b_{2}}{b_{1}}\leq a_{1}  \label{cbm-tdr}
\end{equation}%
on model parameters appearing in the classical Burgers model (\ref{cbm}) are
derived in \cite{OZ-1} by requiring non-negativity of the storage and loss
modulus that is obtained as a consequence of the dissipativity inequality in
the steady state regime, see \cite{b-t}. 

The fractional generalization of Burgers model is derived in \cite{OZ-1} by
considering the Scott-Blair (fractional) element instead of the dash-pot
element in the rheological representation from Figure \ref{rrbm}, with the
orders of fractional differentiation corresponding to the fractional
elements and their sums being replaced by the arbitrary orders of fractional
derivatives $\alpha ,\beta ,\mu ,\gamma ,$ and $\nu .$ Such obtained
fractional Burgers model takes the form%
\begin{equation}
\left( 1+a_{1}\,{}_{0}\mathrm{D}_{t}^{\alpha }+a_{2}\,{}_{0}\mathrm{D}%
_{t}^{\beta }+a_{3}\,{}_{0}\mathrm{D}_{t}^{\gamma }\right) \sigma \left(
t\right) =\left( b_{1}\,{}_{0}\mathrm{D}_{t}^{\mu }+b_{2}\,{}_{0}\mathrm{D}%
_{t}^{\nu }\right) \varepsilon \left( t\right) ,  \label{fbm}
\end{equation}%
where the model parameters are denoted by $a_{1},a_{2},a_{3},b_{1},b_{2}>0,$ 
$\alpha ,\beta ,\mu \in \left[ 0,1\right] ,$ with $\alpha \leq \beta ,$ and $%
\gamma ,\nu \in \left[ 1,2\right] ,$ while ${}_{0}\mathrm{D}_{t}^{\xi }$
denotes the operator of Riemann-Liouville fractional derivative of order $%
\xi \in \left[ n,n+1\right] ,$ $n\in 
\mathbb{N}
_{0},$ defined by%
\begin{equation*}
{}_{0}\mathrm{D}_{t}^{\xi }y\left( t\right) =\frac{\mathrm{d}^{n+1}}{\mathrm{%
d}t^{n+1}}\left( \frac{t^{-\left( \xi -n\right) }}{\Gamma \left( 1-\left(
\xi -n\right) \right) }\ast y\left( t\right) \right) ,\;\;t>0,
\end{equation*}%
see \cite{TAFDE}, with $\ast $ denoting the convolution in time: $f\left(
t\right) \ast g\left( t\right) =\int_{0}^{t}f\left( u\right) g\left(
t-u\right) \mathrm{d}u,$ $t>0.$

Thermodynamical consistency analysis of the fractional Burgers model (\ref%
{fbm}), conducted in \cite{OZ-1} by the use of storage and loss modulus
non-negativity requirement, implied that the orders of fractional
derivatives from interval $\left[ 1,2\right] $ cannot be independent of the
orders of fractional derivatives from interval $\left[ 0,1\right] ,$ and
this led to formulation of eight thermodynamically consistent fractional
Burgers models.

Two classes of thermodynamically consistent fractional Burgers models are
distinguished according to the orders of fractional derivatives acting on
stress and strain. The first class of fractional Burgers models consists of
models having different orders of fractional differentiation of stress and
strain from both intervals $\left[ 0,1\right] $ and $\left[ 1,2\right] .$
Namely, in the case of Model I the highest order of fractional
differentiation of stress is $\gamma \in \left[ 0,1\right] ,$ while the
highest order of fractional differentiation of strain is $\nu =\mu
+\varkappa \in \left[ 1,2\right] ,$ with $0\leq \alpha \leq \beta \leq
\gamma \leq \mu \leq 1$ and $\varkappa \in \left\{ \alpha ,\beta ,\gamma
\right\} ,$ while in the case of Models II - V one has $0\leq \alpha \leq
\beta \leq \mu \leq 1$ and $1\leq \gamma \leq \nu \leq 2,$ with $\left(
\gamma ,\nu \right) \in \left\{ \left( 2\alpha ,\mu +\alpha \right) ,\left(
\alpha +\beta ,\mu +\alpha \right) ,\left( \alpha +\beta ,\mu +\beta \right)
,\left( 2\beta ,\mu +\beta \right) \right\} ,$ see the unified constitutive
equation (\ref{UCE-1-5}) below. Likewise the classical Burgers model (\ref%
{cbm}), Models VI - VIII contain the same orders of fractional derivatives
acting on stress and strain from both intervals $\left[ 0,1\right] $ and $%
\left[ 1,2\right] ,$ such that for Models VI and VII in fractional Burgers
model (\ref{fbm}) one has $\mu =\beta \in \left[ 0,1\right] $ and $\gamma
=\nu =\beta +\eta \in \left[ 1,2\right] ,$ with $\eta \in \left\{ \alpha
,\beta \right\} ,$ while Model VIII is obtained from (\ref{fbm}) for $\beta
=\alpha ,$ $\bar{a}_{1}=a_{1}+a_{2},$ $\bar{a}_{2}=a_{3},$ $\mu =\alpha ,$ $%
\gamma =\nu =2\alpha ,$ see the unified constitutive equation (\ref{UCE})
below.

Models I - VIII, along with corresponding thermodynamical constraints, are
listed below.\smallskip

\noindent \textbf{Model I}: 
\begin{gather}
\left( 1+a_{1}\,{}_{0}\mathrm{D}_{t}^{\alpha }+a_{2}\,{}_{0}\mathrm{D}%
_{t}^{\beta }+a_{3}\,{}_{0}\mathrm{D}_{t}^{\gamma }\right) \sigma \left(
t\right) =\left( b_{1}\,{}_{0}\mathrm{D}_{t}^{\mu }+b_{2}\,{}_{0}\mathrm{D}%
_{t}^{\mu +\varkappa }\right) \varepsilon \left( t\right) ,  \label{Model 1}
\\
0\leq \alpha \leq \beta \leq \gamma \leq \mu \leq 1,\;\;1\leq \mu +\varkappa
\leq 1+\alpha ,\;\;\frac{b_{2}}{b_{1}}\leq a_{i}\frac{\cos \frac{\left( \mu
-\varkappa \right) \pi }{2}}{\left\vert \cos \frac{\left( \mu +\varkappa
\right) \pi }{2}\right\vert },  \label{Model 1 - tdr}
\end{gather}%
with $\left( \varkappa ,i\right) \in \left\{ \left( \alpha ,1\right) ,\left(
\beta ,2\right) ,\left( \gamma ,3\right) \right\} ;$

\noindent \textbf{Model II}:%
\begin{gather}
\left( 1+a_{1}\,{}_{0}\mathrm{D}_{t}^{\alpha }+a_{2}\,{}_{0}\mathrm{D}%
_{t}^{\beta }+a_{3}\,{}_{0}\mathrm{D}_{t}^{2\alpha }\right) \sigma \left(
t\right) =\left( b_{1}\,{}_{0}\mathrm{D}_{t}^{\mu }+b_{2}\,{}_{0}\mathrm{D}%
_{t}^{\mu +\alpha }\right) \varepsilon \left( t\right) ,  \label{Model 2} \\
\frac{1}{2}\leq \alpha \leq \beta \leq \mu \leq 1,\;\;\frac{a_{3}}{a_{1}}%
\frac{\left\vert \sin \frac{\left( \mu -2\alpha \right) \pi }{2}\right\vert 
}{\sin \frac{\mu \pi }{2}}\leq \frac{b_{2}}{b_{1}}\leq a_{1}\frac{\cos \frac{%
\left( \mu -\alpha \right) \pi }{2}}{\left\vert \cos \frac{\left( \mu
+\alpha \right) \pi }{2}\right\vert };  \label{Model 2 - tdr}
\end{gather}

\noindent \textbf{Model III}:%
\begin{gather}
\left( 1+a_{1}\,{}_{0}\mathrm{D}_{t}^{\alpha }+a_{2}\,{}_{0}\mathrm{D}%
_{t}^{\beta }+a_{3}\,{}_{0}\mathrm{D}_{t}^{\alpha +\beta }\right) \sigma
\left( t\right) =\left( b_{1}\,{}_{0}\mathrm{D}_{t}^{\mu }+b_{2}\,{}_{0}%
\mathrm{D}_{t}^{\mu +\alpha }\right) \varepsilon \left( t\right) ,
\label{Model 3} \\
0\leq \alpha \leq \beta \leq \mu \leq 1,\;\;\alpha +\beta \geq 1,\;\;\frac{%
a_{3}}{a_{2}}\frac{\left\vert \sin \frac{\left( \mu -\beta -\alpha \right)
\pi }{2}\right\vert }{\sin \frac{\left( \mu -\beta +\alpha \right) \pi }{2}}%
\leq \frac{b_{2}}{b_{1}}\leq a_{1}\frac{\cos \frac{\left( \mu -\alpha
\right) \pi }{2}}{\left\vert \cos \frac{\left( \mu +\alpha \right) \pi }{2}%
\right\vert };  \label{Model 3 - tdr}
\end{gather}

\noindent \textbf{Model IV}:%
\begin{gather}
\left( 1+a_{1}\,{}_{0}\mathrm{D}_{t}^{\alpha }+a_{2}\,{}_{0}\mathrm{D}%
_{t}^{\beta }+a_{3}\,{}_{0}\mathrm{D}_{t}^{\alpha +\beta }\right) \sigma
\left( t\right) =\left( b_{1}\,{}_{0}\mathrm{D}_{t}^{\mu }+b_{2}\,{}_{0}%
\mathrm{D}_{t}^{\mu +\beta }\right) \varepsilon \left( t\right) ,
\label{Model 4} \\
0\leq \alpha \leq \beta \leq \mu \leq 1,\;\;1-\alpha \leq \beta \leq
1-\left( \mu -\alpha \right) ,\;\;\frac{a_{3}}{a_{1}}\frac{\left\vert \sin 
\frac{\left( \mu -\alpha -\beta \right) \pi }{2}\right\vert }{\sin \frac{%
\left( \mu -\alpha +\beta \right) \pi }{2}}\leq \frac{b_{2}}{b_{1}}\leq a_{2}%
\frac{\cos \frac{\left( \mu -\beta \right) \pi }{2}}{\left\vert \cos \frac{%
\left( \mu +\beta \right) \pi }{2}\right\vert };  \label{Model 4 - tdr}
\end{gather}

\noindent \textbf{Model V}:%
\begin{gather}
\left( 1+a_{1}\,{}_{0}\mathrm{D}_{t}^{\alpha }+a_{2}\,{}_{0}\mathrm{D}%
_{t}^{\beta }+a_{3}\,{}_{0}\mathrm{D}_{t}^{2\beta }\right) \sigma \left(
t\right) =\left( b_{1}\,{}_{0}\mathrm{D}_{t}^{\mu }+b_{2}\,{}_{0}\mathrm{D}%
_{t}^{\mu +\beta }\right) \varepsilon \left( t\right) ,  \label{Model 5} \\
0\leq \alpha \leq \beta \leq \mu \leq 1,\;\;\frac{1}{2}\leq \beta \leq
1-\left( \mu -\alpha \right) ,\;\;\frac{a_{3}}{a_{2}}\frac{\left\vert \sin 
\frac{\left( \mu -2\beta \right) \pi }{2}\right\vert }{\sin \frac{\mu \pi }{2%
}}\leq \frac{b_{2}}{b_{1}}\leq a_{2}\frac{\cos \frac{\left( \mu -\beta
\right) \pi }{2}}{\left\vert \cos \frac{\left( \mu +\beta \right) \pi }{2}%
\right\vert }.  \label{Model 5 - tdr}
\end{gather}

\noindent \textbf{Model VI}:%
\begin{gather}
\left( 1+a_{1}\,{}_{0}\mathrm{D}_{t}^{\alpha }+a_{2}\,{}_{0}\mathrm{D}%
_{t}^{\beta }+a_{3}\,{}_{0}\mathrm{D}_{t}^{\alpha +\beta }\right) \sigma
\left( t\right) =\left( b_{1}\,{}_{0}\mathrm{D}_{t}^{\beta }+b_{2}\,{}_{0}%
\mathrm{D}_{t}^{\alpha +\beta }\right) \varepsilon \left( t\right) ,
\label{Model 6} \\
0\leq \alpha \leq \beta \leq 1,\;\;\alpha +\beta \geq 1,\;\;\frac{a_{3}}{%
a_{2}}\leq \frac{b_{2}}{b_{1}}\leq a_{1}\frac{\cos \frac{\left( \beta
-\alpha \right) \pi }{2}}{\left\vert \cos \frac{\left( \beta +\alpha \right)
\pi }{2}\right\vert };  \label{Model 6 - tdr}
\end{gather}

\noindent \textbf{Model VII}:%
\begin{gather}
\left( 1+a_{1}\,{}_{0}\mathrm{D}_{t}^{\alpha }+a_{2}\,{}_{0}\mathrm{D}%
_{t}^{\beta }+a_{3}\,{}_{0}\mathrm{D}_{t}^{2\beta }\right) \sigma \left(
t\right) =\left( b_{1}\,{}_{0}\mathrm{D}_{t}^{\beta }+b_{2}\,{}_{0}\mathrm{D}%
_{t}^{2\beta }\right) \varepsilon \left( t\right) ,  \label{Model 7} \\
0\leq \alpha \leq \beta \leq 1,\;\;\frac{1}{2}\leq \beta \leq \frac{1+\alpha 
}{2},\;\;\frac{a_{3}}{a_{2}}\leq \frac{b_{2}}{b_{1}}\leq a_{2}\frac{1}{%
\left\vert \cos \left( \beta \pi \right) \right\vert };
\label{Model 7 - tdr}
\end{gather}

\noindent \textbf{Model VIII}:%
\begin{gather}
\left( 1+\bar{a}_{1}\,{}_{0}\mathrm{D}_{t}^{\alpha }+\bar{a}_{2}\,{}_{0}%
\mathrm{D}_{t}^{2\alpha }\right) \sigma \left( t\right) =\left( b_{1}\,{}_{0}%
\mathrm{D}_{t}^{\alpha }+b_{2}\,{}_{0}\mathrm{D}_{t}^{2\alpha }\right)
\varepsilon \left( t\right) ,  \label{Model 8} \\
\frac{1}{2}\leq \alpha \leq 1,\;\;\frac{\bar{a}_{2}}{\bar{a}_{1}}\leq \frac{%
b_{2}}{b_{1}}\leq \bar{a}_{1}\frac{1}{\left\vert \cos \left( \alpha \pi
\right) \right\vert }.  \label{Model 8 - tdr}
\end{gather}

Models I - VIII describe different mechanical behavior, that will be
illustrated by examining the responses in creep and stress relaxation tests.
Recall, creep compliance (relaxation modulus) is the strain (stress) history
function obtained as a response to the imposed sudden and time-constant
stress (strain), i.e., stress (strain) assumed as the Heaviside step
function. The difference between models belonging to the first and second
class reflects in the behavior they describe near the initial time-instant,
since models of the first class predict zero glass compliance and thus
infinite glass modulus, while the glass compliance is non-zero implying the
non-zero glass modulus in the case of models belonging to the second class,
see (\ref{gg}) and (\ref{jg}) below. For both model classes equilibrium
compliance is infinite implying the zero equilibrium modulus.

For both classes of fractional Burgers models as well as for the classical
Burgers model, thermodynamical requirements will prove to be less
restrictive than the conditions guaranteeing that the creep compliance is a
Bernstein function, while the relaxation modulus is a completely monotonic
function. Therefore, if the model parameters fulfill the thermodynamical
requirements but not the restrictive ones, the creep compliance and
relaxation modulus may even not be a monotonic function. Moreover,
conditions on model parameters guaranteeing oscillatory behavior of the
relaxation modulus having amplitudes decreasing in time will be obtained.
Conditions guaranteeing that the creep compliance (relaxation modulus) is a
Bernstein (completely monotonic) function and conditions guaranteeing the
oscillatory behavior of the relaxation modulus are independent. Recall,
completely monotonic function is a positive, monotonically decreasing convex
function, or more precisely function $f$ satisfying $\left( -1\right)
^{n}f^{\left( n\right) }\left( t\right) \geq 0,$ $n\in 
\mathbb{N}
_{0},$ while Bernstein function is a positive, monotonically increasing,
concave function, or more precisely non-negative function having its first
derivative completely monotonic.

The properties of creep compliance as a Bernstein function and relaxation
modulus as a completely monotonic function are discussed in \cite{Mai-10},
while \cite{BazhlekovaBazhlekov} deal with the complete monotonicity of the
relaxation moduli corresponding to distributed-order fractional Zener model.
The review of creep compliances in the frequency domain corresponding to the
integer-order models of viscoelasticity is presented in \cite{Makris}.

Classical (\ref{cbm}), fractional and generalized fractional Burgers models
in the form 
\begin{gather*}
\left( 1+a_{1}\,{}_{0}\mathrm{D}_{t}^{\alpha }+a_{2}\,{}_{0}\mathrm{D}%
_{t}^{2\alpha }\right) \sigma \left( t\right) =\eta \left( 1+b_{1}\,{}_{0}%
\mathrm{D}_{t}^{\beta }\right) \dot{\varepsilon}\left( t\right) , \\
\left( 1+a_{1}\,{}_{0}\mathrm{D}_{t}^{\alpha }+a_{2}\,{}_{0}\mathrm{D}%
_{t}^{2\alpha }\right) \sigma \left( t\right) =\eta \left( 1+b_{1}\,{}_{0}%
\mathrm{D}_{t}^{\beta }+b_{2}\,{}_{0}\mathrm{D}_{t}^{2\beta }\right) \dot{%
\varepsilon}\left( t\right) ,
\end{gather*}%
were extensively used in modeling various fluid flow problems, see
references in \cite{OZ-1}. The thermodynamical consistency of the
generalized fractional Burgers model for viscoelastic fluid 
\begin{equation*}
\left( 1+a_{1}\,{}_{0}\mathrm{D}_{t}^{\alpha }+a_{2}\,{}_{0}\mathrm{D}%
_{t}^{2\alpha }\right) \sigma \left( t\right) =\eta \left( 1+b_{1}\,{}_{0}%
\mathrm{D}_{t}^{\beta }+b_{2}\,{}_{0}\mathrm{D}_{t}^{2\beta }\right) \dot{%
\varepsilon}\left( t\right) ,
\end{equation*}%
is examined in \cite{BazhlekovaTsocheva}.

Some form of the fractional Burgers constitutive equation is used in \cite%
{CelauroFecarottiPirrottaCollop,OeserPellinenScarpasKasbergen,Zbiciak} for
modeling asphalt concrete mixtures according to the experimental data
obtained in creep and creep-recovery experiments, while the fractional
Burgers model 
\begin{equation*}
\left( 1+a_{1}\,{}_{0}\mathrm{D}_{t}^{\nu }+a_{2}\,{}_{0}\mathrm{D}%
_{t}^{1+\nu }\right) \sigma \left( t\right) =\left( b_{1}\,{}_{0}\mathrm{D}%
_{t}^{\nu }+b_{2}\,{}_{0}\mathrm{D}_{t}^{1+\nu }\right) \varepsilon \left(
t\right) ,
\end{equation*}%
is examined in creep and stress relaxation tests in \cite%
{Mai-10,MainardiSpada}. 

Thermodynamical constraints on model parameters appearing in the
fractional-order models of viscoelastic body in the case when orders of
fractional differentiation do not exceed the first order, along with the
material responses in cases of damped oscillations and wave propagation are
considered in \cite{APSZ-1,APSZ-2,R-S-2001,R-S2,R-S-2008}. The behavior in
creep and stress relaxation tests of a distributed-order fractional
viscoelastic material with the inertial effects taken into account is
considered in \cite{APZ-4,APZ-3}.

\section{Classical Burgers model: creep and stress relaxation tests}

The aim is to investigate the behavior of thermodynamically consistent
classical Burgers model (\ref{cbm}) in creep and stress relaxation tests. It
will be shown that the requirement for creep compliance to be the Bernstein
function and for relaxation modulus to be the completely monotonic function
narrows down the thermodynamical restriction (\ref{cbm-tdr}) to%
\begin{equation*}
\frac{a_{2}}{a_{1}}\leq \frac{a_{1}}{2}\left( 1-\sqrt{1-\frac{4a_{2}}{%
a_{1}^{2}}}\right) <\frac{b_{2}}{b_{1}}<\frac{a_{1}}{2}\left( 1+\sqrt{1-%
\frac{4a_{2}}{a_{1}^{2}}}\right) \leq a_{1},
\end{equation*}%
provided that $\frac{a_{1}^{2}}{a_{2}}\geq 4.$ Otherwise, still having the
thermodynamical requirements fulfilled, the creep compliance will prove to
be a non-negative, monotonically increasing, but convex function, lying
above its oblique asymptote, contrary to the case when it is a Bernstein
function when the creep compliance is a concave function lying below the
asymptote. If $\frac{b_{2}}{b_{1}}=\frac{a_{1}}{2}\left( 1\pm \sqrt{1-\frac{%
4a_{2}}{a_{1}^{2}}}\right) ,$ then the creep compliance is a linear,
increasing function having the same form as the asymptote. In any case it
starts from a finite value of strain and tends to infinity.

Apart from being completely monotonic, the relaxation modulus will prove to
be either non-monotonic function having a negative minimum if $\frac{a_{1}}{2%
}\left( 1+\sqrt{1-\frac{4a_{2}}{a_{1}^{2}}}\right) <\frac{b_{2}}{b_{1}}\leq
a_{1},$ or a non-negative, monotonically decreasing function that may change
its convexity if $\frac{a_{2}}{a_{1}}\leq \frac{b_{2}}{b_{1}}<\frac{a_{1}}{2}%
\left( 1-\sqrt{1-\frac{4a_{2}}{a_{1}^{2}}}\right) .$ Moreover, the
relaxation modulus can be an oscillatory function having decreasing
amplitude if $1\leq \frac{a_{1}^{2}}{a_{2}}<4.$ In any case it starts from a
finite value of stress and tends to zero.

The creep compliance in the form 
\begin{equation}
\varepsilon _{cr}\left( t\right) =\frac{a_{1}-\frac{b_{2}}{b_{1}}}{b_{1}}%
\left( 1-\mathrm{e}^{-\frac{b_{1}}{b_{2}}t}\right) +\frac{1}{b_{1}}t+\frac{%
a_{2}}{b_{2}}\mathrm{e}^{-\frac{b_{1}}{b_{2}}t},\;\;t\geq 0,  \label{cbm-cc}
\end{equation}%
having the glass compliance finite and equilibrium compliance infinite,
i.e., 
\begin{equation*}
\varepsilon _{cr}^{\left( g\right) }=\varepsilon _{cr}\left( 0^{+}\right) =%
\frac{a_{2}}{b_{2}}\;\;\text{and}\;\;\varepsilon _{cr}^{\left( e\right)
}=\lim_{t\rightarrow \infty }\varepsilon _{cr}\left( t\right) =\infty ,
\end{equation*}%
is obtained by: assuming $\sigma =H$ ($H$ is the Heaviside function);
applying the Laplace transform%
\begin{equation*}
\tilde{f}\left( s\right) =\mathcal{L}\left[ f\left( t\right) \right] \left(
s\right) =\int_{0}^{\infty }f\left( t\right) \mathrm{e}^{\mathrm{-}st}%
\mathrm{d}t,\;\;\func{Re}s>0,
\end{equation*}%
to the classical Burgers model (\ref{cbm}) yielding%
\begin{equation}
\tilde{\varepsilon}_{cr}\left( s\right) =\frac{1}{s}\frac{1+a_{1}s+a_{2}s^{2}%
}{b_{1}s+b_{2}s^{2}}=\frac{a_{1}-\frac{b_{2}}{b_{1}}}{b_{1}}\frac{1}{s}+%
\frac{1}{b_{1}}\frac{1}{s^{2}}+\left( \frac{a_{2}}{b_{2}}-\frac{a_{1}-\frac{%
b_{2}}{b_{1}}}{b_{1}}\right) \frac{1}{s+\frac{b_{1}}{b_{2}}};
\label{cbm-cr-lt}
\end{equation}%
and by the subsequent inversion of the Laplace transform in (\ref{cbm-cr-lt}%
). Due to the thermodynamical restriction (\ref{cbm-tdr}), the creep
compliance (\ref{cbm-cc}) is a non-negative function and since 
\begin{equation*}
\dot{\varepsilon}_{cr}\left( t\right) =\frac{1}{b_{1}}\left( 1-\mathrm{e}^{-%
\frac{b_{1}}{b_{2}}t}\right) +\frac{a_{1}b_{1}}{b_{2}^{2}}\left( \frac{b_{2}%
}{b_{1}}-\frac{a_{2}}{a_{1}}\right) \mathrm{e}^{-\frac{b_{1}}{b_{2}}%
t}>0,\;\;t\geq 0,
\end{equation*}%
where $\left( \cdot \right) ^{\cdot }=\frac{\mathrm{d}}{\mathrm{d}t}\left(
\cdot \right) ,$ it is also a monotonically increasing function, again due
to (\ref{cbm-tdr}). The second derivative of creep compliance (\ref{cbm-cc})%
\begin{equation}
\ddot{\varepsilon}_{cr}\left( t\right) =\frac{b_{1}}{b_{2}^{2}}\left( \frac{%
a_{2}}{b_{2}}-\frac{a_{1}-\frac{b_{2}}{b_{1}}}{b_{1}}\right) \mathrm{e}^{-%
\frac{b_{1}}{b_{2}}t},\;\;t\geq 0,  \label{cbm-cr-2-izvod}
\end{equation}%
is either non-negative or negative function for all $t\geq 0$ depending on
the sign of the term in brackets. The creep compliance is Bernstein function
if $\frac{a_{2}}{b_{2}}<\frac{1}{b_{1}}\left( a_{1}-\frac{b_{2}}{b_{1}}%
\right) ,$ since $\varepsilon _{cr}\left( t\right) >0,$ $\dot{\varepsilon}%
_{cr}\left( t\right) >0,$ $\ddot{\varepsilon}_{cr}\left( t\right) <0$ and
higher order derivatives are of alternating sign due to the exponential
function in (\ref{cbm-cr-2-izvod}). If $\frac{a_{2}}{b_{2}}>\frac{1}{b_{1}}%
\left( a_{1}-\frac{b_{2}}{b_{1}}\right) ,$ then the creep compliance is a
non-negative, monotonically increasing, convex function. In the case $\frac{%
a_{2}}{b_{2}}=\frac{1}{b_{1}}\left( a_{1}-\frac{b_{2}}{b_{1}}\right) ,$ the
creep compliance (\ref{cbm-cc}) becomes%
\begin{equation}
\varepsilon _{cr}\left( t\right) =\frac{a_{1}-\frac{b_{2}}{b_{1}}}{b_{1}}+%
\frac{1}{b_{1}}t=\frac{a_{2}}{b_{2}}+\frac{1}{b_{1}}t,\;\;t\geq 0.
\label{cbm-cc-lin}
\end{equation}

The oblique asymptote of creep compliance (\ref{cbm-cc}) takes the form 
\begin{equation}
g\left( t\right) =\frac{a_{1}-\frac{b_{2}}{b_{1}}}{b_{1}}+\frac{1}{b_{1}}%
t,\;\;\text{as}\;\;t\rightarrow \infty .  \label{obl-asympt}
\end{equation}%
If $\varepsilon _{cr}^{\left( g\right) }<g\left( 0\right) ,$ i.e., $\frac{%
a_{2}}{b_{2}}<\frac{1}{b_{1}}\left( a_{1}-\frac{b_{2}}{b_{1}}\right) ,$ ($%
\varepsilon _{cr}^{\left( g\right) }>g\left( 0\right) ,$ i.e., $\frac{a_{2}}{%
b_{2}}>\frac{1}{b_{1}}\left( a_{1}-\frac{b_{2}}{b_{1}}\right) $), then the
creep compliance is below (above) the oblique asymptote for all $t\geq 0,$
since it is a monotonically increasing concave (convex) function for all $%
t\geq 0.$ Note that if $\varepsilon _{cr}^{\left( g\right) }=g\left(
0\right) ,$ i.e. $,\frac{a_{2}}{b_{2}}=\frac{1}{b_{1}}\left( a_{1}-\frac{%
b_{2}}{b_{1}}\right) ,$ then the creep compliance coincides with the
asymptote for all $t\geq 0.$

The relaxation modulus, corresponding to the classical Burgers model (\ref%
{cbm}), is obtained by the Laplace transform method either as a
non-oscillatory%
\begin{eqnarray}
\sigma _{sr}\left( t\right) &=&\frac{b_{1}}{2a_{2}\nu }\left( \left(
1-\left( \mu -\nu \right) \frac{b_{2}}{b_{1}}\right) \mathrm{e}^{-\left( \mu
-\nu \right) t}+\left( \left( \mu +\nu \right) \frac{b_{2}}{b_{1}}-1\right) 
\mathrm{e}^{-\left( \mu +\nu \right) t}\right) ,\;\;\text{i.e.},
\label{cbm-sr-0} \\
\sigma _{sr}\left( t\right) &=&\frac{b_{2}}{a_{2}}\left( \cosh \left( \nu
t\right) +\left( \frac{b_{1}}{b_{2}}-\mu \right) \frac{\sinh \left( \nu
t\right) }{\nu }\right) \mathrm{e}^{-\mu t},\;\;t\geq 0,  \label{cbm-sr-1}
\end{eqnarray}%
or as an oscillatory function having decreasing amplitude%
\begin{equation}
\sigma _{sr}\left( t\right) =\frac{b_{2}}{a_{2}}\left( \cos \left( \omega
t\right) +\left( \frac{b_{1}}{b_{2}}-\mu \right) \frac{\sin \left( \omega
t\right) }{\omega }\right) \mathrm{e}^{-\mu t},\;\;t\geq 0,  \label{cbm-sr-2}
\end{equation}%
with 
\begin{equation}
\mu =\frac{a_{1}}{2a_{2}},\;\;\nu =\sqrt{\mu ^{2}-\frac{1}{a_{2}}},\;\;\text{%
and}\;\;\omega =\frac{\nu }{\mathrm{i}}=\sqrt{\frac{1}{a_{2}}-\mu ^{2}}.
\label{mi,ni,omega}
\end{equation}%
The relaxation moduli (\ref{cbm-sr-1}) and (\ref{cbm-sr-2}) yield finite
glass modulus and zero equilibrium modulus, i.e., 
\begin{equation*}
\sigma _{sr}^{\left( g\right) }=\sigma _{sr}\left( 0\right) =\frac{1}{%
\varepsilon _{cr}^{\left( g\right) }}=\frac{b_{2}}{a_{2}}\;\;\text{and}%
\;\;\sigma _{sr}^{\left( e\right) }=\lim_{t\rightarrow \infty }\sigma
_{sr}\left( t\right) =\frac{1}{\varepsilon _{cr}^{\left( e\right) }}=0.
\end{equation*}%
The Laplace transform applied to the classical Burgers model (\ref{cbm}),
with the assumption $\varepsilon =H,$ yields%
\begin{eqnarray}
\tilde{\sigma}_{sr}\left( s\right) &=&\frac{1}{a_{2}}\frac{b_{1}+b_{2}s}{%
s^{2}+\frac{a_{1}}{a_{2}}s+\frac{1}{a_{2}}}=\frac{1}{a_{2}}\frac{b_{1}+b_{2}s%
}{\left( s+\mu -\nu \right) \left( s+\mu +\nu \right) }  \notag \\
&=&\frac{1}{a_{2}}\left( \left( \frac{b_{2}}{2}+\frac{b_{1}-\mu b_{2}}{2\nu }%
\right) \frac{1}{s+\mu -\nu }+\left( \frac{b_{2}}{2}-\frac{b_{1}-\mu b_{2}}{%
2\nu }\right) \frac{1}{s+\mu +\nu }\right) ,  \label{cbm-sr-lt}
\end{eqnarray}%
so that (\ref{cbm-sr-0}) is obtained by the Laplace transform inversion in (%
\ref{cbm-sr-lt}) if $\mu ^{2}>\frac{1}{a_{2}},$ while (\ref{cbm-sr-2})
follows from (\ref{cbm-sr-1}) if $\mu ^{2}<\frac{1}{a_{2}}.$ If $\mu ^{2}=%
\frac{1}{a_{2}},$ then (\ref{cbm-sr-lt}) yields 
\begin{eqnarray}
\tilde{\sigma}_{sr}\left( s\right) &=&\frac{1}{a_{2}}\left( \frac{b_{2}}{%
s+\mu }+\frac{b_{1}-\mu b_{2}}{\left( s+\mu \right) ^{2}}\right) ,\;\;\text{%
i.e.,}  \notag \\
\sigma _{sr}\left( t\right) &=&\frac{b_{2}}{a_{2}}\left( 1+\left( \frac{b_{1}%
}{b_{2}}-\mu \right) t\right) \mathrm{e}^{-\mu t},\;\;t\geq 0.
\label{cbm-sr-1-prim}
\end{eqnarray}

Thermodynamical restriction (\ref{cbm-tdr}), rewritten as $\frac{a_{1}^{2}}{%
a_{2}}\geq 1,$ allows for both non-oscillatory and oscillatory forms of the
relaxation moduli (\ref{cbm-sr-1}) and (\ref{cbm-sr-2}), since in (\ref%
{mi,ni,omega}), the condition $\nu ^{2}=\frac{1}{4a_{2}}\left( \frac{%
a_{1}^{2}}{a_{2}}-4\right) \geq 0$ for obtaining (\ref{cbm-sr-1}) and (\ref%
{cbm-sr-1-prim}), along with the thermodynamical restriction $\frac{a_{1}^{2}%
}{a_{2}}\geq 1$, implies $\frac{a_{1}^{2}}{a_{2}}\geq 4,$ while the
condition $\nu ^{2}=\frac{1}{4a_{2}}\left( \frac{a_{1}^{2}}{a_{2}}-4\right)
<0$ for obtaining (\ref{cbm-sr-2}) implies $1\leq \frac{a_{1}^{2}}{a_{2}}<4.$

The non-oscillatory relaxation modulus (\ref{cbm-sr-0}), transformed into%
\begin{equation}
\sigma _{sr}\left( t\right) =\frac{b_{1}}{2a_{2}\nu }\left( \left( \mu -\nu
\right) \left( a_{2}\left( \mu +\nu \right) -\frac{b_{2}}{b_{1}}\right) 
\mathrm{e}^{-\left( \mu -\nu \right) t}+\left( \mu +\nu \right) \left( \frac{%
b_{2}}{b_{1}}-a_{2}\left( \mu -\nu \right) \right) \mathrm{e}^{-\left( \mu
+\nu \right) t}\right) ,  \label{cbm-sr-0-prim}
\end{equation}%
using $a_{2}\left( \mu ^{2}-\nu ^{2}\right) =1$ from (\ref{mi,ni,omega}), is
a completely monotonic function if the creep compliance (\ref{cbm-cc}) is a
Bernstein function. Namely, the condition $\frac{a_{2}}{b_{2}}<\frac{1}{b_{1}%
}\left( a_{1}-\frac{b_{2}}{b_{1}}\right) $ solved with respect to $\frac{%
b_{2}}{b_{1}}$ yields%
\begin{equation*}
a_{2}\left( \mu -\nu \right) <\frac{b_{2}}{b_{1}}<a_{2}\left( \mu +\nu
\right) ,
\end{equation*}%
which for all $t\geq 0$ implies $\sigma _{sr}\left( t\right) >0$ in (\ref%
{cbm-sr-0-prim}) and also implies that the derivatives of $\sigma _{sr}$
have the alternating sign due to the exponential function. The condition 
\begin{equation*}
\frac{a_{2}}{a_{1}}\leq a_{2}\left( \mu -\nu \right) <\frac{b_{2}}{b_{1}}%
<a_{2}\left( \mu +\nu \right) \leq a_{1}
\end{equation*}%
for creep compliance (\ref{cbm-cc}) to be the Bernstein function and
relaxation modulus (\ref{cbm-sr-1}) to be the completely monotonic function
is narrower than the thermodynamical restriction (\ref{cbm-tdr}) since%
\begin{eqnarray*}
a_{2}\left( \mu +\nu \right) &=&\frac{a_{1}}{2}\left( 1+\sqrt{1-\frac{4a_{2}%
}{a_{1}^{2}}}\right) \leq a_{1}\;\;\text{and} \\
a_{2}\left( \mu -\nu \right) &=&\frac{a_{1}}{2}\left( 1-\sqrt{1-\frac{4a_{2}%
}{a_{1}^{2}}}\right) =\frac{a_{1}}{2}\left( \frac{1}{2}\frac{4a_{2}}{%
a_{1}^{2}}+\sum_{k=2}^{\infty }\frac{\left( 2k-3\right) !!}{2^{k}k!}\left( 
\frac{4a_{2}}{a_{1}^{2}}\right) ^{k}\right) \geq \frac{a_{2}}{a_{1}},
\end{eqnarray*}%
where the binomial formula%
\begin{equation}
\sqrt{1-x}=1-\frac{1}{2}x-\sum_{k=2}^{\infty }\frac{\left( 2k-3\right) !!}{%
2^{k}k!}x^{k},\;\;\text{for}\;\;0<x<1,  \label{bin-form}
\end{equation}%
is used.

If the creep compliance is a linear function (\ref{cbm-cc-lin}), i.e., if $%
\frac{a_{2}}{b_{2}}=\frac{1}{b_{1}}\left( a_{1}-\frac{b_{2}}{b_{1}}\right) ,$
i.e., $\frac{b_{2}}{b_{1}}=a_{2}\left( \mu -\nu \right) ,$ or $\frac{b_{2}}{%
b_{1}}=a_{2}\left( \mu +\nu \right) ,$ then the relaxation modulus (\ref%
{cbm-sr-0}) reduces to%
\begin{equation*}
\sigma _{sr}\left( t\right) =\frac{b_{2}}{a_{2}}\mathrm{e}^{-\frac{b_{2}}{%
a_{2}}\frac{1}{b_{1}}t}.
\end{equation*}

Assume $a_{2}\left( \mu +\nu \right) <\frac{b_{2}}{b_{1}}\leq a_{1}.$ The
first derivative of relaxation modulus (\ref{cbm-sr-0-prim}) reads%
\begin{equation*}
\dot{\sigma}_{sr}\left( t\right) =-\frac{b_{1}}{2a_{2}\nu }\left( \mu +\nu
\right) ^{2}\left( \frac{b_{2}}{b_{1}}-a_{2}\left( \mu -\nu \right) \right)
\left( \mathrm{e}^{-2\nu t}-\left( \frac{\mu -\nu }{\mu +\nu }\right) ^{2}%
\frac{\frac{b_{2}}{b_{1}}-a_{2}\left( \mu +\nu \right) }{\frac{b_{2}}{b_{1}}%
-a_{2}\left( \mu -\nu \right) }\right) \mathrm{e}^{-\left( \mu -\nu \right)
t},
\end{equation*}%
with $0<\left( \frac{\mu -\nu }{\mu +\nu }\right) ^{2}\frac{\frac{b_{2}}{%
b_{1}}-a_{2}\left( \mu +\nu \right) }{\frac{b_{2}}{b_{1}}-a_{2}\left( \mu
-\nu \right) }\leq 1,$ due to the thermodynamical restriction $\frac{a_{2}}{%
a_{1}}\leq \frac{b_{2}}{b_{1}},$ see (\ref{cbm-tdr}). Since the exponential
function in the previous expression decreases from one to zero, at the
time-instant%
\begin{equation*}
t^{\ast }=\frac{1}{2\nu }\ln \left( \left( \frac{\mu +\nu }{\mu -\nu }%
\right) ^{2}\frac{\frac{b_{2}}{b_{1}}-a_{2}\left( \mu -\nu \right) }{\frac{%
b_{2}}{b_{1}}-a_{2}\left( \mu +\nu \right) }\right)
\end{equation*}%
the relaxation modulus has a minimum, since $\dot{\sigma}_{sr}$ changes the
sign from non-negative to negative at $t^{\ast }.$ This fact, along with the
finite glass and zero equilibrium modulus, implies that the relaxation
modulus decreases from $\sigma _{sr}^{\left( g\right) }=\frac{b_{2}}{a_{2}}$
to a negative minimum and further, being negative, increases to $\sigma
_{sr}^{\left( e\right) }=0.$

Assume $\frac{a_{2}}{a_{1}}\leq \frac{b_{2}}{b_{1}}<a_{2}\left( \mu -\nu
\right) .$ The first derivative of relaxation modulus (\ref{cbm-sr-0-prim})
is%
\begin{equation*}
\dot{\sigma}_{sr}\left( t\right) =-\frac{b_{1}}{2a_{2}\nu }\left( \mu -\nu
\right) ^{2}\left( a_{2}\left( \mu +\nu \right) -\frac{b_{2}}{b_{1}}\right)
\left( \mathrm{e}^{2\nu t}-\left( \frac{\mu +\nu }{\mu -\nu }\right) ^{2}%
\frac{a_{2}\left( \mu -\nu \right) -\frac{b_{2}}{b_{1}}}{a_{2}\left( \mu
+\nu \right) -\frac{b_{2}}{b_{1}}}\right) \mathrm{e}^{-\left( \mu +\nu
\right) t}<0,
\end{equation*}%
since, for $t>0,$ $\mathrm{e}^{2\nu t}>1$ and $0<\left( \frac{\mu +\nu }{\mu
-\nu }\right) ^{2}\frac{a_{2}\left( \mu -\nu \right) -\frac{b_{2}}{b_{1}}}{%
a_{2}\left( \mu +\nu \right) -\frac{b_{2}}{b_{1}}}\leq 1,$ due to the
thermodynamical restriction $\frac{a_{2}}{a_{1}}\leq \frac{b_{2}}{b_{1}},$
see (\ref{cbm-tdr}). Thus, the relaxation modulus, being a non-negative
function, monotonically decreases from $\sigma _{sr}^{\left( g\right) }=%
\frac{b_{2}}{a_{2}}$ to $\sigma _{sr}^{\left( e\right) }=0.$ However, the
relaxation modulus may change its convexity, due to 
\begin{equation*}
\left( \frac{\mu +\nu }{\mu -\nu }\right) ^{3}\frac{a_{2}\left( \mu -\nu
\right) -\frac{b_{2}}{b_{1}}}{a_{2}\left( \mu +\nu \right) -\frac{b_{2}}{%
b_{1}}}\geq 1\;\;\text{if}\;\;\frac{b_{2}}{b_{1}}\leq \frac{\frac{a_{2}}{%
a_{1}}}{1-\frac{a_{2}}{a_{1}^{2}}},
\end{equation*}%
that appears in the second derivative of the relaxation modulus%
\begin{equation*}
\ddot{\sigma}_{sr}\left( t\right) =\frac{b_{1}}{2a_{2}\nu }\left( \mu -\nu
\right) ^{3}\left( a_{2}\left( \mu +\nu \right) -\frac{b_{2}}{b_{1}}\right)
\left( \mathrm{e}^{2\nu t}-\left( \frac{\mu +\nu }{\mu -\nu }\right) ^{3}%
\frac{a_{2}\left( \mu -\nu \right) -\frac{b_{2}}{b_{1}}}{a_{2}\left( \mu
+\nu \right) -\frac{b_{2}}{b_{1}}}\right) \mathrm{e}^{-\left( \mu +\nu
\right) t}.
\end{equation*}

\section{Fractional Burgers models: creep and stress relaxation tests}

The fractional Burgers models I - VIII will be examined in creep and stress
relaxation tests. Models I - V, respectively given by (\ref{Model 1}), (\ref%
{Model 2}), (\ref{Model 3}), (\ref{Model 4}), and (\ref{Model 5}), have zero
glass compliance and thus infinite glass modulus, while Models VI - VIII,
respectively given by (\ref{Model 6}), (\ref{Model 7}), and (\ref{Model 8}),
behave similarly as the classical Burgers model (\ref{cbm}), having non-zero
glass compliance and thus glass modulus as well. In the case of both
classical and fractional Burgers models, the equilibrium compliance is
infinite and therefore equilibrium modulus is zero.

The Laplace transform method will be used in calculating the creep
compliance and relaxation modulus, so that both will be obtained in an
integral form using the definition of inverse Laplace transform, while the
creep compliance will additionally be expressed in terms of the
Mittag-Leffler function. In Section \ref{rrmp}, the integral forms will
prove to be useful in showing that the thermodynamical requirements (\ref%
{Model 1 - tdr}), (\ref{Model 2 - tdr}), (\ref{Model 3 - tdr}), (\ref{Model
4 - tdr}), (\ref{Model 5 - tdr}), (\ref{Model 6 - tdr}), (\ref{Model 7 - tdr}%
), and (\ref{Model 8 - tdr}) allow wider range of model parameters than the
range in which the creep compliance is a Bernstein function and the
relaxation modulus is a completely monotonic function, see (\ref{pos-K-MD1}%
), (\ref{pos-K-MD2}), (\ref{pos-K-MD3}), (\ref{pos-K-MD4}), (\ref{pos-K-MD5}%
), (\ref{Model 6 - nerov}), (\ref{Model 7 - nerov}), and (\ref{Model 8 -
nerov}). Hence, still being non-oscillatory, the creep compliance does not
have to be a monotonic function, while the relaxation modulus may even be
oscillatory function having decreasing amplitude if the model parameters
still fulfill the thermodynamical requirements.

Models I - V, i.e., models having zero glass compliance, can be written in
an unified manner as 
\begin{equation}
\left( 1+a_{1}\,{}_{0}\mathrm{D}_{t}^{\alpha }+a_{2}\,{}_{0}\mathrm{D}%
_{t}^{\beta }+a_{3}\,{}_{0}\mathrm{D}_{t}^{\gamma }\right) \sigma \left(
t\right) =\left( b_{1}\,{}_{0}\mathrm{D}_{t}^{\mu }+b_{2}\,{}_{0}\mathrm{D}%
_{t}^{\mu +\eta }\right) \varepsilon \left( t\right) ,  \label{UCE-1-5}
\end{equation}%
while Models VI - VIII, i.e., models having non-zero glass compliance, take
the following unified form 
\begin{equation}
\left( 1+a_{1}\,{}_{0}\mathrm{D}_{t}^{\alpha }+a_{2}\,{}_{0}\mathrm{D}%
_{t}^{\beta }+a_{3}\,{}_{0}\mathrm{D}_{t}^{\beta +\eta }\right) \sigma
\left( t\right) =\left( b_{1}\,{}_{0}\mathrm{D}_{t}^{\beta }+b_{2}\,{}_{0}%
\mathrm{D}_{t}^{\beta +\eta }\right) \varepsilon \left( t\right) ,
\label{UCE}
\end{equation}%
where in (\ref{UCE-1-5}) the highest order of fractional differentiation of
strain is $\mu +\eta \in \left[ 1,2\right] $ with $\eta \in \left\{ \alpha
,\beta \right\} ,$ while the highest order of fractional differentiation of
stress is either $\gamma \in \left[ 0,1\right] $ in the case of Model I (\ref%
{Model 1}), with $0\leq \alpha \leq \beta \leq \gamma \leq \mu \leq 1$ and $%
\eta =\varkappa \in \left\{ \alpha ,\beta ,\gamma \right\} ,$ or $\gamma \in %
\left[ 1,2\right] $ in the case of Models II - V, with $0\leq \alpha \leq
\beta \leq \mu \leq 1$ and $\left( \eta ,\gamma \right) \in \left\{ \left(
\alpha ,2\alpha \right) ,\left( \alpha ,\alpha +\beta \right) ,\left( \beta
,\alpha +\beta \right) ,\left( \beta ,2\beta \right) \right\} ,$ see (\ref%
{Model 2}), (\ref{Model 3}), (\ref{Model 4}), and (\ref{Model 5}), while in (%
\ref{UCE}) one has $0\leq \alpha \leq \beta \leq 1$ and $\beta +\eta \in %
\left[ 1,2\right] ,$ with $\eta =\alpha $ in the case of Model VI (\ref%
{Model 6}) and $\eta =\beta $ in the case of Model VII (\ref{Model 7}),
while Model VIII (\ref{Model 8}) is obtained for $\eta =\beta =\alpha ,$ $%
\bar{a}_{1}=a_{1}+a_{2},$ and $\bar{a}_{2}=a_{3}.$

\subsection{Creep compliance}

The creep compliances in complex domain%
\begin{eqnarray}
\tilde{\varepsilon}_{cr}\left( s\right) &=&\frac{1}{s}\frac{1+a_{1}s^{\alpha
}+a_{2}s^{\beta }+a_{3}s^{\gamma }}{b_{1}s^{\mu }+b_{2}s^{\mu +\eta }}=\frac{%
1}{s^{1+\mu }}\frac{\Psi \left( s\right) }{b_{1}+b_{2}s^{\eta }},
\label{cr-LT} \\
\tilde{\varepsilon}_{cr}\left( s\right) &=&\frac{1}{s}\frac{1+a_{1}s^{\alpha
}+a_{2}s^{\beta }+a_{3}s^{\beta +\eta }}{b_{1}s^{\beta }+b_{2}s^{\beta +\eta
}}=\frac{a_{3}}{b_{2}}\frac{1}{s}\left( 1+\frac{1}{s^{\beta }}\frac{\psi
\left( s\right) }{\frac{b_{1}}{b_{2}}+s^{\eta }}\right) ,  \label{cr-LT-1} \\
\tilde{\varepsilon}_{cr}\left( s\right) &=&\frac{1}{s}\frac{1+\bar{a}%
_{1}s^{\alpha }+\bar{a}_{2}s^{2\alpha }}{b_{1}s^{\alpha }+b_{2}s^{2\alpha }}=%
\frac{\bar{a}_{2}}{b_{2}}\frac{1}{s}\left( 1+\frac{1}{s^{\alpha }}\frac{\bar{%
\psi}\left( s\right) }{\frac{b_{1}}{b_{2}}+s^{\alpha }}\right) ,
\label{cr-LT-2}
\end{eqnarray}%
respectively corresponding to Models I - V, Models VI and VII, and Model
VIII, with functions $\Psi ,$ $\psi ,$ and $\bar{\psi},$ defined for $s\in 
\mathbb{C}
$ by respective expressions%
\begin{eqnarray}
\Psi \left( s\right) &=&1+a_{1}s^{\alpha }+a_{2}s^{\beta }+a_{3}s^{\gamma },
\label{psi} \\
\psi \left( s\right) &=&\frac{1}{a_{3}}+\frac{a_{1}}{a_{3}}s^{\alpha
}+\left( \frac{a_{2}}{a_{3}}-\frac{b_{1}}{b_{2}}\right) s^{\beta },
\label{psi1} \\
\bar{\psi}\left( s\right) &=&\frac{1}{\bar{a}_{2}}+\left( \frac{\bar{a}_{1}}{%
\bar{a}_{2}}-\frac{b_{1}}{b_{2}}\right) s^{\alpha },  \label{psi1-bar}
\end{eqnarray}%
are obtained by applying the Laplace transform to the unified constitutive
equations (\ref{UCE-1-5}) and (\ref{UCE}) having assumed that $\sigma =H.$
Note that $\frac{a_{2}}{a_{3}}-\frac{b_{1}}{b_{2}}\geq 0$ is due to the
thermodynamical restrictions (\ref{Model 6 - tdr}) and (\ref{Model 7 - tdr}%
), while $\frac{\bar{a}_{1}}{\bar{a}_{2}}-\frac{b_{1}}{b_{2}}\geq 0$ is due
to (\ref{Model 8 - tdr}).

The creep compliance at initial time-instant either starts at zero
deformation or has a jump, while it tends to an infinite deformation for
large time, as observed from the glass and equilibrium compliances%
\begin{eqnarray}
\varepsilon _{cr}^{\left( g\right) } &=&\varepsilon _{cr}\left( 0\right)
=\lim_{s\rightarrow \infty }\left( s\tilde{\varepsilon}_{cr}\left( s\right)
\right) =\left\{ 
\begin{tabular}{ll}
$0,$ & for Models I - V, \\ 
$\frac{a_{3}}{b_{2}},$ & for Models VI and VII, \\ 
$\frac{\bar{a}_{2}}{b_{2}},$ & for Model VIII,%
\end{tabular}%
\right.  \label{gg} \\
\varepsilon _{cr}^{\left( e\right) } &=&\lim_{t\rightarrow \infty
}\varepsilon _{cr}\left( t\right) =\lim_{s\rightarrow 0}\left( s\tilde{%
\varepsilon}_{cr}\left( s\right) \right) =\infty ,  \notag
\end{eqnarray}%
obtained from the Tauberian theorem with $\tilde{\varepsilon}_{cr}$ being
the creep compliance in complex domain (\ref{cr-LT}), or (\ref{cr-LT-1}), or
(\ref{cr-LT-2}).

Rewriting the creep compliances in complex domain (\ref{cr-LT}), (\ref%
{cr-LT-1}), and (\ref{cr-LT-2}) respectively as%
\begin{eqnarray*}
\tilde{\varepsilon}_{cr}\left( s\right) &=&\frac{1}{b_{2}}\frac{s^{\eta
-\left( 1+\mu +\eta \right) }}{\frac{b_{1}}{b_{2}}+s^{\eta }}+\frac{a_{1}}{%
b_{2}}\frac{s^{\eta -\left( 1+\mu +\eta -\alpha \right) }}{\frac{b_{1}}{b_{2}%
}+s^{\eta }}+\frac{a_{2}}{b_{2}}\frac{s^{\eta -\left( 1+\mu +\eta -\beta
\right) }}{\frac{b_{1}}{b_{2}}+s^{\eta }}+\frac{a_{3}}{b_{2}}\frac{s^{\eta
-\left( 1+\mu +\eta -\gamma \right) }}{\frac{b_{1}}{b_{2}}+s^{\eta }}, \\
\tilde{\varepsilon}_{cr}\left( s\right) &=&\frac{a_{3}}{b_{2}}\frac{1}{s}+%
\frac{1}{b_{2}}\frac{s^{\eta -\left( 1+\eta +\beta \right) }}{\frac{b_{1}}{%
b_{2}}+s^{\eta }}+\frac{a_{1}}{b_{2}}\frac{s^{\eta -\left( 1+\eta +\beta
-\alpha \right) }}{\frac{b_{1}}{b_{2}}+s^{\eta }}+\frac{a_{3}}{b_{2}}\left( 
\frac{a_{2}}{a_{3}}-\frac{b_{1}}{b_{2}}\right) \frac{s^{\eta -\left( 1+\eta
\right) }}{\frac{b_{1}}{b_{2}}+s^{\eta }}, \\
\tilde{\varepsilon}_{cr}\left( s\right) &=&\frac{\bar{a}_{2}}{b_{2}}\frac{1}{%
s}+\frac{1}{b_{2}}\frac{s^{\alpha -\left( 1+2\alpha \right) }}{\frac{b_{1}}{%
b_{2}}+s^{\alpha }}+\frac{\bar{a}_{2}}{b_{2}}\left( \frac{\bar{a}_{1}}{\bar{a%
}_{2}}-\frac{b_{1}}{b_{2}}\right) \frac{s^{\alpha -\left( 1+\alpha \right) }%
}{\frac{b_{1}}{b_{2}}+s^{\alpha }},
\end{eqnarray*}%
the creep compliance is expressed in terms of the two-parameter
Mittag-Leffler function\footnote{%
If $\zeta =1$ in (\ref{MLF}), the two-parameter Mittag-Leffler function
reduces to the one-parameter Mittag-Leffler function 
\begin{equation*}
e_{\eta ,\lambda }\left( t\right) =e_{\eta ,1,\lambda }\left( t\right)
=E_{_{\eta }}\left( -\lambda t^{\eta }\right) =\mathcal{L}^{-1}\left[ \frac{%
s^{\eta -1}}{s^{\eta }+\lambda }\right] \left( t\right) ,\;\;\text{with}%
\;\;E_{_{\eta }}\left( z\right) =\sum_{k=0}^{\infty }\frac{z^{k}}{\Gamma
\left( \eta k+1\right) }.
\end{equation*}%
}%
\begin{equation}
e_{\eta ,\zeta ,\lambda }\left( t\right) =t^{\zeta -1}E_{_{\eta ,\zeta
}}\left( -\lambda t^{\eta }\right) =\mathcal{L}^{-1}\left[ \frac{s^{\eta
-\zeta }}{s^{\eta }+\lambda }\right] \left( t\right) ,\;\;\text{with}%
\;\;E_{_{\eta ,\zeta }}\left( z\right) =\sum_{k=0}^{\infty }\frac{z^{k}}{%
\Gamma \left( \eta k+\zeta \right) },  \label{MLF}
\end{equation}%
see \cite{GoreMai}, as 
\begin{eqnarray}
\varepsilon _{cr}\left( t\right) &=&\frac{1}{b_{2}}e_{\eta ,1+\mu +\eta ,%
\frac{b_{1}}{b_{2}}}\left( t\right) +\frac{a_{1}}{b_{2}}e_{\eta ,1+\mu +\eta
-\alpha ,\frac{b_{1}}{b_{2}}}\left( t\right) +\frac{a_{2}}{b_{2}}e_{\eta
,1+\mu +\eta -\beta ,\frac{b_{1}}{b_{2}}}\left( t\right) +\frac{a_{3}}{b_{2}}%
e_{\eta ,1+\mu +\eta -\gamma ,\frac{b_{1}}{b_{2}}}\left( t\right) ,
\label{CR-1-5} \\
\varepsilon _{cr}\left( t\right) &=&\frac{a_{3}}{b_{2}}+\frac{1}{b_{2}}%
e_{\eta ,1+\eta +\beta ,\frac{b_{1}}{b_{2}}}\left( t\right) +\frac{a_{1}}{%
b_{2}}e_{\eta ,1+\eta +\beta -\alpha ,\frac{b_{1}}{b_{2}}}\left( t\right) +%
\frac{a_{3}}{b_{2}}\left( \frac{a_{2}}{a_{3}}-\frac{b_{1}}{b_{2}}\right)
e_{\eta ,1+\eta ,\frac{b_{1}}{b_{2}}}\left( t\right) ,  \label{CR} \\
\varepsilon _{cr}\left( t\right) &=&\frac{\bar{a}_{2}}{b_{2}}+\frac{1}{b_{2}}%
e_{\alpha ,1+2\alpha ,\frac{b_{1}}{b_{2}}}\left( t\right) +\frac{\bar{a}_{2}%
}{b_{2}}\left( \frac{\bar{a}_{1}}{\bar{a}_{2}}-\frac{b_{1}}{b_{2}}\right)
e_{\alpha ,1+\alpha ,\frac{b_{1}}{b_{2}}}\left( t\right) ,  \label{CR-1}
\end{eqnarray}%
respectively corresponding to Models I - V, Models VI and VII, and Model
VIII.

In the case of Models I - V, the integral form of creep compliance takes the
form 
\begin{equation}
\varepsilon _{cr}\left( t\right) =\frac{1}{\pi }\int_{0}^{\infty }\frac{%
K\left( \rho \right) }{\left\vert b_{1}+b_{2}\rho ^{\eta }\mathrm{e}^{%
\mathrm{i}\eta \pi }\right\vert ^{2}}\frac{1-\mathrm{e}^{-\rho t}}{\rho
^{1+\mu }}\mathrm{d}\rho ,  \label{CR-ILT-1-5}
\end{equation}%
where%
\begin{eqnarray}
K\left( \rho \right) &=&b_{1}K_{1}\left( \rho \right) +b_{2}\rho ^{\eta
}K_{2}\left( \rho \right) ,\;\;\text{with}  \label{funct-K} \\
K_{1}\left( \rho \right) &=&\sin \left( \mu \pi \right) +a_{1}\rho ^{\alpha
}\sin \left( \left( \mu -\alpha \right) \pi \right) +a_{2}\rho ^{\beta }\sin
\left( \left( \mu -\beta \right) \pi \right) +a_{3}\rho ^{\gamma }\sin
\left( \left( \mu -\gamma \right) \pi \right) ,  \notag \\
K_{2}\left( \rho \right) &=&\sin \left( \left( \mu +\eta \right) \pi \right)
+a_{1}\rho ^{\alpha }\sin \left( \left( \mu +\eta -\alpha \right) \pi
\right) +a_{2}\rho ^{\beta }\sin \left( \left( \mu +\eta -\beta \right) \pi
\right) +a_{3}\rho ^{\gamma }\sin \left( \left( \mu +\eta -\gamma \right)
\pi \right) ,  \notag
\end{eqnarray}%
while the creep compliance corresponding to Models VI and VII is given by 
\begin{equation}
\varepsilon _{cr}\left( t\right) =\frac{a_{3}}{b_{2}}+\frac{a_{3}}{b_{2}}%
\int_{0}^{t}f_{cr}\left( \tau \right) \mathrm{d}\tau ,\;\;\text{with}%
\;\;f_{cr}\left( t\right) =\frac{1}{\pi }\int_{0}^{\infty }\frac{Q\left(
\rho \right) }{\left\vert \frac{b_{1}}{b_{2}}+\rho ^{\eta }\mathrm{e}^{%
\mathrm{i}\eta \pi }\right\vert ^{2}}\frac{\mathrm{e}^{-\rho t}}{\rho
^{\beta }}\mathrm{d}\rho ,  \label{CR-ILT}
\end{equation}%
where $\varepsilon _{cr}^{\left( g\right) }=\frac{a_{3}}{b_{2}}$ is the
glass compliance (\ref{gg}), and where 
\begin{eqnarray}
Q\left( \rho \right) &=&\frac{b_{1}}{b_{2}}Q_{1}\left( \rho \right) +\rho
^{\eta }Q_{2}\left( \rho \right) ,\;\;\text{with}  \label{Q} \\
Q_{1}\left( \rho \right) &=&\frac{1}{a_{3}}\sin \left( \beta \pi \right) +%
\frac{a_{1}}{a_{3}}\rho ^{\alpha }\sin \left( \left( \beta -\alpha \right)
\pi \right) ,  \notag \\
Q_{2}\left( \rho \right) &=&\frac{1}{a_{3}}\sin \left( \left( \beta +\eta
\right) \pi \right) +\frac{a_{1}}{a_{3}}\rho ^{\alpha }\sin \left( \left(
\beta +\eta -\alpha \right) \pi \right) +\left( \frac{a_{2}}{a_{3}}-\frac{%
b_{1}}{b_{2}}\right) \rho ^{\beta }\sin \left( \eta \pi \right) .  \notag
\end{eqnarray}%
In the case of Model VIII, the creep compliance in integral form is obtained
as%
\begin{equation}
\varepsilon _{cr}\left( t\right) =\frac{\bar{a}_{2}}{b_{2}}+\frac{\bar{a}_{2}%
}{b_{2}}\int_{0}^{t}\bar{f}_{cr}\left( \tau \right) \mathrm{d}\tau ,\;\;%
\text{with}\;\;\bar{f}_{cr}\left( t\right) =\frac{1}{\pi }\int_{0}^{\infty }%
\frac{\bar{Q}\left( \rho \right) }{\left\vert \frac{b_{1}}{b_{2}}+\rho
^{\alpha }\mathrm{e}^{\mathrm{i}\alpha \pi }\right\vert ^{2}}\frac{\mathrm{e}%
^{-\rho t}}{\rho ^{\alpha }}\mathrm{d}\rho ,  \label{CR-ILT-1}
\end{equation}%
where $\varepsilon _{cr}^{\left( g\right) }=\frac{\bar{a}_{2}}{b_{2}}$ is
the glass compliance (\ref{gg}), and where 
\begin{gather}
\bar{Q}\left( \rho \right) =\frac{b_{1}}{b_{2}}\bar{Q}_{1}\left( \rho
\right) +\rho ^{\alpha }\bar{Q}_{2}\left( \rho \right) ,\;\;\text{with}
\label{Q-bar} \\
\bar{Q}_{1}\left( \rho \right) =\frac{1}{\bar{a}_{2}}\sin \left( \alpha \pi
\right) ,\;\;\bar{Q}_{2}\left( \rho \right) =\frac{1}{\bar{a}_{2}}\sin
\left( 2\alpha \pi \right) +\left( \frac{\bar{a}_{1}}{\bar{a}_{2}}-\frac{%
b_{1}}{b_{2}}\right) \rho ^{\alpha }\sin \left( \alpha \pi \right) .  \notag
\end{gather}

The the creep compliances in integral form (\ref{CR-ILT-1-5}), (\ref{CR-ILT}%
), and (\ref{CR-ILT-1}), respectively corresponding to Models I - V, Models
VI and VII, and Model VIII will be calculated in Section \ref{Cr-Calc} by
the definition of inverse Laplace transform using the creep compliance in
complex domain (\ref{cr-LT}), (\ref{cr-LT-1}), and (\ref{cr-LT-2}).

In order to prove that the thermodynamical restrictions are less restrictive
than the conditions for creep compliance to be the Bernstein function, the
creep compliances in integral form (\ref{CR-ILT-1-5}), (\ref{CR-ILT}), and (%
\ref{CR-ILT-1}) will be used. Namely, in the case of Models I - V, by
requiring the non-negativity of kernel $K$ (\ref{funct-K}), appearing in (%
\ref{CR-ILT-1-5}), one has 
\begin{equation}
\varepsilon _{cr}\left( t\right) \geq 0\;\;\text{and}\;\;\left( -1\right)
^{k}\frac{\mathrm{d}^{k}}{\mathrm{d}t^{k}}\dot{\varepsilon}_{cr}\left(
t\right) \geq 0,\;\;k\in 
\mathbb{N}
_{0},\;t>0,  \label{BF}
\end{equation}%
i.e., the creep compliance (\ref{CR-ILT-1-5}) is a Bernstein function. The
conditions for non-negativity of the kernel $K$ will be derived in Sections %
\ref{mdl-1} - \ref{mdl-5} and it will be proved that these conditions are
more restrictive than the corresponding thermodynamical restrictions. In the
case of Models VI and VII from (\ref{CR-ILT}) and for Model VIII from (\ref%
{CR-ILT-1}) one respectively has%
\begin{equation*}
\dot{\varepsilon}_{cr}\left( t\right) =\frac{a_{3}}{b_{2}}f_{cr}\left(
t\right) \;\;\text{and}\;\;\dot{\varepsilon}_{cr}\left( t\right) =\frac{\bar{%
a}_{2}}{b_{2}}\bar{f}_{cr}\left( t\right) .
\end{equation*}%
If functions $f_{cr}$ (\ref{CR-ILT})$_{2}$ and $\bar{f}_{cr}$ (\ref{CR-ILT-1}%
)$_{2}$ are completely monotonic, then (\ref{BF}) holds, i.e., the creep
compliances (\ref{CR-ILT}) and (\ref{CR-ILT-1}) are Bernstein functions. The
conditions for completely monotonicity of functions $f_{cr}$ and $\bar{f}%
_{cr}$ will be derived in Sections \ref{mdl-6} - \ref{mdl-8} by requiring
that the kernels $Q$ (\ref{Q}) and $\bar{Q}$ (\ref{Q-bar}) are non-negative
and it will be proved that these conditions are narrower than the
thermodynamical restrictions.

\subsection{Relaxation modulus}

Assuming $\varepsilon =H$ and by applying the Laplace transform to the
unified constitutive equation (\ref{UCE-1-5}) in the case of Models I - V,
as well as to the unified constitutive equation (\ref{UCE}) in the case of
Models VI and VII, and Model VII, the relaxation moduli in complex domain
take the respective forms%
\begin{eqnarray}
\tilde{\sigma}_{sr}\left( s\right) &=&\frac{1}{s}\frac{b_{1}s^{\mu
}+b_{2}s^{\mu +\eta }}{1+a_{1}s^{\alpha }+a_{2}s^{\beta }+a_{3}s^{\gamma }}=%
\frac{1}{s^{1-\mu }}\frac{b_{1}+b_{2}s^{\eta }}{\Psi \left( s\right) },
\label{sr-lap} \\
\tilde{\sigma}_{sr}\left( s\right) &=&\frac{1}{s}\frac{b_{1}s^{\beta
}+b_{2}s^{\beta +\eta }}{1+a_{1}s^{\alpha }+a_{2}s^{\beta }+a_{3}s^{\beta
+\eta }}=\frac{b_{2}}{a_{3}}\frac{1}{s}\left( 1-\frac{\psi \left( s\right) }{%
\psi \left( s\right) +s^{\beta }\left( \frac{b_{1}}{b_{2}}+s^{\eta }\right) }%
\right) ,  \label{sr-lap-1} \\
\tilde{\sigma}_{sr}\left( s\right) &=&\frac{1}{s}\frac{b_{1}s^{\alpha
}+b_{2}s^{2\alpha }}{1+\bar{a}_{1}s^{\alpha }+\bar{a}_{2}s^{2\alpha }}=\frac{%
b_{2}}{\bar{a}_{2}}\frac{1}{s}\left( 1-\frac{\bar{\psi}\left( s\right) }{%
\bar{\psi}\left( s\right) +s^{\alpha }\left( \frac{b_{1}}{b_{2}}+s^{\alpha
}\right) }\right) ,  \label{sr-lap-2}
\end{eqnarray}%
with functions $\Psi ,$ $\psi ,$ and $\bar{\psi}$ given by (\ref{psi}), (\ref%
{psi1}), (\ref{psi1-bar}), respectively. Note that functions in the
denominator of relaxation moduli in complex domain (\ref{sr-lap-1}) and (\ref%
{sr-lap-2}) are (up to multiplication with constant $a_{3}$ or $\bar{a}_{2}$%
) either function $\Psi ,$ with $\gamma =\beta +\eta ,$ given by (\ref{psi}%
), or the following function of complex variable%
\begin{equation}
\Phi \left( s\right) =1+\bar{a}_{1}s^{\alpha }+\bar{a}_{2}s^{2\alpha }.
\label{fi}
\end{equation}

The relaxation modulus at initial time-instant either tends to infinity, or
has a jump, while it tends to zero for large time, since by the Tauberian
theorem glass and equilibrium moduli are%
\begin{eqnarray}
\sigma _{sr}^{\left( g\right) } &=&\sigma _{sr}\left( 0\right)
=\lim_{s\rightarrow \infty }\left( s\tilde{\sigma}_{sr}\left( s\right)
\right) =\left\{ 
\begin{tabular}{ll}
$\infty $ & for Models I - V, \\ 
$\frac{b_{2}}{a_{3}},$ & for Models VI and VII, \\ 
$\frac{b_{2}}{\bar{a}_{2}},$ & for Model VIII,%
\end{tabular}%
\right.  \label{jg} \\
\sigma _{sr}^{\left( e\right) } &=&\lim_{t\rightarrow \infty }\sigma
_{sr}\left( t\right) =\lim_{s\rightarrow 0}\left( s\tilde{\sigma}_{sr}\left(
s\right) \right) =0,  \label{je}
\end{eqnarray}%
where $\tilde{\sigma}_{sr}$ is the relaxation modulus in complex domain,
given by (\ref{sr-lap}), or (\ref{sr-lap-1}), or (\ref{sr-lap-2}).

In the case of Models I - V, by inverting the Laplace transform in (\ref%
{sr-lap}), the relaxation modulus is obtained in the integral form as%
\begin{eqnarray}
\sigma _{sr}\left( t\right) &=&\frac{1}{\pi }\int_{0}^{\infty }\frac{K\left(
\rho \right) }{\left\vert \Psi \left( \rho \mathrm{e}^{\mathrm{i}\pi
}\right) \right\vert ^{2}}\frac{\mathrm{e}^{-\rho t}}{\rho ^{1-\mu }}\mathrm{%
d}\rho +g_{sr}\left( t\right) ,\;\;\text{with}  \label{SR} \\
g\left( t\right) &=&\left\{ 
\begin{tabular}{ll}
$0,$ & in Case 1, \\ 
$f_{sr}^{\ast }\left( \rho ^{\ast }\right) \mathrm{e}^{-\rho ^{\ast }t},$ & 
in Case 2, \\ 
$f_{sr}^{\left( r\right) }\left( t\right) ,$ & in Case 3,%
\end{tabular}%
\right.  \label{SRF-1-5}
\end{eqnarray}%
where functions $K$ and $\Psi $ are given by (\ref{funct-K}) and (\ref{psi}%
), $\rho ^{\ast }$ is determined from the equation%
\begin{equation}
\frac{a_{1}\sin \left( \alpha \pi \right) }{a_{3}\left\vert \sin \left(
\gamma \pi \right) \right\vert }+\frac{a_{2}\sin \left( \beta \pi \right) }{%
a_{3}\left\vert \sin \left( \gamma \pi \right) \right\vert }\left( \rho
^{\ast }\right) ^{\beta -\alpha }=\left( \rho ^{\ast }\right) ^{\gamma
-\alpha },  \label{rho-zvezda-1-5}
\end{equation}%
while functions $f_{sr}^{\ast }$ and $f_{sr}^{\left( r\right) }$ are given by%
\begin{eqnarray}
f_{sr}^{\ast }\left( \rho ^{\ast }\right) &=&\func{Re}\left( \frac{%
b_{1}+b_{2}\left( \rho ^{\ast }\right) ^{\eta }\mathrm{e}^{\mathrm{i}\eta
\pi }}{\alpha a_{1}\left( \rho ^{\ast }\right) ^{\alpha }\mathrm{e}^{\mathrm{%
i}\alpha \pi }+\beta a_{2}\left( \rho ^{\ast }\right) ^{\beta }\mathrm{e}^{%
\mathrm{i}\beta \pi }+\gamma a_{3}\left( \rho ^{\ast }\right) ^{\gamma }%
\mathrm{e}^{\mathrm{i}\gamma \pi }}\mathrm{e}^{\mathrm{i}\mu \pi }\right)
\left( \rho ^{\ast }\right) ^{\mu },  \label{f-sr-zvezda-1-5} \\
f_{sr}^{\left( r\right) }\left( t\right) &=&2\mathrm{e}^{\rho _{0}t\cos
\varphi _{0}}\func{Re}\left( \left. \frac{b_{1}+b_{2}s^{\eta }}{\alpha
a_{1}s^{\alpha -1}+\beta a_{2}s^{\beta -1}+\gamma a_{3}s^{\gamma -1}}%
\right\vert _{s=\rho _{0}\mathrm{e}^{\mathrm{i}\varphi _{0}}}\frac{\mathrm{e}%
^{-\mathrm{i}\left( 1-\mu \right) \varphi _{0}}}{\rho _{0}^{1-\mu }}\mathrm{e%
}^{\mathrm{i}\rho _{0}t\sin \varphi _{0}}\right) ,  \label{f-sr-res-1-5}
\end{eqnarray}%
with $s_{0}=\rho _{0}\mathrm{e}^{\mathrm{i}\varphi _{0}},$ $\varphi _{0}\in
\left( \frac{\pi }{2},\pi \right) ,$ being one of the complex conjugated
zeros of function $\Psi ,$ given by (\ref{psi}), while the relaxation
modulus in integral form corresponding to Models VI and VII is obtained
using the relaxation modulus in complex domain (\ref{sr-lap-1}) as 
\begin{eqnarray}
\sigma _{sr}\left( t\right) &=&\frac{b_{2}}{a_{3}}-\frac{b_{2}}{a_{3}}%
\int_{0}^{t}f_{sr}\left( \tau \right) \mathrm{d}\tau -\frac{b_{2}}{a_{3}}%
g_{sr}\left( t\right) ,\;\;\text{with}  \label{SR-ILT} \\
f_{sr}\left( t\right) &=&\frac{1}{\pi }\int_{0}^{\infty }\frac{Q\left( \rho
\right) }{\left\vert \psi \left( \rho \mathrm{e}^{\mathrm{i}\pi }\right)
+\rho ^{\beta }\mathrm{e}^{\mathrm{i}\beta \pi }\left( \frac{b_{1}}{b_{2}}%
+\rho ^{\eta }\mathrm{e}^{\mathrm{i}\eta \pi }\right) \right\vert ^{2}}\rho
^{\beta }\mathrm{e}^{-\rho t}\mathrm{d}\rho \;\;\text{and}  \label{f-sr} \\
g_{sr}\left( t\right) &=&\left\{ 
\begin{tabular}{ll}
$0,$ & in Case 1, \\ 
$-f_{sr}^{\ast }\left( \rho ^{\ast }\right) \left( 1-\mathrm{e}^{-\rho
^{\ast }t}\right) ,$ & in Case 2, \\ 
$\int_{0}^{t}f_{sr}^{\left( r\right) }\left( \tau \right) \mathrm{d}\tau ,$
& in Case 3,%
\end{tabular}%
\right.  \label{SRF}
\end{eqnarray}%
where $\sigma _{sr}^{\left( g\right) }=\frac{1}{\varepsilon _{cr}^{\left(
g\right) }}=\frac{b_{2}}{a_{3}}$ is the glass modulus (\ref{jg}), function $%
Q $ is given by (\ref{Q}), $\rho ^{\ast }$ is determined from the equation (%
\ref{rho-zvezda-1-5}) with $\gamma =\beta +\eta ,$ while functions $%
f_{sr}^{\ast }$ and $f_{sr}^{\left( r\right) }$ are given by%
\begin{eqnarray}
f_{sr}^{\ast }\left( \rho ^{\ast }\right) &=&\func{Re}\left( \frac{%
1+a_{1}\left( \rho ^{\ast }\right) ^{\alpha }\mathrm{e}^{\mathrm{i}\alpha
\pi }+a_{3}\left( \frac{a_{2}}{a_{3}}-\frac{b_{1}}{b_{2}}\right) \left( \rho
^{\ast }\right) ^{\beta }\mathrm{e}^{\mathrm{i}\beta \pi }}{\alpha
a_{1}\left( \rho ^{\ast }\right) ^{\alpha }\mathrm{e}^{\mathrm{i}\alpha \pi
}+\beta a_{2}\left( \rho ^{\ast }\right) ^{\beta }\mathrm{e}^{\mathrm{i}%
\beta \pi }+\left( \beta +\eta \right) a_{3}\left( \rho ^{\ast }\right)
^{\beta +\eta }\mathrm{e}^{\mathrm{i}\left( \beta +\eta \right) \pi }}%
\right) ,  \label{f-sr-zvezda} \\
f_{sr}^{\left( r\right) }\left( t\right) &=&2\mathrm{e}^{\rho _{0}t\cos
\varphi _{0}}\func{Re}\left( \left. \frac{1+a_{1}s^{\alpha }+a_{3}\left( 
\frac{a_{2}}{a_{3}}-\frac{b_{1}}{b_{2}}\right) s^{\beta }}{\alpha
a_{1}s^{\alpha -1}+\beta a_{2}s^{\beta -1}+\left( \beta +\eta \right)
a_{3}s^{\beta +\eta -1}}\right\vert _{s=\rho _{0}\mathrm{e}^{\mathrm{i}%
\varphi _{0}}}\mathrm{e}^{\mathrm{i}\rho _{0}t\sin \varphi _{0}}\right) ,
\label{f-sr-res}
\end{eqnarray}%
with $s_{0}=\rho _{0}\mathrm{e}^{\mathrm{i}\varphi _{0}},$ $\varphi _{0}\in
\left( \frac{\pi }{2},\pi \right) ,$ being one of the complex conjugated
zeros of function $\Psi ,$ given by (\ref{psi}). In the case of Model VIII,
the relaxation modulus is also obtained by the Laplace transform inversion
of the relaxation modulus in complex domain (\ref{sr-lap-2}), and it takes
the following form%
\begin{eqnarray}
\sigma _{sr}\left( t\right) &=&\frac{b_{2}}{\bar{a}_{2}}-\frac{b_{2}}{\bar{a}%
_{2}}\int_{0}^{t}\bar{f}_{sr}\left( \tau \right) \mathrm{d}\tau -\frac{b_{2}%
}{a_{3}}\bar{g}_{sr}\left( t\right) ,\;\;\text{with}  \label{SR-ILT-1} \\
\bar{f}_{sr}\left( t\right) &=&\frac{1}{\pi }\int_{0}^{\infty }\frac{\bar{Q}%
\left( \rho \right) }{\left\vert \bar{\psi}\left( \rho \mathrm{e}^{\mathrm{i}%
\pi }\right) +\rho ^{\alpha }\mathrm{e}^{\mathrm{i}\alpha \pi }\left( \frac{%
b_{1}}{b_{2}}+\rho ^{\alpha }\mathrm{e}^{\mathrm{i}\alpha \pi }\right)
\right\vert ^{2}}\rho ^{\alpha }\mathrm{e}^{-\rho t}\mathrm{d}\rho \;\;\text{%
and}  \label{f-sr-1} \\
\bar{g}_{sr}\left( t\right) &=&\left\{ 
\begin{tabular}{ll}
$0,$ & in Case 1, \\ 
$-\bar{f}_{sr}^{\ast }\left( \rho ^{\ast }\right) \left( 1-\mathrm{e}^{-\rho
^{\ast }t}\right) ,$ & in Case 2, \\ 
$\int_{0}^{t}\bar{f}_{sr}^{\left( r\right) }\left( \tau \right) \mathrm{d}%
\tau ,$ & in Case 3,%
\end{tabular}%
\right.  \label{SRF-1}
\end{eqnarray}%
where $\sigma _{sr}^{\left( g\right) }=\frac{1}{\varepsilon _{cr}^{\left(
g\right) }}=\frac{b_{2}}{\bar{a}_{2}}$ is the glass modulus (\ref{jg}),
function $\bar{Q}$ is given by (\ref{Q-bar}), $\rho ^{\ast }$ is determined
by%
\begin{equation*}
\rho ^{\ast }=\left( \frac{1}{\sin \left( \alpha \pi \right) }\sqrt{\frac{1}{%
\bar{a}_{2}}-\left( \frac{\bar{a}_{1}}{2\bar{a}_{2}}\right) ^{2}}\right) ^{%
\frac{1}{\alpha }},
\end{equation*}%
while functions $\bar{f}_{sr}^{\ast }$ and $\bar{f}_{sr}^{\left( r\right) }$
are given by%
\begin{eqnarray}
\bar{f}_{sr}^{\ast }\left( \rho ^{\ast }\right) &=&\frac{1}{\alpha \left(
\rho ^{\ast }\right) ^{\alpha }}\func{Re}\left( \frac{1+\bar{a}_{2}\left( 
\frac{\bar{a}_{1}}{\bar{a}_{2}}-\frac{b_{1}}{b_{2}}\right) \left( \rho
^{\ast }\right) ^{\alpha }\mathrm{e}^{\mathrm{i}\alpha \pi }}{\bar{a}_{1}+2%
\bar{a}_{2}\left( \rho ^{\ast }\right) ^{\alpha }\mathrm{e}^{\mathrm{i}%
\alpha \pi }}\mathrm{e}^{-\mathrm{i}\alpha \pi }\right) ,  \notag \\
\bar{f}_{sr}^{\left( r\right) }\left( t\right) &=&2\mathrm{e}^{\rho
_{0}t\cos \varphi _{0}}\func{Re}\left( \left. \frac{1+\bar{a}_{2}\left( 
\frac{\bar{a}_{1}}{\bar{a}_{2}}-\frac{b_{1}}{b_{2}}\right) s^{\alpha }}{%
\alpha s^{\alpha -1}\left( \bar{a}_{1}+2\bar{a}_{2}s^{\alpha }\right) }%
\right\vert _{s=\rho _{0}\mathrm{e}^{\mathrm{i}\varphi _{0}}}\mathrm{e}^{%
\mathrm{i}\rho _{0}t\sin \varphi _{0}}\right) ,  \label{f-sr-res-1}
\end{eqnarray}%
with $s_{0}=\rho _{0}\mathrm{e}^{\mathrm{i}\varphi _{0}},$ $\varphi _{0}\in
\left( \frac{\pi }{2},\pi \right) ,$ being one of the complex conjugated
zeros of function $\Phi ,$ given by (\ref{fi}). The calculation of the
relaxation moduli (\ref{SR}), (\ref{SR-ILT}), and (\ref{SR-ILT-1}) by the
definition of the inverse Laplace transform and integration in the complex
plane will be given in Section \ref{SR-Calc}.

In Cases 1 and 2, the relaxation moduli (\ref{SR}), (\ref{SR-ILT}), and (\ref%
{SR-ILT-1}) may have a non-monotonic behavior, due to possibly non-monotonic
behavior of functions $f_{sr}$ (\ref{f-sr}) and $\bar{f}_{sr}$ (\ref{f-sr-1}%
), even though functions $g_{sr}$ in (\ref{SRF-1-5}), $g_{sr}$ in (\ref{SRF}%
), and $\bar{g}_{sr}$ in (\ref{SRF-1}) are monotonic. However, for Models I
- V the relaxation modulus (\ref{SR}) in Case 1 is a completely monotonic
function in the range of model parameters narrower than the thermodynamical
requirements, which is the same range as for the creep compliance (\ref%
{CR-ILT-1-5}) to be a Bernstein function, since%
\begin{equation*}
\sigma _{sr}\left( t\right) \geq 0\;\;\text{and}\;\;\left( -1\right) ^{k}%
\frac{\mathrm{d}^{k}}{\mathrm{d}t^{k}}\dot{\sigma}_{sr}\left( t\right) \leq
0,\;\;k\in 
\mathbb{N}
_{0},\;t>0,
\end{equation*}%
provided that kernel $K,$ given by (\ref{funct-K}), is non-negative. Also,
for each of Models VI - VIII the relaxation moduli (\ref{SR-ILT}) and (\ref%
{SR-ILT-1}) in Case 1 are completely monotonic functions in the same, more
restrictive domain of model parameters when the creep compliance is the
Bernstein function. Namely, in the case of Models VI and VII from (\ref%
{SR-ILT}) and in the case of Model VIII from (\ref{SR-ILT-1}), one
respectively has%
\begin{equation*}
\dot{\sigma}_{sr}\left( t\right) =-\frac{b_{2}}{a_{3}}f_{sr}\left( t\right)
\;\;\text{and}\;\;\dot{\sigma}_{sr}\left( t\right) =-\frac{b_{2}}{\bar{a}_{2}%
}\bar{f}_{sr}\left( t\right) .
\end{equation*}%
If functions $f_{sr}$ (\ref{f-sr}) and $\bar{f}_{cr}$ (\ref{f-sr-1}) are
completely monotonic, then $\left( -1\right) ^{k}\frac{\mathrm{d}^{k}}{%
\mathrm{d}t^{k}}\dot{\sigma}_{sr}\left( t\right) \leq 0,$ $k\in 
\mathbb{N}
_{0},$ $t>0,$ and also $\sigma _{sr}\left( t\right) \geq 0,$ $t>0,$ since $%
\sigma _{sr}$ monotonically decreases from $\sigma _{sr}^{\left( g\right) }=%
\frac{b_{2}}{a_{3}},$ in the case of Models VI and VII, or from $\sigma
_{sr}^{\left( g\right) }=\frac{b_{2}}{\bar{a}_{2}}$ in the case of Model
VIII, to $\sigma _{sr}^{\left( e\right) }=0,$ see (\ref{jg}) and (\ref{je}),
implying that the relaxation moduli (\ref{SR-ILT}) and (\ref{SR-ILT-1}) in
Case 1 are completely monotonic functions. The conditions for complete
monotonicity of functions $f_{sr}$ and $\bar{f}_{sr}$ are the same as for
the functions $f_{cr}$ and $\bar{f}_{cr},$ since they depend on the same
kernels, so that in the same range of model parameters, narrower than the
thermodynamical restrictions, the relaxation modulus is a completely
monotonic function and the creep compliance is a Bernstein function. In Case
3, the relaxation moduli (\ref{SR}), (\ref{SR-ILT}), and (\ref{SR-ILT-1})
have damped oscillatory behavior, since functions $g_{sr}$ in (\ref{SRF-1-5}%
), $g_{sr}$ in (\ref{SRF}), and $\bar{g}_{sr}$ in (\ref{SRF-1}) are
oscillatory with decreasing amplitude, due to functions $f_{sr}^{\left(
r\right) }$ (\ref{f-sr-res-1-5}), $f_{sr}^{\left( r\right) }$ (\ref{f-sr-res}%
), and $\bar{f}_{sr}^{\left( r\right) }$ (\ref{f-sr-res-1}).

Cases in functions $g_{sr}$ (\ref{SRF-1-5}), $g_{sr}$ (\ref{SRF}), and $\bar{%
g}_{sr}$ (\ref{SRF-1}) that appear in the relaxation moduli (\ref{SR}), (\ref%
{SR-ILT}), and (\ref{SR-ILT-1}) are determined according to the number and
position of zeros of function $\Psi $\ (\ref{psi}) in cases of Models I -
VII and of function $\Phi \ $(\ref{fi}) in the case of Model VIII. Namely,
in Case 1 function $\Psi $ ($\Phi $)\ has no zeros in the complex plane,
while in Case 2 it has a negative real zero $-\rho ^{\ast },$\ and in Case 3
function $\Psi $\ ($\Phi $) has a pair of complex conjugated zeros $s_{0}$\
and $\bar{s}_{0}$ having negative real part. In the case of Model I, since $%
\gamma \in \left[ 0,1\right] ,$ function $\Psi $ has no zeros in the complex
plane, implying only non-oscillatory behavior of the relaxation modulus in
time domain, i.e., there exists only Case 1 in (\ref{SRF-1-5}). In cases of
Models II - V, since $\gamma \in \left[ 1,2\right] ,$ and in cases of Models
VI and VII, since $\gamma =\beta +\eta \in \left[ 1,2\right] ,$ the number
and position of zeros of function $\Psi $\ is as follows%
\begin{equation*}
\begin{tabular}{lll}
Case 1: & if $\func{Re}\Psi \left( \rho ^{\ast }\right) <0,$ & 
\begin{tabular}{l}
then $\Psi $ has no zeros in the complex plane;%
\end{tabular}
\\ 
Case 2: & if $\func{Re}\Psi \left( \rho ^{\ast }\right) =0,$ & 
\begin{tabular}{l}
then $\Psi $ has one negative real zero $-\rho ^{\ast }$;%
\end{tabular}
\\ 
Case 3: & if $\func{Re}\Psi \left( \rho ^{\ast }\right) >0,$ & 
\begin{tabular}{l}
then $\Psi $ has a pair of complex conjugated \\ 
zeros $s_{0}$ and $\bar{s}_{0}$ having negative real part;%
\end{tabular}%
\end{tabular}%
\end{equation*}%
where%
\begin{equation*}
\func{Re}\Psi \left( \rho ^{\ast }\right) =1+a_{1}\left( \rho ^{\ast
}\right) ^{\alpha }\cos \left( \alpha \pi \right) +a_{2}\left( \rho ^{\ast
}\right) ^{\beta }\cos \left( \beta \pi \right) +a_{3}\left( \rho ^{\ast
}\right) ^{\gamma }\cos \left( \gamma \pi \right) ,
\end{equation*}%
with $\rho ^{\ast }$ determined from (\ref{rho-zvezda-1-5}), allowing for
both non-oscillatory and oscillatory behavior of the relaxation modulus. The
analysis of the number and position of zeros of function $\Psi $ (\ref{psi})
using the argument principle is performed in Section \ref{CFpsi}. In the
case of Model VIII it will be shown in Section \ref{CFfi} that the following
holds true for function $\Phi ,$ given by (\ref{fi}):%
\begin{equation*}
\begin{tabular}{lll}
Case 1: & 
\begin{tabular}{l}
if $\left( \frac{\bar{a}_{1}}{2\bar{a}_{2}}\right) ^{2}\geq \frac{1}{\bar{a}%
_{2}},$ or \\ 
if $\left( \frac{\bar{a}_{1}}{2\bar{a}_{2}}\right) ^{2}<\frac{1}{\bar{a}_{2}}
$ and $a<b\frac{\left\vert \cos \left( \alpha \pi \right) \right\vert }{\sin
\left( \alpha \pi \right) },$%
\end{tabular}
& 
\begin{tabular}{l}
then $\Phi $ has no zeros in the complex plane;%
\end{tabular}
\\ 
Case 2: & 
\begin{tabular}{l}
if $\left( \frac{\bar{a}_{1}}{2\bar{a}_{2}}\right) ^{2}<\frac{1}{\bar{a}_{2}}
$ and $a=b\frac{\left\vert \cos \left( \alpha \pi \right) \right\vert }{\sin
\left( \alpha \pi \right) },$%
\end{tabular}
& 
\begin{tabular}{l}
then $\Phi $ has one negative real zero $-\rho ^{\ast }$ \\ 
determined by $\rho ^{\ast }=\left( \frac{b}{\sin \left( \alpha \pi \right) }%
\right) ^{\frac{1}{\alpha }}$;%
\end{tabular}
\\ 
Case 3: & 
\begin{tabular}{l}
if $\left( \frac{\bar{a}_{1}}{2\bar{a}_{2}}\right) ^{2}<\frac{1}{\bar{a}_{2}}
$ and $a>b\frac{\left\vert \cos \left( \alpha \pi \right) \right\vert }{\sin
\left( \alpha \pi \right) },$%
\end{tabular}
& 
\begin{tabular}{l}
then $\Phi $ has a pair of complex conjugated \\ 
zeros $s_{0}$ and $\bar{s}_{0}$ having negative real part;%
\end{tabular}%
\end{tabular}%
\end{equation*}%
where $a=\frac{\bar{a}_{1}}{2\bar{a}_{2}},$ and $b=\sqrt{\frac{1}{\bar{a}_{2}%
}-\left( \frac{\bar{a}_{1}}{2\bar{a}_{2}}\right) ^{2}}.$

\section{Restrictions on range of model parameters and asymptotics \label%
{rrmp}}

Kernels $K,$ $Q,$ and $\bar{Q},$ respectively given by (\ref{funct-K}), (\ref%
{Q}), and (\ref{Q-bar}), that appear in creep compliances (\ref{CR-ILT-1-5}%
), (\ref{CR-ILT}), and (\ref{CR-ILT-1}) and relaxation moduli (\ref{SR}), (%
\ref{SR-ILT}), and (\ref{SR-ILT-1}), respectively corresponding to Models I
- V, Models VI and VII, and Model VIII, will be examined. Namely, by
requiring kernels' non-negativity, the range of model parameters in which
the creep compliance is a Bernstein function, while the relaxation modulus
is completely monotonic, will be explicitly obtained. Moreover, it will be
proved that such obtained range is narrower than the corresponding
thermodynamical restriction.

The asymptotic analysis will reveal that in the vicinity of initial
time-instant creep compliance starts from the zero value of deformation and
increases: proportionally to $t^{\mu -\gamma +\varkappa },$ with $\varkappa
\in \left\{ \alpha ,\beta ,\gamma \right\} ,$ in the case of Model I, see (%
\ref{MD1-krip-0}); proportionally to $t^{\mu -\alpha }$ in the case of
Models II and IV, see (\ref{MD2-krip-0}) and (\ref{MD4-krip-0}); and
proportionally to $t^{\mu -\beta }$ in the case of Models III and V, see (%
\ref{MD3-krip-0}) and (\ref{MD5-krip-0}), while the creep compliance starts
from non-zero value of deformation and increases: proportionally to $%
t^{\alpha }$ in the case of Models VI and VIII, see (\ref{MD6-krip-0}) and (%
\ref{MD8-krip-0}); and proportionally to $t^{\beta }$ in the case of Model
VII, see (\ref{MD7-krip-0}).

The relaxation modulus for small time either decreases from infinity:
proportionally to $t^{-\left( \mu -\gamma \right) -\varkappa },$ with $%
\varkappa \in \left\{ \alpha ,\beta ,\gamma \right\} ,$ in the case of Model
I, see (\ref{MD1-sr-0}); proportionally to $t^{-\left( \mu -\alpha \right) }$
in the case of Models II and IV, see (\ref{MD2-sr-0}) and (\ref{MD4-sr-0});
and proportionally to $t^{-\left( \mu -\beta \right) }$ in the case of
Models III and V, see (\ref{MD3-sr-0}) and (\ref{MD5-sr-0}), or decreases
from finite glass modulus: proportionally to $t^{\alpha }$ in the case of
Models VI and VIII, see (\ref{MD6-sr-0}) and (\ref{MD8-sr-0}); and
proportionally to $t^{\beta }$ in the case of Model VII, see (\ref{MD7-sr-0}%
).

The growth of creep compliance for large time is governed: by $t^{\mu }$ in
the case of all Models I - V, see (\ref{MD1-krip-besk}), (\ref{MD2-krip-besk}%
), (\ref{MD3-krip-besk}), (\ref{MD4-krip-besk}), and (\ref{MD5-krip-besk});
by $t^{\beta }$ in the case of Models VI and VII, see (\ref{MD6-krip-besk})
and (\ref{MD7-krip-besk}); and by $t^{\alpha }$ in the case of Model VIII,
see (\ref{MD8-krip-besk}). For all fractional Burgers models, the growth of
creep compliance for large time is slower than in the case of classical
Burgers model when the growth in infinity is linear, see (\ref{obl-asympt}).

The relaxation modulus for large time tends to zero: proportionally to $%
t^{-\mu }$ in the case of all Models I - V, see (\ref{MD1-sr-besk}), (\ref%
{MD2-sr-besk}), (\ref{MD3-sr-besk}), (\ref{MD4-sr-besk}), and (\ref%
{MD5-sr-besk}); proportionally to $t^{-\beta }$ in the case of Models VI and
VII, see (\ref{MD6-sr-besk}) and (\ref{MD7-sr-besk}); and proportionally to $%
t^{-\alpha }$ in the case of Model VIII, see (\ref{MD8-sr-besk}).

The asymptotic analysis will be performed using the property of Laplace
transform that if $\tilde{f}(s)\sim \tilde{g}(s)$ as $s\rightarrow \infty $ (%
$s\rightarrow 0$), then $f\left( t\right) \sim g\left( t\right) $ as $%
t\rightarrow 0$ ($t\rightarrow \infty $), where $\tilde{f}=\mathcal{L}\left[
f\right] $ and $\tilde{g}=\mathcal{L}\left[ g\right] ,$ i.e., if the
function $\tilde{g}$ is asymptotic expansion of Laplace image $\tilde{f},$
then its inverse Laplace transform $g$ is asymptotic expansion of original $%
f.$

\subsection{Model I \label{mdl-1}}

Model I, given by (\ref{Model 1}) and subject to thermodynamical
restrictions (\ref{Model 1 - tdr}), is obtained from the unified model (\ref%
{UCE-1-5}) for $\eta =\varkappa \in \left\{ \alpha ,\beta ,\gamma \right\} .$

\subsubsection{Restrictions on range of model parameters}

The requirement that the creep compliance (\ref{CR-ILT-1-5}) is a Bernstein
function, while the relaxation modulus (\ref{SR}) is completely monotonic
narrows down the thermodynamical restriction (\ref{Model 1 - tdr}) to%
\begin{equation}
\frac{b_{2}}{b_{1}}\leq a_{i}\frac{\sin \frac{\left( \mu -\varkappa \right)
\pi }{2}}{\sin \frac{\left( \mu +\varkappa \right) \pi }{2}}\frac{\cos \frac{%
\left( \mu -\varkappa \right) \pi }{2}}{\left\vert \cos \frac{\left( \mu
+\varkappa \right) \pi }{2}\right\vert },  \label{pos-K-MD1}
\end{equation}%
with $\left( \varkappa ,i\right) \in \left\{ \left( \alpha ,1\right) ,\left(
\beta ,2\right) ,\left( \gamma ,3\right) \right\} ,$ since%
\begin{equation}
\frac{\sin \frac{\left( \mu -\varkappa \right) \pi }{2}}{\sin \frac{\left(
\mu +\varkappa \right) \pi }{2}}\leq 1.  \label{ineqs-MD1}
\end{equation}

The requirement (\ref{pos-K-MD1}) is obtained by insuring the non-negativity
of the terms appearing in brackets in function $K,$ given by (\ref{funct-K})
and having for $\varkappa \in \left\{ \alpha ,\beta ,\gamma \right\} $ the
respective forms 
\begin{eqnarray*}
K\left( \rho \right) \!\!\!\! &=&\!\!\!\!b_{1}\sin \left( \mu \pi \right) \\
&&+\left\{ 
\begin{tabular}{l}
\begin{tabular}{l}
$\!\!\!\!b_{1}\rho ^{\alpha }\left\vert \sin \left( \left( \mu +\alpha
\right) \pi \right) \right\vert \left( a_{1}\frac{\sin \left( \left( \mu
-\alpha \right) \pi \right) }{\left\vert \sin \left( \left( \mu +\alpha
\right) \pi \right) \right\vert }-\frac{b_{2}}{b_{1}}\right) +a_{2}b_{1}\rho
^{\beta }\sin \left( \left( \mu -\beta \right) \pi \right) +a_{3}b_{1}\rho
^{\gamma }\sin \left( \left( \mu -\gamma \right) \pi \right) $ \\ 
$+a_{1}b_{2}\rho ^{2\alpha }\sin \left( \mu \pi \right) +a_{2}b_{2}\rho
^{\alpha +\beta }\sin \left( \left( \mu -\beta +\alpha \right) \pi \right)
+a_{3}b_{2}\rho ^{\alpha +\gamma }\sin \left( \left( \mu -\gamma +\alpha
\right) \pi \right) ,\smallskip $%
\end{tabular}
\\ 
\begin{tabular}{l}
$\!\!\!\!a_{1}b_{1}\rho ^{\alpha }\sin \left( \left( \mu -\alpha \right) \pi
\right) +b_{1}\rho ^{\beta }\left\vert \sin \left( \left( \mu +\beta \right)
\pi \right) \right\vert \left( a_{2}\frac{\sin \left( \left( \mu -\beta
\right) \pi \right) }{\left\vert \sin \left( \left( \mu +\beta \right) \pi
\right) \right\vert }-\frac{b_{2}}{b_{1}}\right) +a_{3}b_{1}\rho ^{\gamma
}\sin \left( \left( \mu -\gamma \right) \pi \right) $ \\ 
$+a_{1}b_{2}\rho ^{\alpha +\beta }\sin \left( \left( \mu -\alpha +\beta
\right) \pi \right) +a_{2}b_{2}\rho ^{2\beta }\sin \left( \mu \pi \right)
+a_{3}b_{2}\rho ^{\beta +\gamma }\sin \left( \left( \mu -\gamma +\beta
\right) \pi \right) ,\smallskip $%
\end{tabular}
\\ 
\begin{tabular}{l}
$\!\!\!\!a_{1}b_{1}\rho ^{\alpha }\sin \left( \left( \mu -\alpha \right) \pi
\right) +a_{2}b_{1}\rho ^{\beta }\sin \left( \left( \mu -\beta \right) \pi
\right) +b_{1}\rho ^{\gamma }\left\vert \sin \left( \left( \mu +\gamma
\right) \pi \right) \right\vert \left( a_{3}\frac{\sin \left( \left( \mu
-\gamma \right) \pi \right) }{\left\vert \sin \left( \left( \mu +\gamma
\right) \pi \right) \right\vert }-\frac{b_{2}}{b_{1}}\right) $ \\ 
$+a_{1}b_{2}\rho ^{\alpha +\gamma }\sin \left( \left( \mu -\alpha +\gamma
\right) \pi \right) +a_{2}b_{2}\rho ^{\beta +\gamma }\sin \left( \left( \mu
-\beta +\gamma \right) \pi \right) +a_{3}b_{2}\rho ^{2\gamma }\sin \left(
\mu \pi \right) ,$%
\end{tabular}%
\end{tabular}%
\right.
\end{eqnarray*}%
since all other terms in $K$ are non-negative.

Rewriting the left-hand-side of (\ref{ineqs-MD1}), one obtains%
\begin{equation*}
\frac{\sin \frac{\left( \mu -\varkappa \right) \pi }{2}}{\sin \frac{\left(
\mu +\varkappa \right) \pi }{2}}=\frac{\tan \frac{\mu \pi }{2}-\tan \frac{%
\varkappa \pi }{2}}{\tan \frac{\mu \pi }{2}+\tan \frac{\varkappa \pi }{2}}%
\leq 1,
\end{equation*}%
with $\varkappa \in \left\{ \alpha ,\beta ,\gamma \right\} .$ According to
thermodynamical restriction (\ref{Model 1 - tdr}), one has $0\leq \frac{%
\varkappa \pi }{2}\leq \frac{\mu \pi }{2}\leq \frac{\pi }{2}$ implying $%
0\leq \tan \frac{\varkappa \pi }{2}\leq \tan \frac{\mu \pi }{2},$ that
yields the inequality (\ref{ineqs-MD1}).

\subsubsection{Asymptotic expansions}

The asymptotic expansion of creep compliance (\ref{CR-ILT-1-5}) for Model I
near initial time-instant is obtained in the form%
\begin{equation}
\varepsilon _{cr}\left( t\right) =\left\{ 
\begin{tabular}{ll}
$\frac{a_{3}}{b_{2}}\frac{t^{\mu -\gamma +\alpha }}{\Gamma \left( 1+\mu
-\gamma +\alpha \right) }+O\left( t^{\mu +\alpha -\rho _{1}}\right)
,\smallskip $ & $\text{for }\varkappa =\alpha ,$ \\ 
$\frac{a_{3}}{b_{2}}\frac{t^{\mu -\gamma +\beta }}{\Gamma \left( 1+\mu
-\gamma +\beta \right) }+O\left( t^{\mu +\beta -\rho _{2}}\right)
,\smallskip $ & $\text{for }\varkappa =\beta ,$ \\ 
\begin{tabular}{l}
$\!\!\!\!\frac{a_{3}}{b_{2}}\frac{t^{\mu }}{\Gamma \left( 1+\mu \right) }+%
\frac{a_{2}}{b_{2}}\frac{t^{\mu +\gamma -\beta }}{\Gamma \left( 1+\mu
+\gamma -\beta \right) }+\frac{a_{1}}{b_{2}}\frac{t^{\mu +\gamma -\alpha }}{%
\Gamma \left( 1+\mu +\gamma -\alpha \right) }\smallskip $ \\ 
$-\frac{b_{1}}{b_{2}}\left( a_{3}-\frac{b_{2}}{b_{1}}\right) \frac{t^{\mu
+\gamma }}{\Gamma \left( 1+\mu +\gamma \right) }+O\left( t^{\mu +2\gamma
-\beta }\right) ,$%
\end{tabular}
& $\text{for }\varkappa =\gamma ,$%
\end{tabular}%
\right. \;\;\text{as}\;\;t\rightarrow 0,  \label{MD1-krip-0}
\end{equation}%
with $\rho _{1}=\max \left\{ \beta ,\gamma -\alpha \right\} $ and $\rho
_{2}=\max \left\{ \beta ,\gamma -\beta \right\} ,$ as the inverse Laplace
transform of the creep compliance in complex domain (\ref{cr-LT}) rewritten
as%
\begin{equation*}
\tilde{\varepsilon}_{cr}\left( s\right) =\left\{ 
\begin{tabular}{ll}
$\frac{1}{b_{2}}\frac{1}{s^{1+\mu +\alpha }}\left( a_{3}s^{\gamma }+O\left(
s^{\rho _{1}}\right) \right) ,\smallskip $ & $\text{for }\varkappa =\alpha ,$
\\ 
$\frac{1}{b_{2}}\frac{1}{s^{1+\mu +\beta }}\left( a_{3}s^{\gamma }+O\left(
s^{\rho _{2}}\right) \right) ,\smallskip $ & $\text{for }\varkappa =\beta ,$
\\ 
$\frac{1}{b_{2}}\frac{1}{s^{1+\mu +\gamma }}\left( a_{3}s^{\gamma
}+a_{2}s^{\beta }+a_{1}s^{\alpha }+1-\frac{a_{3}b_{1}}{b_{2}}+O\left(
s^{-\gamma +\beta }\right) \right) ,$ & $\text{for }\varkappa =\gamma ,$%
\end{tabular}%
\right. \;\;\text{as}\;\;s\rightarrow \infty ,
\end{equation*}%
using the binomial formula%
\begin{equation}
\frac{1}{1+x}=1-x+x^{2}+O\left( x^{3}\right) ,\;\;\text{as}\;\;x\rightarrow
0,  \label{wke}
\end{equation}%
while the asymptotic expansion of creep compliance (\ref{CR-ILT-1-5}) for
large time takes the form%
\begin{equation}
\varepsilon _{cr}\left( t\right) =\left\{ 
\begin{tabular}{ll}
$\frac{1}{b_{1}}\frac{t^{\mu }}{\Gamma \left( 1+\mu \right) }+\frac{1}{b_{1}}%
\left( a_{1}-\frac{b_{2}}{b_{1}}\right) \frac{t^{\mu -\alpha }}{\Gamma
\left( 1+\mu -\alpha \right) }+O\left( t^{\mu -\rho _{1}}\right) ,\smallskip 
$ & $\text{for }\varkappa =\alpha ,$ \\ 
$\frac{1}{b_{1}}\frac{t^{\mu }}{\Gamma \left( 1+\mu \right) }+\frac{a_{1}}{%
b_{1}}\frac{t^{\mu -\alpha }}{\Gamma \left( 1+\mu -\alpha \right) }+\frac{1}{%
b_{1}}\left( a_{2}-\frac{b_{2}}{b_{1}}\right) \frac{t^{\mu -\beta }}{\Gamma
\left( 1+\mu -\beta \right) }+O\left( t^{\mu -\rho _{2}}\right) ,\smallskip $
& $\text{for }\varkappa =\beta ,$ \\ 
\begin{tabular}{l}
$\!\!\!\!\frac{1}{b_{1}}\frac{t^{\mu }}{\Gamma \left( 1+\mu \right) }+\frac{%
a_{1}}{b_{1}}\frac{t^{\mu -\alpha }}{\Gamma \left( 1+\mu -\alpha \right) }+%
\frac{a_{2}}{b_{1}}\frac{t^{\mu -\beta }}{\Gamma \left( 1+\mu -\beta \right) 
}\smallskip $ \\ 
$+\frac{1}{b_{1}}\left( a_{3}-\frac{b_{2}}{b_{1}}\right) \frac{t^{\mu
-\gamma }}{\Gamma \left( 1+\mu -\gamma \right) }+O\left( t^{\mu -\gamma
-\alpha }\right) ,$%
\end{tabular}
& $\text{for }\varkappa =\gamma ,$%
\end{tabular}%
\right. \;\;\text{as}\;\;t\rightarrow \infty ,  \label{MD1-krip-besk}
\end{equation}%
with $\rho _{1}=\min \left\{ 2\alpha ,\beta \right\} $ and $\rho _{2}=\min
\left\{ \alpha +\beta ,\gamma \right\} ,$ and it is obtained as the inverse
Laplace transform of (\ref{CR-ILT-1-5}):%
\begin{equation*}
\tilde{\varepsilon}_{cr}\left( s\right) =\left\{ 
\begin{tabular}{ll}
$\frac{1}{b_{1}}\frac{1}{s^{1+\mu }}\left( 1+\frac{a_{1}b_{1}-b_{2}}{b_{1}}%
s^{\alpha }+O\left( s^{\rho _{1}}\right) \right) ,\smallskip $ & $\text{for }%
\varkappa =\alpha ,$ \\ 
$\frac{1}{b_{1}}\frac{1}{s^{1+\mu }}\left( 1+a_{1}s^{\alpha }+\frac{%
a_{2}b_{1}-b_{2}}{b_{1}}s^{\beta }+O\left( s^{\rho _{2}}\right) \right)
,\smallskip $ & $\text{for }\varkappa =\beta ,$ \\ 
$\frac{1}{b_{1}}\frac{1}{s^{1+\mu }}\left( 1+a_{1}s^{\alpha }+a_{2}s^{\beta
}+\frac{a_{3}b_{1}-b_{2}}{b_{1}}s^{\gamma }+O\left( s^{\gamma +\alpha
}\right) \right) ,$ & $\text{for }\varkappa =\gamma ,$%
\end{tabular}%
\right. \;\;\text{as}\;\;s\rightarrow 0,
\end{equation*}%
where the binomial formula (\ref{wke}) is used.

The asymptotic expansion of relaxation modulus (\ref{SR}) for Model I near
initial time-instant is obtained in the form%
\begin{equation}
\sigma _{sr}\left( t\right) =\left\{ 
\begin{tabular}{ll}
$\frac{b_{2}}{a_{3}}\frac{t^{-\left( \mu -\gamma \right) -\alpha }}{\Gamma
\left( 1-\left( \mu -\gamma \right) -\alpha \right) }+O\left( t^{-\mu
+\gamma +\rho _{1}}\right) ,\smallskip $ & $\text{for }\varkappa =\alpha ,$
\\ 
$\frac{b_{2}}{a_{3}}\frac{t^{-\left( \mu -\gamma \right) -\beta }}{\Gamma
\left( 1-\left( \mu -\gamma \right) -\beta \right) }+O\left( t^{-\mu +\gamma
+\rho _{2}}\right) ,\smallskip $ & $\text{for }\varkappa =\beta ,$ \\ 
$\frac{b_{2}}{a_{3}}\frac{t^{-\mu }}{\Gamma \left( 1-\mu \right) }-\frac{%
a_{2}b_{2}}{a_{3}^{2}}\frac{t^{-\left( \mu -\gamma \right) -\beta }}{\Gamma
\left( 1-\left( \mu -\gamma \right) -\beta \right) }+O\left( t^{-\mu +\gamma
-\rho _{3}}\right) ,$ & $\text{for }\varkappa =\gamma ,$%
\end{tabular}%
\right. \;\;\text{as}\;\;t\rightarrow 0,  \label{MD1-sr-0}
\end{equation}%
with $\rho _{1}=\min \left\{ 0,\gamma -\beta -\alpha \right\} ,$ $\rho
_{2}=\min \left\{ 0,\gamma -2\beta \right\} ,$ and $\rho _{3}=\max \left\{
\alpha ,\gamma -2\beta \right\} ,$ as the inverse Laplace transform of the
relaxation modulus in complex domain (\ref{sr-lap}) rewritten as 
\begin{equation*}
\tilde{\sigma}_{sr}\left( s\right) =\left\{ 
\begin{tabular}{ll}
$\frac{1}{a_{3}}\frac{1}{s^{1-\mu +\gamma }}\left( b_{2}s^{\alpha }+O\left( 
\frac{1}{s^{\rho _{1}}}\right) \right) ,\smallskip $ & $\text{for }\varkappa
=\alpha ,$ \\ 
$\frac{1}{a_{3}}\frac{1}{s^{1-\mu +\gamma }}\left( b_{2}s^{\beta }+O\left( 
\frac{1}{s^{\rho _{2}}}\right) \right) ,\smallskip $ & $\text{for }\varkappa
=\beta ,$ \\ 
$\frac{1}{a_{3}}\frac{1}{s^{1-\mu +\gamma }}\left( b_{2}s^{\gamma }-\frac{%
a_{2}b_{2}}{a_{3}}s^{\beta }+O\left( s^{\rho _{3}}\right) \right) ,$ & $%
\text{for }\varkappa =\gamma ,$%
\end{tabular}%
\right. \;\;\text{as}\;\;s\rightarrow \infty ,
\end{equation*}%
using the binomial formula (\ref{wke}), while the asymptotic expansion of
the relaxation modulus (\ref{SR}) for large time takes the form%
\begin{equation}
\sigma _{sr}\left( t\right) =\left\{ 
\begin{tabular}{ll}
$b_{1}\frac{t^{-\mu }}{\Gamma \left( 1-\mu \right) }-b_{1}\left( a_{1}-\frac{%
b_{2}}{b_{1}}\right) \frac{t^{-\mu -\alpha }}{\Gamma \left( 1-\mu -\alpha
\right) }+O\left( t^{-\mu -\rho }\right) ,\smallskip $ & $\text{for }%
\varkappa =\alpha ,$ \\ 
$b_{1}\frac{t^{-\mu }}{\Gamma \left( 1-\mu \right) }-a_{1}b_{1}\frac{t^{-\mu
-\alpha }}{\Gamma \left( 1-\mu -\alpha \right) }+O\left( t^{-\mu -\rho
}\right) ,$ & for$\text{ }\varkappa =\left\{ \beta ,\gamma \right\} ,$%
\end{tabular}%
\right. \;\;\text{as}\;\;t\rightarrow \infty ,  \label{MD1-sr-besk}
\end{equation}%
with $\rho =\min \left\{ 2\alpha ,\beta \right\} ,$ and it is obtained as
the inverse Laplace transform of (\ref{sr-lap}):%
\begin{equation*}
\tilde{\sigma}_{sr}\left( s\right) =\left\{ 
\begin{tabular}{ll}
$\frac{1}{s^{1-\mu }}\left( b_{1}+\left( b_{2}-a_{1}b_{1}\right) s^{\alpha
}+O\left( s^{\rho }\right) \right) ,\smallskip $ & $\text{for }\varkappa
=\alpha ,$ \\ 
$\frac{1}{s^{1-\mu }}\left( b_{1}-a_{1}b_{1}s^{\alpha }+O\left( s^{\rho
}\right) \right) ,$ & for$\text{ }\varkappa =\left\{ \beta ,\gamma \right\}
, $%
\end{tabular}%
\right. \;\;\text{as}\;\;s\rightarrow 0,
\end{equation*}%
where the binomial formula (\ref{wke}) is used.

\subsection{Model II \label{mdl-2}}

Model II, given by (\ref{Model 2}) and subject to thermodynamical
restrictions (\ref{Model 2 - tdr}), is obtained from the unified model (\ref%
{UCE-1-5}) for $\eta =\alpha $ and $\gamma =2\alpha .$

\subsubsection{Restrictions on range of model parameters}

The requirement that the creep compliance (\ref{CR-ILT-1-5}) is a Bernstein
function, while the relaxation modulus (\ref{SR}) is completely monotonic
narrows down the thermodynamical restriction (\ref{Model 2 - tdr}) to%
\begin{equation}
\frac{a_{3}}{a_{1}}\frac{\left\vert \sin \frac{\left( \mu -2\alpha \right)
\pi }{2}\right\vert }{\sin \frac{\mu \pi }{2}}\frac{\cos \frac{\left( \mu
-2\alpha \right) \pi }{2}}{\cos \frac{\mu \pi }{2}}\leq \frac{b_{2}}{b_{1}}%
\leq a_{1}\frac{\sin \frac{\left( \mu -\alpha \right) \pi }{2}}{\sin \frac{%
\left( \mu +\alpha \right) \pi }{2}}\frac{\cos \frac{\left( \mu -\alpha
\right) \pi }{2}}{\left\vert \cos \frac{\left( \mu +\alpha \right) \pi }{2}%
\right\vert },  \label{pos-K-MD2}
\end{equation}%
since%
\begin{equation}
\frac{\cos \frac{\left( \mu -2\alpha \right) \pi }{2}}{\cos \frac{\mu \pi }{2%
}}\geq 1\;\;\text{and}\;\;\frac{\sin \frac{\left( \mu -\alpha \right) \pi }{2%
}}{\sin \frac{\left( \mu +\alpha \right) \pi }{2}}\leq 1.  \label{ineqs-MD2}
\end{equation}

The requirement (\ref{pos-K-MD2}) is obtained by insuring the non-negativity
of the terms appearing in brackets in function $K,$ given by (\ref{funct-K})
and having the form%
\begin{eqnarray*}
K\left( \rho \right) &=&b_{1}\sin \left( \mu \pi \right) +b_{1}\rho ^{\alpha
}\left\vert \sin \left( \left( \mu +\alpha \right) \pi \right) \right\vert
\left( a_{1}\frac{\sin \left( \left( \mu -\alpha \right) \pi \right) }{%
\left\vert \sin \left( \left( \mu +\alpha \right) \pi \right) \right\vert }-%
\frac{b_{2}}{b_{1}}\right) \\
&&+a_{2}b_{1}\rho ^{\beta }\sin \left( \left( \mu -\beta \right) \pi \right)
+a_{1}b_{1}\rho ^{2\alpha }\sin \left( \mu \pi \right) \left( \frac{b_{2}}{%
b_{1}}-\frac{a_{3}}{a_{1}}\frac{\left\vert \sin \left( \left( \mu -2\alpha
\right) \pi \right) \right\vert }{\sin \left( \mu \pi \right) }\right) \\
&&+a_{2}b_{2}\rho ^{\alpha +\beta }\sin \left( \left( \mu -\beta +\alpha
\right) \pi \right) +a_{3}b_{2}\rho ^{3\alpha }\sin \left( \left( \mu
-\alpha \right) \pi \right) ,
\end{eqnarray*}%
since all other terms in $K$ are non-negative.

According to thermodynamical restriction (\ref{Model 2 - tdr}), one has $%
0\leq 2\alpha -\mu \leq \alpha \leq \mu \leq 1,$ so that $0\leq \frac{\left(
2\alpha -\mu \right) \pi }{2}\leq \frac{\mu \pi }{2}\leq \frac{\pi }{2}$
implies the first inequality in (\ref{ineqs-MD2}), while the second
inequality holds since%
\begin{equation*}
\frac{\sin \frac{\left( \mu -\alpha \right) \pi }{2}}{\sin \frac{\left( \mu
+\alpha \right) \pi }{2}}=\frac{\tan \frac{\mu \pi }{2}-\tan \frac{\alpha
\pi }{2}}{\tan \frac{\mu \pi }{2}+\tan \frac{\alpha \pi }{2}}\leq 1,
\end{equation*}%
due to $0\leq \frac{\alpha \pi }{2}\leq \frac{\mu \pi }{2}\leq \frac{\pi }{2}
$ implying $0\leq \tan \frac{\alpha \pi }{2}\leq \tan \frac{\mu \pi }{2}.$

\subsubsection{Asymptotic expansions}

The asymptotic expansion of creep compliance (\ref{CR-ILT-1-5}) for Model II
near initial time-instant is obtained in the form%
\begin{eqnarray}
\varepsilon _{cr}\left( t\right) &=&\frac{a_{3}}{b_{2}}\frac{t^{\mu -\alpha }%
}{\Gamma \left( 1+\mu -\alpha \right) }+\frac{a_{2}}{b_{2}}\frac{t^{\mu
-\left( \beta -\alpha \right) }}{\Gamma \left( 1+\mu -\left( \beta -\alpha
\right) \right) }+\frac{a_{1}b_{1}}{b_{2}^{2}}\left( \frac{b_{2}}{b_{1}}-%
\frac{a_{3}}{a_{1}}\right) \frac{t^{\mu }}{\Gamma \left( 1+\mu \right) } 
\notag \\
&&-\frac{a_{2}b_{1}}{b_{2}}\frac{t^{\mu +2\alpha -\beta }}{\Gamma \left(
1+\mu +2\alpha -\beta \right) }+O\left( t^{\mu +\alpha }\right) ,\;\;\text{as%
}\;\;t\rightarrow 0,  \label{MD2-krip-0}
\end{eqnarray}%
as the inverse Laplace transform of (\ref{cr-LT}) rewritten as 
\begin{eqnarray*}
\tilde{\varepsilon}_{cr}\left( s\right) &=&\frac{1}{b_{2}s^{1+\mu +\alpha }}%
\frac{1+a_{1}s^{\alpha }+a_{2}s^{\beta }+a_{3}s^{2\alpha }}{1+\frac{b_{1}}{%
b_{2}}\frac{1}{s^{\alpha }}} \\
\tilde{\varepsilon}_{cr}\left( s\right) &=&\frac{1}{b_{2}s^{1+\mu +\alpha }}%
\left( a_{3}s^{2\alpha }+a_{2}s^{\beta }+\frac{a_{1}b_{2}-a_{3}b_{1}}{b_{2}}%
s^{\alpha }-\frac{a_{2}b_{1}}{b_{2}}s^{\beta -\alpha }+O\left( 1\right)
\right) ,\;\;\text{as}\;\;s\rightarrow \infty ,
\end{eqnarray*}%
using the binomial formula (\ref{wke}), while the asymptotic expansion of
creep compliance (\ref{CR-ILT-1-5}) for large time takes the form%
\begin{equation}
\varepsilon _{cr}\left( t\right) =\frac{1}{b_{1}}\frac{t^{\mu }}{\Gamma
\left( 1+\mu \right) }+\frac{1}{b_{1}}\left( a_{1}-\frac{b_{2}}{b_{1}}%
\right) \frac{t^{\mu -\alpha }}{\Gamma \left( 1+\mu -\alpha \right) }+\frac{%
a_{2}}{b_{1}}\frac{t^{\mu -\beta }}{\Gamma \left( 1+\mu -\beta \right) }%
+O\left( t^{\mu -2\alpha }\right) ,\;\;\text{as}\;\;t\rightarrow \infty ,
\label{MD2-krip-besk}
\end{equation}%
and it is obtained as the inverse Laplace transform of (\ref{CR-ILT-1-5}):%
\begin{eqnarray*}
\tilde{\varepsilon}_{cr}\left( s\right) &=&\frac{1}{b_{1}s^{1+\mu }}\frac{%
1+a_{1}s^{\alpha }+a_{2}s^{\beta }+a_{3}s^{2\alpha }}{1+\frac{b_{2}}{b_{1}}%
s^{\alpha }} \\
\tilde{\varepsilon}_{cr}\left( s\right) &=&\frac{1}{b_{1}}\frac{1}{s^{1+\mu }%
}\left( 1+\frac{a_{1}b_{1}-b_{2}}{b_{1}}s^{\alpha }+a_{2}s^{\beta }+O\left(
s^{2\alpha }\right) \right) ,\;\;\text{as}\;\;s\rightarrow 0,
\end{eqnarray*}%
where the binomial formula (\ref{wke}) is used.

The asymptotic expansion of relaxation modulus (\ref{SR}) for Model II near
initial time-instant is obtained in the form%
\begin{equation}
\sigma _{sr}\left( t\right) =\frac{b_{2}}{a_{3}}\frac{t^{-\left( \mu -\alpha
\right) }}{\Gamma \left( 1-\left( \mu -\alpha \right) \right) }+O\left(
t^{2\alpha -\mu -\rho }\right) ,\;\;\text{as}\;\;t\rightarrow 0,
\label{MD2-sr-0}
\end{equation}%
with $\rho =\max \left\{ \beta -\alpha ,3\alpha -2\beta \right\} ,$ as the
inverse Laplace transform of the relaxation modulus in complex domain (\ref%
{sr-lap}) rewritten as 
\begin{eqnarray*}
\tilde{\sigma}_{sr}\left( s\right) &=&\frac{1}{a_{3}s^{1+2\alpha -\mu }}%
\frac{b_{1}+b_{2}s^{\alpha }}{1+\frac{a_{1}}{a_{3}}\frac{1}{s^{\alpha }}+%
\frac{a_{2}}{a_{3}}\frac{1}{s^{2\alpha -\beta }}+\frac{1}{a_{3}}\frac{1}{%
s^{2\alpha }}} \\
\tilde{\sigma}_{sr}\left( s\right) &=&\frac{1}{a_{3}s^{1+2\alpha -\mu }}%
\left( b_{2}s^{\alpha }+O\left( s^{\rho }\right) \right) ,\;\;\text{as}%
\;\;s\rightarrow \infty ,
\end{eqnarray*}%
using the binomial formula (\ref{wke}), while the asymptotic expansion of
the relaxation modulus (\ref{SR}) for large time takes the form%
\begin{equation}
\sigma _{sr}\left( t\right) =b_{1}\frac{t^{-\mu }}{\Gamma \left( 1-\mu
\right) }-b_{1}\left( a_{1}-\frac{b_{2}}{b_{1}}\right) \frac{t^{-\mu -\alpha
}}{\Gamma \left( 1-\mu -\alpha \right) }-a_{2}b_{1}\frac{t^{-\mu -\beta }}{%
\Gamma \left( 1-\mu -\beta \right) }+O\left( t^{-\mu -2\alpha }\right) ,\;\;%
\text{as}\;\;t\rightarrow \infty ,  \label{MD2-sr-besk}
\end{equation}%
and it is obtained as the inverse Laplace transform of (\ref{sr-lap}):%
\begin{eqnarray*}
\tilde{\sigma}_{sr}\left( s\right) &=&\frac{1}{s^{1-\mu }}\frac{%
b_{1}+b_{2}s^{\alpha }}{1+a_{1}s^{\alpha }+a_{2}s^{\beta }+a_{3}s^{2\alpha }}
\\
\tilde{\sigma}_{sr}\left( s\right) &=&\frac{1}{s^{1-\mu }}\left(
b_{1}+\left( b_{2}-a_{1}b_{1}\right) s^{\alpha }-a_{2}b_{1}s^{\beta
}+O\left( s^{2\alpha }\right) \right) ,\;\;\text{as}\;\;s\rightarrow 0,
\end{eqnarray*}%
where the binomial formula (\ref{wke}) is used.

\subsection{Model III \label{mdl-3}}

Model III, given by (\ref{Model 3}) and subject to thermodynamical
restrictions (\ref{Model 3 - tdr}), is obtained from the unified model (\ref%
{UCE-1-5}) for $\eta =\alpha $ and $\gamma =\alpha +\beta .$

\subsubsection{Restrictions on range of model parameters}

The requirement that the creep compliance (\ref{CR-ILT-1-5}) is a Bernstein
function, while the relaxation modulus (\ref{SR}) is completely monotonic
narrows down the thermodynamical restriction (\ref{Model 3 - tdr}) to%
\begin{equation}
\frac{a_{3}}{a_{2}}\frac{\left\vert \sin \frac{\left( \mu -\beta -\alpha
\right) \pi }{2}\right\vert }{\sin \frac{\left( \mu -\beta +\alpha \right)
\pi }{2}}\frac{\cos \frac{\left( \mu -\beta -\alpha \right) \pi }{2}}{\cos 
\frac{\left( \mu -\beta +\alpha \right) \pi }{2}}\leq \frac{b_{2}}{b_{1}}%
\leq a_{1}\frac{\sin \frac{\left( \mu -\alpha \right) \pi }{2}}{\sin \frac{%
\left( \mu +\alpha \right) \pi }{2}}\frac{\cos \frac{\left( \mu -\alpha
\right) \pi }{2}}{\left\vert \cos \frac{\left( \mu +\alpha \right) \pi }{2}%
\right\vert },  \label{pos-K-MD3}
\end{equation}%
since%
\begin{equation}
\frac{\cos \frac{\left( \mu -\beta -\alpha \right) \pi }{2}}{\cos \frac{%
\left( \mu -\beta +\alpha \right) \pi }{2}}\geq 1\;\;\text{and}\;\;\frac{%
\sin \frac{\left( \mu -\alpha \right) \pi }{2}}{\sin \frac{\left( \mu
+\alpha \right) \pi }{2}}\leq 1.  \label{ineqs-MD3}
\end{equation}

The requirement (\ref{pos-K-MD3}) is obtained by insuring the non-negativity
of the terms appearing in brackets in function $K,$ given by (\ref{funct-K})
and having the form%
\begin{eqnarray*}
K\left( \rho \right) &=&b_{1}\sin \left( \mu \pi \right) +b_{1}\rho ^{\alpha
}\left\vert \sin \left( \left( \mu +\alpha \right) \pi \right) \right\vert
\left( a_{1}\frac{\sin \left( \left( \mu -\alpha \right) \pi \right) }{%
\left\vert \sin \left( \left( \mu +\alpha \right) \pi \right) \right\vert }-%
\frac{b_{2}}{b_{1}}\right) +a_{2}b_{1}\rho ^{\beta }\sin \left( \left( \mu
-\beta \right) \pi \right) \\
&&+a_{1}b_{2}\rho ^{2\alpha }\sin \left( \mu \pi \right) +a_{2}b_{1}\rho
^{\alpha +\beta }\sin \left( \left( \mu -\beta +\alpha \right) \pi \right)
\left( \frac{b_{2}}{b_{1}}-\frac{a_{3}}{a_{2}}\frac{\left\vert \sin \left(
\left( \mu -\beta -\alpha \right) \pi \right) \right\vert }{\sin \left(
\left( \mu -\beta +\alpha \right) \pi \right) }\right) \\
&&+a_{3}b_{2}\rho ^{2\alpha +\beta }\sin \left( \left( \mu -\beta \right)
\pi \right) ,
\end{eqnarray*}%
since all other terms in $K$ are non-negative.

Rewriting the left-hand-sides of (\ref{ineqs-MD3}), one obtains%
\begin{equation*}
\frac{\cos \frac{\left( \mu -\beta -\alpha \right) \pi }{2}}{\cos \frac{%
\left( \mu -\beta +\alpha \right) \pi }{2}}=\frac{1+\tan \frac{\left( \mu
-\beta \right) \pi }{2}\tan \frac{\alpha \pi }{2}}{1-\tan \frac{\left( \mu
-\beta \right) \pi }{2}\tan \frac{\alpha \pi }{2}}\geq 1\;\;\text{and}\;\;%
\frac{\sin \frac{\left( \mu -\alpha \right) \pi }{2}}{\sin \frac{\left( \mu
+\alpha \right) \pi }{2}}=\frac{\tan \frac{\mu \pi }{2}-\tan \frac{\alpha
\pi }{2}}{\tan \frac{\mu \pi }{2}+\tan \frac{\alpha \pi }{2}}\leq 1.
\end{equation*}%
According to thermodynamical restriction (\ref{Model 3 - tdr}), one has $%
0\leq \mu -\beta +\alpha \leq 1,$ so that $0\leq \frac{\left( \mu -\beta
+\alpha \right) \pi }{2}\leq \frac{\pi }{2}$ implies $\cos \frac{\left( \mu
-\beta +\alpha \right) \pi }{2}\geq 0,$ thus yielding the first inequality
in (\ref{ineqs-MD3}), while the second inequality holds since $0\leq \frac{%
\alpha \pi }{2}\leq \frac{\mu \pi }{2}\leq \frac{\pi }{2}$ implies $0\leq
\tan \frac{\alpha \pi }{2}\leq \tan \frac{\mu \pi }{2}.$

\subsubsection{Asymptotic expansions}

The asymptotic expansion of creep compliance (\ref{CR-ILT-1-5}) for Model
III near initial time-instant is obtained in the form%
\begin{equation}
\varepsilon _{cr}\left( t\right) =\frac{a_{3}}{b_{2}}\frac{t^{\mu -\beta }}{%
\Gamma \left( 1+\mu -\beta \right) }+\frac{b_{1}a_{2}}{b_{2}^{2}}\left( 
\frac{b_{2}}{b_{1}}-\frac{a_{3}}{a_{2}}\right) \frac{t^{\mu -\left( \beta
-\alpha \right) }}{\Gamma \left( 1+\mu -\left( \beta -\alpha \right) \right) 
}+O\left( t^{\mu +\alpha -\rho }\right) ,\;\;\text{as}\;\;t\rightarrow 0,
\label{MD3-krip-0}
\end{equation}%
with $\rho =\max \left\{ \alpha ,\beta -\alpha \right\} ,$ as the inverse
Laplace transform of (\ref{cr-LT}) rewritten as 
\begin{eqnarray*}
\tilde{\varepsilon}_{cr}\left( s\right) &=&\frac{1}{b_{2}s^{1+\mu +\alpha }}%
\frac{1+a_{1}s^{\alpha }+a_{2}s^{\beta }+a_{3}s^{\alpha +\beta }}{1+\frac{%
b_{1}}{b_{2}}\frac{1}{s^{\alpha }}} \\
\tilde{\varepsilon}_{cr}\left( s\right) &=&\frac{1}{b_{2}s^{1+\mu +\alpha }}%
\left( a_{3}s^{\alpha +\beta }+\frac{a_{2}b_{2}-a_{3}b_{1}}{b_{2}}s^{\beta
}+O\left( s^{\rho }\right) \right) ,\;\;\text{as}\;\;s\rightarrow \infty ,
\end{eqnarray*}%
using the binomial formula (\ref{wke}), while the asymptotic expansion of
creep compliance (\ref{CR-ILT-1-5}) for large time takes the form 
\begin{equation}
\varepsilon _{cr}\left( t\right) =\frac{1}{b_{1}}\frac{t^{\mu }}{\Gamma
\left( 1+\mu \right) }+\frac{1}{b_{1}}\left( a_{1}-\frac{b_{2}}{b_{1}}%
\right) \frac{t^{\mu -\alpha }}{\Gamma \left( 1+\mu -\alpha \right) }%
+O\left( t^{\mu -\rho }\right) ,\;\;\text{as}\;\;t\rightarrow \infty ,
\label{MD3-krip-besk}
\end{equation}%
with $\rho =\min \left\{ 2\alpha ,\beta \right\} ,$ and it is obtained as
the inverse Laplace transform of (\ref{CR-ILT-1-5}): 
\begin{eqnarray*}
\tilde{\varepsilon}_{cr}\left( s\right) &=&\frac{1}{b_{1}s^{1+\mu }}\frac{%
1+a_{1}s^{\alpha }+a_{2}s^{\beta }+a_{3}s^{\alpha +\beta }}{1+\frac{b_{2}}{%
b_{1}}s^{\alpha }} \\
\tilde{\varepsilon}_{cr}\left( s\right) &=&\frac{1}{b_{1}}\frac{1}{s^{1+\mu }%
}\left( 1+\frac{a_{1}b_{1}-b_{2}}{b_{1}}s^{\alpha }+O\left( s^{\rho }\right)
\right) ,\;\;\text{as}\;\;s\rightarrow 0,
\end{eqnarray*}%
where the binomial formula (\ref{wke}) is used.

The asymptotic expansion of relaxation modulus (\ref{SR}) for Model III near
initial time-instant is obtained in the form%
\begin{equation}
\sigma _{sr}\left( t\right) =\frac{b_{2}}{a_{3}}\frac{t^{-\left( \mu -\beta
\right) }}{\Gamma \left( 1-\left( \mu -\beta \right) \right) }-\frac{%
a_{2}b_{1}}{a_{3}^{2}}\left( \frac{b_{2}}{b_{1}}-\frac{a_{3}}{a_{2}}\right) 
\frac{t^{\alpha +\beta -\mu }}{\Gamma \left( 1+\alpha +\beta -\mu \right) }%
+O\left( t^{\alpha +\beta +\rho -\mu }\right) ,\;\;\text{as}\;\;t\rightarrow
0,  \label{MD3-sr-0}
\end{equation}%
with $\rho =\min \left\{ \alpha ,\beta -\alpha \right\} ,$ as the inverse
Laplace transform of the relaxation modulus in complex domain (\ref{sr-lap})
rewritten as%
\begin{eqnarray*}
\tilde{\sigma}_{sr}\left( s\right) &=&\frac{1}{a_{3}s^{1+\alpha +\beta -\mu }%
}\frac{b_{1}+b_{2}s^{\alpha }}{1+\frac{a_{1}}{a_{3}}\frac{1}{s^{\beta }}+%
\frac{a_{2}}{a_{3}}\frac{1}{s^{\alpha }}+\frac{1}{a_{3}}\frac{1}{s^{\alpha
+\beta }}} \\
\tilde{\sigma}_{sr}\left( s\right) &=&\frac{1}{a_{3}s^{1+\alpha +\beta -\mu }%
}\left( b_{2}s^{\alpha }+\frac{a_{3}b_{1}-a_{2}b_{2}}{a_{3}}+O\left( \frac{1%
}{s^{\rho }}\right) \right) ,\;\;\text{as}\;\;s\rightarrow \infty ,
\end{eqnarray*}%
using the binomial formula (\ref{wke}), while the asymptotic expansion of
the relaxation modulus (\ref{SR}) for large time takes the form%
\begin{equation}
\sigma _{sr}\left( t\right) =b_{1}\frac{t^{-\mu }}{\Gamma \left( 1-\mu
\right) }-b_{1}\left( a_{1}-\frac{b_{2}}{b_{1}}\right) \frac{t^{-\mu -\alpha
}}{\Gamma \left( 1-\mu -\alpha \right) }+O\left( t^{-\mu -\rho }\right) ,\;\;%
\text{as}\;\;t\rightarrow \infty ,  \label{MD3-sr-besk}
\end{equation}%
with $\rho =\min \left\{ 2\alpha ,\beta \right\} ,$ and it is obtained as
the inverse Laplace transform of (\ref{sr-lap}):%
\begin{eqnarray*}
\tilde{\sigma}_{sr}\left( s\right) &=&\frac{1}{s^{1-\mu }}\frac{%
b_{1}+b_{2}s^{\alpha }}{1+a_{1}s^{\alpha }+a_{2}s^{\beta }+a_{3}s^{\alpha
+\beta }} \\
\tilde{\sigma}_{sr}\left( s\right) &=&\frac{1}{s^{1-\mu }}\left(
b_{1}+\left( b_{2}-a_{1}b_{1}\right) s^{\alpha }+O\left( s^{\rho }\right)
\right) ,\;\;\text{as}\;\;s\rightarrow 0,
\end{eqnarray*}%
where the binomial formula (\ref{wke}) is used.

\subsection{Model IV \label{mdl-4}}

Model IV, given by (\ref{Model 4}) and subject to thermodynamical
restrictions (\ref{Model 4 - tdr}), is obtained from the unified model (\ref%
{UCE-1-5}) for $\eta =\beta $ and $\gamma =\alpha +\beta .$

\subsubsection{Restrictions on range of model parameters}

The requirement that the creep compliance (\ref{CR-ILT-1-5}) is a Bernstein
function, while the relaxation modulus (\ref{SR}) is completely monotonic
narrows down the thermodynamical restriction (\ref{Model 4 - tdr}) to%
\begin{equation}
\frac{a_{3}}{a_{1}}\frac{\left\vert \sin \frac{\left( \mu -\alpha -\beta
\right) \pi }{2}\right\vert }{\sin \frac{\left( \mu -\alpha +\beta \right)
\pi }{2}}\frac{\cos \frac{\left( \mu -\alpha -\beta \right) \pi }{2}}{\cos 
\frac{\left( \mu -\alpha +\beta \right) \pi }{2}}\leq \frac{b_{2}}{b_{1}}%
\leq a_{2}\frac{\sin \frac{\left( \mu -\beta \right) \pi }{2}}{\sin \frac{%
\left( \mu +\beta \right) \pi }{2}}\frac{\cos \frac{\left( \mu -\beta
\right) \pi }{2}}{\left\vert \cos \frac{\left( \mu +\beta \right) \pi }{2}%
\right\vert },  \label{pos-K-MD4}
\end{equation}%
since%
\begin{equation}
\frac{\cos \frac{\left( \mu -\alpha -\beta \right) \pi }{2}}{\cos \frac{%
\left( \mu -\alpha +\beta \right) \pi }{2}}\geq 1\;\;\text{and}\;\;\frac{%
\sin \frac{\left( \mu -\beta \right) \pi }{2}}{\sin \frac{\left( \mu +\beta
\right) \pi }{2}}\leq 1.  \label{ineqs-MD4}
\end{equation}

The requirement (\ref{pos-K-MD4}) is obtained by insuring the non-negativity
of the terms appearing in brackets in function $K,$ given by (\ref{funct-K})
and having the form%
\begin{eqnarray*}
K\left( \rho \right) &=&b_{1}\sin \left( \mu \pi \right) +a_{1}b_{1}\rho
^{\alpha }\sin \left( \left( \mu -\alpha \right) \pi \right) +b_{1}\rho
^{\beta }\left\vert \sin \left( \left( \mu +\beta \right) \pi \right)
\right\vert \left( a_{2}\frac{\sin \left( \left( \mu -\beta \right) \pi
\right) }{\left\vert \sin \left( \left( \mu +\beta \right) \pi \right)
\right\vert }-\frac{b_{2}}{b_{1}}\right) \\
&&+a_{1}b_{1}\rho ^{\alpha +\beta }\sin \left( \left( \mu +\beta -\alpha
\right) \pi \right) \left( \frac{b_{2}}{b_{1}}-\frac{a_{3}}{a_{1}}\frac{%
\left\vert \sin \left( \left( \mu -\alpha -\beta \right) \pi \right)
\right\vert }{\sin \left( \left( \mu -\alpha +\beta \right) \pi \right) }%
\right) \\
&&+a_{2}b_{2}\rho ^{2\beta }\sin \left( \mu \pi \right) +a_{3}b_{2}\rho
^{\alpha +2\beta }\sin \left( \left( \mu -\alpha \right) \pi \right) ,
\end{eqnarray*}%
since all other terms in $K$ are non-negative.

Rewriting the left-hand-sides of (\ref{ineqs-MD4}), one obtains%
\begin{equation*}
\frac{\cos \frac{\left( \mu -\alpha -\beta \right) \pi }{2}}{\cos \frac{%
\left( \mu -\alpha +\beta \right) \pi }{2}}=\frac{1+\tan \frac{\left( \mu
-\alpha \right) \pi }{2}\tan \frac{\beta \pi }{2}}{1-\tan \frac{\left( \mu
-\alpha \right) \pi }{2}\tan \frac{\beta \pi }{2}}\geq 1\;\;\text{and}\;\;%
\frac{\sin \frac{\left( \mu -\beta \right) \pi }{2}}{\sin \frac{\left( \mu
+\beta \right) \pi }{2}}=\frac{\tan \frac{\mu \pi }{2}-\tan \frac{\beta \pi 
}{2}}{\tan \frac{\mu \pi }{2}+\tan \frac{\beta \pi }{2}}\leq 1.
\end{equation*}%
According to thermodynamical restriction (\ref{Model 4 - tdr}), one has $%
0\leq \mu -\alpha +\beta \leq 1,$ so that $0\leq \frac{\left( \mu -\alpha
+\beta \right) \pi }{2}\leq \frac{\pi }{2}$ implies $\cos \frac{\left( \mu
-\alpha +\beta \right) \pi }{2}\geq 0,$ thus yielding the first inequality
in (\ref{ineqs-MD4}), while the second inequality holds since $0\leq \frac{%
\beta \pi }{2}\leq \frac{\mu \pi }{2}\leq \frac{\pi }{2}$ implies $0\leq
\tan \frac{\beta \pi }{2}\leq \tan \frac{\mu \pi }{2}.$

\subsubsection{Asymptotic expansions}

The asymptotic expansion of creep compliance (\ref{CR-ILT-1-5}) for Model IV
near initial time-instant is obtained in the form%
\begin{eqnarray}
\varepsilon _{cr}\left( t\right) &=&\frac{a_{3}}{b_{2}}\frac{t^{\mu -\alpha }%
}{\Gamma \left( 1+\mu -\alpha \right) }+\frac{a_{2}}{b_{2}}\frac{t^{\mu }}{%
\Gamma \left( 1+\mu \right) }+\frac{a_{1}b_{1}}{b_{2}^{2}}\left( \frac{b_{2}%
}{b_{1}}-\frac{a_{3}}{a_{1}}\right) \frac{t^{\mu +\beta -\alpha }}{\Gamma
\left( 1+\mu +\beta -\alpha \right) }  \notag \\
&&-\frac{b_{1}}{b_{2}^{2}}\left( a_{2}-\frac{b_{2}}{b_{1}}\right) \frac{%
t^{\mu +\beta }}{\Gamma \left( 1+\mu +\beta \right) }+O\left( t^{\mu +2\beta
-\alpha }\right) ,\;\;\text{as}\;\;t\rightarrow 0,  \label{MD4-krip-0}
\end{eqnarray}%
as the inverse Laplace transform of (\ref{cr-LT}) rewritten as 
\begin{eqnarray*}
\tilde{\varepsilon}_{cr}\left( s\right) &=&\frac{1}{b_{2}s^{1+\mu +\beta }}%
\frac{1+a_{1}s^{\alpha }+a_{2}s^{\beta }+a_{3}s^{\alpha +\beta }}{1+\frac{%
b_{1}}{b_{2}}\frac{1}{s^{\beta }}} \\
\tilde{\varepsilon}_{cr}\left( s\right) &=&\frac{1}{b_{2}s^{1+\mu +\beta }}%
\left( a_{3}s^{\alpha +\beta }+a_{2}s^{\beta }+\frac{a_{1}b_{2}-a_{3}b_{1}}{%
b_{2}}s^{\alpha }+1-\frac{a_{2}b_{1}}{b_{2}}+O\left( s^{-\beta +\alpha
}\right) \right) ,\;\;\text{as}\;\;s\rightarrow \infty ,
\end{eqnarray*}%
using the binomial formula (\ref{wke}), while the asymptotic expansion of
creep compliance (\ref{CR-ILT-1-5}) for large time takes the form 
\begin{eqnarray}
\varepsilon _{cr}\left( t\right) &=&\frac{1}{b_{1}}\frac{t^{\mu }}{\Gamma
\left( 1+\mu \right) }+\frac{a_{1}}{b_{1}}\frac{t^{\mu -\alpha }}{\Gamma
\left( 1+\mu -\alpha \right) }+\frac{1}{b_{1}}\left( a_{2}-\frac{b_{2}}{b_{1}%
}\right) \frac{t^{\mu -\beta }}{\Gamma \left( 1+\mu -\beta \right) }  \notag
\\
&&-\frac{a_{1}}{b_{1}}\left( \frac{b_{2}}{b_{1}}-\frac{a_{3}}{a_{1}}\right) 
\frac{t^{-\left( \alpha +\beta -\mu \right) }}{\Gamma \left( 1-\left( \alpha
+\beta -\mu \right) \right) }+O\left( t^{\mu -2\beta }\right) ,\;\;\text{as}%
\;\;t\rightarrow \infty ,  \label{MD4-krip-besk}
\end{eqnarray}%
and it is obtained as the inverse Laplace transform of (\ref{CR-ILT-1-5}): 
\begin{eqnarray*}
\tilde{\varepsilon}_{cr}\left( s\right) &=&\frac{1}{b_{1}s^{1+\mu }}\frac{%
1+a_{1}s^{\alpha }+a_{2}s^{\beta }+a_{3}s^{\alpha +\beta }}{1+\frac{b_{2}}{%
b_{1}}s^{\beta }} \\
\tilde{\varepsilon}_{cr}\left( s\right) &=&\frac{1}{b_{1}}\frac{1}{s^{1+\mu }%
}\left( 1+a_{1}s^{\alpha }+\frac{a_{2}b_{1}-b_{2}}{b_{1}}s^{\beta }+\frac{%
a_{3}b_{1}-a_{1}b_{2}}{b_{1}}s^{\alpha +\beta }+O\left( s^{2\beta }\right)
\right) ,\;\;\text{as}\;\;s\rightarrow 0,
\end{eqnarray*}%
where the binomial formula (\ref{wke}) is used.

The asymptotic expansion of relaxation modulus (\ref{SR}) for Model IV near
initial time-instant is obtained in the form%
\begin{equation}
\sigma _{sr}\left( t\right) =\frac{b_{2}}{a_{3}}\frac{t^{-\left( \mu -\alpha
\right) }}{\Gamma \left( 1-\left( \mu -\alpha \right) \right) }-\frac{%
a_{2}b_{2}}{a_{3}^{2}}\frac{t^{2\alpha -\mu }}{\Gamma \left( 1+2\alpha -\mu
\right) }-\frac{a_{1}b_{1}}{a_{3}^{2}}\left( \frac{b_{2}}{b_{1}}-\frac{a_{3}%
}{a_{1}}\right) \frac{t^{\alpha +\beta -\mu }}{\Gamma \left( 1+\alpha +\beta
-\mu \right) }+O\left( t^{3\alpha -\mu }\right) ,\;\;\text{as}%
\;\;t\rightarrow 0,  \label{MD4-sr-0}
\end{equation}%
as the inverse Laplace transform of the relaxation modulus in complex domain
(\ref{sr-lap}) rewritten as%
\begin{eqnarray*}
\tilde{\sigma}_{sr}\left( s\right) &=&\frac{1}{a_{3}s^{1+\alpha +\beta -\mu }%
}\frac{b_{1}+b_{2}s^{\beta }}{1+\frac{a_{1}}{a_{3}}\frac{1}{s^{\beta }}+%
\frac{a_{2}}{a_{3}}\frac{1}{s^{\alpha }}+\frac{1}{a_{3}}\frac{1}{s^{\alpha
+\beta }}} \\
\tilde{\sigma}_{sr}\left( s\right) &=&\frac{1}{a_{3}s^{1+\alpha +\beta -\mu }%
}\left( b_{2}s^{\beta }-\frac{a_{2}b_{2}}{a_{3}}s^{\beta -\alpha }+\frac{%
a_{3}b_{1}-a_{1}b_{2}}{a_{3}}+O\left( \frac{1}{s^{2\alpha -\beta }}\right)
\right) ,\;\;\text{as}\;\;s\rightarrow \infty ,
\end{eqnarray*}%
using the binomial formula (\ref{wke}), while the asymptotic expansion of
the relaxation modulus (\ref{SR}) for large time takes the form%
\begin{equation}
\sigma _{sr}\left( t\right) =b_{1}\frac{t^{-\mu }}{\Gamma \left( 1-\mu
\right) }-a_{1}b_{1}\frac{t^{-\mu -\alpha }}{\Gamma \left( 1-\mu -\alpha
\right) }+O\left( t^{-\mu -\rho }\right) ,\;\;\text{as}\;\;t\rightarrow
\infty ,  \label{MD4-sr-besk}
\end{equation}%
with $\rho =\min \left\{ 2\alpha ,\beta \right\} ,$ and it is obtained as
the inverse Laplace transform of (\ref{sr-lap}):%
\begin{eqnarray*}
\tilde{\sigma}_{sr}\left( s\right) &=&\frac{1}{s^{1-\mu }}\frac{%
b_{1}+b_{2}s^{\beta }}{1+a_{1}s^{\alpha }+a_{2}s^{\beta }+a_{3}s^{\alpha
+\beta }} \\
\tilde{\sigma}_{sr}\left( s\right) &=&\frac{1}{s^{1-\mu }}\left(
b_{1}-a_{1}b_{1}s^{\alpha }+O\left( s^{\rho }\right) \right) ,\;\;\text{as}%
\;\;s\rightarrow 0,
\end{eqnarray*}%
where the binomial formula (\ref{wke}) is used.

\subsection{Model V \label{mdl-5}}

Model V, given by (\ref{Model 5}) and subject to thermodynamical
restrictions (\ref{Model 5 - tdr}), is obtained from the unified model (\ref%
{UCE-1-5}) for $\eta =\beta $ and $\gamma =2\beta .$

\subsubsection{Restrictions on range of model parameters}

The requirement that the creep compliance (\ref{CR-ILT-1-5}) is a Bernstein
function, while the relaxation modulus (\ref{SR}) is completely monotonic
narrows down the thermodynamical restriction (\ref{Model 5 - tdr}) to%
\begin{equation}
\frac{a_{3}}{a_{2}}\frac{\left\vert \sin \frac{\left( \mu -2\beta \right)
\pi }{2}\right\vert }{\sin \frac{\mu \pi }{2}}\frac{\cos \frac{\left( \mu
-2\beta \right) \pi }{2}}{\cos \frac{\mu \pi }{2}}\leq \frac{b_{2}}{b_{1}}%
\leq a_{2}\frac{\sin \frac{\left( \mu -\beta \right) \pi }{2}}{\sin \frac{%
\left( \mu +\beta \right) \pi }{2}}\frac{\cos \frac{\left( \mu -\beta
\right) \pi }{2}}{\left\vert \cos \frac{\left( \mu +\beta \right) \pi }{2}%
\right\vert },  \label{pos-K-MD5}
\end{equation}%
since%
\begin{equation}
\frac{\cos \frac{\left( \mu -2\beta \right) \pi }{2}}{\cos \frac{\mu \pi }{2}%
}\geq 1\;\;\text{and}\;\;\frac{\sin \frac{\left( \mu -\beta \right) \pi }{2}%
}{\sin \frac{\left( \mu +\beta \right) \pi }{2}}\leq 1.  \label{ineqs-MD5}
\end{equation}

The requirement (\ref{pos-K-MD5}) is obtained by insuring the non-negativity
of the terms appearing in brackets in function $K,$ given by (\ref{funct-K})
and having the form%
\begin{eqnarray*}
K\left( \rho \right) &=&b_{1}\sin \left( \mu \pi \right) +a_{1}b_{1}\rho
^{\alpha }\sin \left( \left( \mu -\alpha \right) \pi \right) \\
&&+b_{1}\rho ^{\beta }\left\vert \sin \left( \left( \mu +\beta \right) \pi
\right) \right\vert \left( a_{2}\frac{\sin \left( \left( \mu -\beta \right)
\pi \right) }{\left\vert \sin \left( \left( \mu +\beta \right) \pi \right)
\right\vert }-\frac{b_{2}}{b_{1}}\right) +a_{1}b_{2}\rho ^{\alpha +\beta
}\sin \left( \left( \mu +\beta -\alpha \right) \pi \right) \\
&&+a_{2}b_{1}\rho ^{2\beta }\sin \left( \mu \pi \right) \left( \frac{b_{2}}{%
b_{1}}-\frac{a_{3}}{a_{2}}\frac{\left\vert \sin \left( \left( \mu -2\beta
\right) \pi \right) \right\vert }{\sin \left( \mu \pi \right) }\right)
+a_{3}b_{2}\rho ^{3\beta }\sin \left( \left( \mu -\beta \right) \pi \right) ,
\end{eqnarray*}%
since all other terms in $K$ are non-negative.

According to thermodynamical restriction (\ref{Model 5 - tdr}), one has $%
0\leq 2\beta -\mu \leq \beta \leq \mu \leq 1,$ so that $0\leq \frac{\left(
2\beta -\mu \right) \pi }{2}\leq \frac{\mu \pi }{2}\leq \frac{\pi }{2}$
implies the first inequality in (\ref{ineqs-MD5}), while the second
inequality holds since 
\begin{equation*}
\frac{\sin \frac{\left( \mu -\beta \right) \pi }{2}}{\sin \frac{\left( \mu
+\beta \right) \pi }{2}}=\frac{\tan \frac{\mu \pi }{2}-\tan \frac{\beta \pi 
}{2}}{\tan \frac{\mu \pi }{2}+\tan \frac{\beta \pi }{2}}\leq 1,
\end{equation*}%
due to $0\leq \frac{\beta \pi }{2}\leq \frac{\mu \pi }{2}\leq \frac{\pi }{2}$
implying $0\leq \tan \frac{\beta \pi }{2}\leq \tan \frac{\mu \pi }{2}.$

\subsubsection{Asymptotic expansions}

The asymptotic expansion of creep compliance (\ref{CR-ILT-1-5}) for Model V
near initial time-instant is obtained in the form%
\begin{equation}
\varepsilon _{cr}\left( t\right) =\frac{a_{3}}{b_{2}}\frac{t^{\mu -\beta }}{%
\Gamma \left( 1+\mu -\beta \right) }+\frac{a_{2}b_{1}}{b_{2}^{2}}\left( 
\frac{b_{2}}{b_{1}}-\frac{a_{3}}{a_{2}}\right) \frac{t^{\mu }}{\Gamma \left(
1+\mu \right) }+\frac{a_{1}}{b_{2}}\frac{t^{\mu +\beta -\alpha }}{\Gamma
\left( 1+\mu +\beta -\alpha \right) }+O\left( t^{\mu +\beta }\right) ,\;\;%
\text{as}\;\;t\rightarrow 0,  \label{MD5-krip-0}
\end{equation}%
as the inverse Laplace transform of (\ref{cr-LT}) rewritten as 
\begin{eqnarray*}
\tilde{\varepsilon}_{cr}\left( s\right) &=&\frac{1}{b_{2}s^{1+\mu +\beta }}%
\frac{1+a_{1}s^{\alpha }+a_{2}s^{\beta }+a_{3}s^{2\beta }}{1+\frac{b_{1}}{%
b_{2}}\frac{1}{s^{\beta }}} \\
\tilde{\varepsilon}_{cr}\left( s\right) &=&\frac{1}{b_{2}s^{1+\mu +\beta }}%
\left( a_{3}s^{2\beta }+\frac{a_{2}b_{2}-a_{3}b_{1}}{b_{2}}s^{\beta
}+a_{1}s^{\alpha }+O\left( 1\right) \right) ,\;\;\text{as}\;\;s\rightarrow
\infty ,
\end{eqnarray*}%
using the binomial formula (\ref{wke}), while the asymptotic expansion of
creep compliance (\ref{CR-ILT-1-5}) for large time takes the form 
\begin{eqnarray}
\varepsilon _{cr}\left( t\right) &=&\frac{1}{b_{1}}\frac{t^{\mu }}{\Gamma
\left( 1+\mu \right) }+\frac{a_{1}}{b_{1}}\frac{t^{\mu -\alpha }}{\Gamma
\left( 1+\mu -\alpha \right) }+\frac{1}{b_{1}}\left( a_{2}-\frac{b_{2}}{b_{1}%
}\right) \frac{t^{\mu -\beta }}{\Gamma \left( 1+\mu -\beta \right) }  \notag
\\
&&-\frac{a_{1}b_{2}}{b_{1}^{2}}\frac{t^{-\left( \alpha +\beta -\mu \right) }%
}{\Gamma \left( 1-\left( \alpha +\beta -\mu \right) \right) }+O\left( t^{\mu
-2\beta }\right) ,\;\;\text{as}\;\;t%
\begin{tabular}{l}
$\rightarrow $%
\end{tabular}%
\infty ,  \label{MD5-krip-besk}
\end{eqnarray}%
and it is obtained as the inverse Laplace transform of (\ref{CR-ILT-1-5}):%
\begin{eqnarray*}
\tilde{\varepsilon}_{cr}\left( s\right) &=&\frac{1}{b_{1}s^{1+\mu }}\frac{%
1+a_{1}s^{\alpha }+a_{2}s^{\beta }+a_{3}s^{2\beta }}{1+\frac{b_{2}}{b_{1}}%
s^{\beta }} \\
\tilde{\varepsilon}_{cr}\left( s\right) &=&\frac{1}{b_{1}}\frac{1}{s^{1+\mu }%
}\left( 1+a_{1}s^{\alpha }+\frac{a_{2}b_{1}-b_{2}}{b_{1}}s^{\beta }-\frac{%
a_{1}b_{2}}{b_{1}}s^{\alpha +\beta }+O\left( s^{2\beta }\right) \right) ,\;\;%
\text{as}\;\;s\rightarrow 0,
\end{eqnarray*}%
where the binomial formula (\ref{wke}) is used.

The asymptotic expansion of relaxation modulus (\ref{SR}) for Model V near
initial time-instant is obtained in the form%
\begin{equation}
\sigma _{sr}\left( t\right) =\frac{b_{2}}{a_{3}}\frac{t^{-\left( \mu -\beta
\right) }}{\Gamma \left( 1-\left( \mu -\beta \right) \right) }-\frac{%
a_{2}b_{1}}{a_{3}^{2}}\left( \frac{b_{2}}{b_{1}}-\frac{a_{3}}{a_{2}}\right) 
\frac{t^{2\beta -\mu }}{\Gamma \left( 1+2\beta -\mu \right) }-\frac{%
a_{1}b_{2}}{a_{3}^{2}}\frac{t^{3\beta +\alpha -\mu }}{\Gamma \left( 1+3\beta
+\alpha -\mu \right) }+O\left( t^{3\beta -\mu }\right) ,\;\;\text{as}%
\;\;t\rightarrow 0,  \label{MD5-sr-0}
\end{equation}%
as the inverse Laplace transform of the relaxation modulus in complex domain
(\ref{sr-lap}) rewritten as%
\begin{eqnarray*}
\tilde{\sigma}_{sr}\left( s\right) &=&\frac{1}{a_{3}s^{1+2\beta -\mu }}\frac{%
b_{1}+b_{2}s^{\beta }}{1+\frac{a_{1}}{a_{3}}\frac{1}{s^{2\beta -\alpha }}+%
\frac{a_{2}}{a_{3}}\frac{1}{s^{\beta }}+\frac{1}{a_{3}}\frac{1}{s^{2\beta }}}
\\
\tilde{\sigma}_{sr}\left( s\right) &=&\frac{1}{a_{3}s^{1+2\beta -\mu }}%
\left( b_{2}s^{\beta }+\frac{a_{3}b_{1}-a_{2}b_{2}}{a_{3}}-\frac{a_{1}b_{2}}{%
a_{3}}\frac{1}{s^{\beta -\alpha }}+O\left( \frac{1}{s^{\beta }}\right)
\right) ,\;\;\text{as}\;\;s\rightarrow \infty ,
\end{eqnarray*}%
using the binomial formula (\ref{wke}), while the asymptotic expansion of
the relaxation modulus (\ref{SR}) for large time takes the form%
\begin{equation}
\sigma _{sr}\left( t\right) =b_{1}\frac{t^{-\mu }}{\Gamma \left( 1-\mu
\right) }-a_{1}b_{1}\frac{t^{-\mu -\alpha }}{\Gamma \left( 1-\mu -\alpha
\right) }+O\left( t^{-\mu -\rho }\right) ,\;\;\text{as}\;\;t\rightarrow
\infty ,  \label{MD5-sr-besk}
\end{equation}%
with $\rho =\min \left\{ 2\alpha ,\beta \right\} ,$ and it is obtained as
the inverse Laplace transform of (\ref{sr-lap}):%
\begin{eqnarray*}
\tilde{\sigma}_{sr}\left( s\right) &=&\frac{1}{s^{1-\mu }}\frac{%
b_{1}+b_{2}s^{\beta }}{1+a_{1}s^{\alpha }+a_{2}s^{\beta }+a_{3}s^{2\beta }}
\\
\tilde{\sigma}_{sr}\left( s\right) &=&\frac{1}{s^{1-\mu }}\left(
b_{1}-a_{1}b_{1}s^{\alpha }+O\left( s^{\rho }\right) \right) ,\;\;\text{as}%
\;\;s\rightarrow 0,
\end{eqnarray*}%
where the binomial formula (\ref{wke}) is used.

\subsection{Model VI \label{mdl-6}}

Model VI, given by (\ref{Model 6}) and subject to thermodynamical
restrictions (\ref{Model 6 - tdr}), is obtained from the unified
constitutive model (\ref{UCE}) for $\eta =\alpha .$

\subsubsection{Restrictions on range of model parameters}

The requirement that the creep compliance (\ref{CR-ILT}) is a Bernstein
function, while the relaxation modulus (\ref{SR-ILT}) is completely
monotonic narrows down the thermodynamical restriction (\ref{Model 6 - tdr})
to%
\begin{equation}
\frac{a_{3}}{a_{2}}\leq \frac{b_{2}}{b_{1}}\leq a_{1}\frac{\sin \frac{\left(
\beta -\alpha \right) \pi }{2}}{\sin \frac{\left( \beta +\alpha \right) \pi 
}{2}}\frac{\cos \frac{\left( \beta -\alpha \right) \pi }{2}}{\left\vert \cos 
\frac{\left( \beta +\alpha \right) \pi }{2}\right\vert }\leq a_{1}\frac{\cos 
\frac{\left( \beta -\alpha \right) \pi }{2}}{\left\vert \cos \frac{\left(
\beta +\alpha \right) \pi }{2}\right\vert }.  \label{Model 6 - nerov}
\end{equation}%
The requirement (\ref{Model 6 - nerov}) is obtained by insuring the
non-negativity of function $Q,$ given by (\ref{Q}) and having the form%
\begin{eqnarray*}
Q\left( \rho \right) &=&\frac{1}{a_{3}}\frac{b_{1}}{b_{2}}\sin \left( \beta
\pi \right) +\frac{1}{a_{3}}\frac{b_{1}}{b_{2}}\rho ^{\alpha }\left\vert
\sin \left( \left( \beta +\alpha \right) \pi \right) \right\vert \left( a_{1}%
\frac{\sin \left( \left( \beta -\alpha \right) \pi \right) }{\left\vert \sin
\left( \left( \beta +\alpha \right) \pi \right) \right\vert }-\frac{b_{2}}{%
b_{1}}\right) \\
&&+\frac{a_{1}}{a_{3}}\rho ^{2\alpha }\sin \left( \beta \pi \right) +\left( 
\frac{a_{2}}{a_{3}}-\frac{b_{1}}{b_{2}}\right) \rho ^{\alpha +\beta }\sin
\left( \alpha \pi \right)
\end{eqnarray*}%
in the case of Model VI. Due to thermodynamical requirement (\ref{Model 6 -
tdr}), one has $\frac{a_{2}}{a_{3}}-\frac{b_{1}}{b_{2}}\geq 0,$ while the
non-negativity of $Q$ is guaranteed if the term in brackets is non-negative,
yielding%
\begin{equation*}
\frac{b_{2}}{b_{1}}\leq a_{1}\frac{\sin \frac{\left( \beta -\alpha \right)
\pi }{2}}{\sin \frac{\left( \beta +\alpha \right) \pi }{2}}\frac{\cos \frac{%
\left( \beta -\alpha \right) \pi }{2}}{\left\vert \cos \frac{\left( \beta
+\alpha \right) \pi }{2}\right\vert }\leq a_{1}\frac{\cos \frac{\left( \beta
-\alpha \right) \pi }{2}}{\left\vert \cos \frac{\left( \beta +\alpha \right)
\pi }{2}\right\vert },
\end{equation*}%
since 
\begin{equation*}
\frac{\sin \frac{\left( \beta -\alpha \right) \pi }{2}}{\sin \frac{\left(
\beta +\alpha \right) \pi }{2}}=\frac{\tan \frac{\beta \pi }{2}-\tan \frac{%
\alpha \pi }{2}}{\tan \frac{\beta \pi }{2}+\tan \frac{\alpha \pi }{2}}\leq 1,
\end{equation*}%
due to $0\leq \frac{\alpha \pi }{2}\leq \frac{\beta \pi }{2}\leq \frac{\pi }{%
2}$ implying $0\leq \tan \frac{\alpha \pi }{2}\leq \tan \frac{\beta \pi }{2}%
. $

\subsubsection{Asymptotic expansions}

The asymptotic expansion of creep compliance (\ref{CR-ILT}) for Model VI
near initial time-instant is obtained in the form%
\begin{equation}
\varepsilon _{cr}\left( t\right) =\frac{a_{3}}{b_{2}}+\frac{a_{2}b_{1}}{%
b_{2}^{2}}\left( \frac{b_{2}}{b_{1}}-\frac{a_{3}}{a_{2}}\right) \frac{%
t^{\alpha }}{\Gamma \left( 1+\alpha \right) }+O\left( t^{\alpha +\beta -\rho
}\right) ,\;\;\text{as}\;\;t\rightarrow 0,  \label{MD6-krip-0}
\end{equation}%
with $\rho =\max \left\{ \alpha ,\beta -\alpha \right\} ,$ as the inverse
Laplace transform of the creep compliance in complex domain (\ref{cr-LT-1})
rewritten as 
\begin{eqnarray*}
\tilde{\varepsilon}_{cr}\left( s\right) &=&\frac{1}{b_{2}s^{1+\alpha +\beta }%
}\frac{1+a_{1}s^{\alpha }+a_{2}s^{\beta }+a_{3}s^{\alpha +\beta }}{1+\frac{%
b_{1}}{b_{2}}\frac{1}{s^{\alpha }}} \\
\tilde{\varepsilon}_{cr}\left( s\right) &=&\frac{1}{b_{2}s^{1+\alpha +\beta }%
}\left( a_{3}s^{\alpha +\beta }+\frac{a_{2}b_{2}-a_{3}b_{1}}{b_{2}}s^{\beta
}+O\left( s^{\rho }\right) \right) ,\;\;\text{as}\;\;s\rightarrow \infty ,
\end{eqnarray*}%
using the binomial formula (\ref{wke}), while the asymptotic expansion of
creep compliance (\ref{CR-ILT}) for large time takes the form 
\begin{equation}
\varepsilon _{cr}\left( t\right) =\frac{1}{b_{1}}\frac{t^{\beta }}{\Gamma
\left( 1+\beta \right) }+\frac{1}{b_{1}}\left( a_{1}-\frac{b_{2}}{b_{1}}%
\right) \frac{t^{\beta -\alpha }}{\Gamma \left( 1+\beta -\alpha \right) }%
+O\left( t^{\beta -\rho }\right) ,\;\;\text{as}\;\;t\rightarrow \infty ,
\label{MD6-krip-besk}
\end{equation}%
with $\rho =\min \left\{ 2\alpha ,\beta \right\} ,$ and it is obtained as
the inverse Laplace transform of (\ref{cr-LT-1}):%
\begin{eqnarray*}
\tilde{\varepsilon}_{cr}\left( s\right) &=&\frac{1}{b_{1}s^{1+\beta }}\frac{%
1+a_{1}s^{\alpha }+a_{2}s^{\beta }+a_{3}s^{\alpha +\beta }}{1+\frac{b_{2}}{%
b_{1}}s^{\alpha }} \\
\tilde{\varepsilon}_{cr}\left( s\right) &=&\frac{1}{b_{1}s^{1+\beta }}\left(
1+\frac{a_{1}b_{1}-b_{2}}{b_{1}}s^{\alpha }+O\left( s^{\rho }\right) \right)
,\;\;\text{as}\;\;s\rightarrow 0,
\end{eqnarray*}%
where the binomial formula (\ref{wke}) is used.

The asymptotic expansion of relaxation modulus (\ref{SR-ILT}) for Model VI
near initial time-instant is obtained in the form%
\begin{equation}
\sigma _{sr}\left( t\right) =\frac{b_{2}}{a_{3}}-\frac{a_{2}b_{1}}{a_{3}^{2}}%
\left( \frac{b_{2}}{b_{1}}-\frac{a_{3}}{a_{2}}\right) \frac{t^{\alpha }}{%
\Gamma \left( 1+\alpha \right) }+O\left( t^{\alpha +\rho }\right) ,\;\;\text{%
as}\;\;t\rightarrow 0,  \label{MD6-sr-0}
\end{equation}%
with $\rho =\min \left\{ \alpha ,\beta -\alpha \right\} ,$ as the inverse
Laplace transform of the relaxation modulus in complex domain (\ref{sr-lap-1}%
) rewritten as 
\begin{eqnarray*}
\tilde{\sigma}_{sr}\left( s\right) &=&\frac{1}{a_{3}s^{1+\alpha }}\frac{%
b_{1}+b_{2}s^{\alpha }}{1+\frac{a_{1}}{a_{3}}\frac{1}{s^{\beta }}+\frac{a_{2}%
}{a_{3}}\frac{1}{s^{\alpha }}+\frac{1}{a_{3}}\frac{1}{s^{\alpha +\beta }}} \\
\tilde{\sigma}_{sr}\left( s\right) &=&\frac{1}{a_{3}s^{1+\alpha }}\left(
b_{2}s^{\alpha }+\frac{a_{3}b_{1}-a_{2}b_{2}}{a_{3}}+O\left( \frac{1}{%
s^{\rho }}\right) \right) ,\;\;\text{as}\;\;s\rightarrow \infty ,
\end{eqnarray*}%
using the binomial formula (\ref{wke}), while the asymptotic expansion of
the relaxation modulus (\ref{SR-ILT}) for large time takes the form%
\begin{equation}
\sigma _{sr}\left( t\right) =b_{1}\frac{t^{-\beta }}{\Gamma \left( 1-\beta
\right) }-b_{1}\left( a_{1}-\frac{b_{2}}{b_{1}}\right) \frac{t^{-\beta
-\alpha }}{\Gamma \left( 1-\beta -\alpha \right) }+O\left( t^{-\beta -\rho
}\right) ,\;\;\text{as}\;\;t\rightarrow \infty ,  \label{MD6-sr-besk}
\end{equation}%
with $\rho =\min \left\{ 2\alpha ,\beta \right\} ,$ and it is obtained as
the inverse Laplace transform of (\ref{sr-lap-1}):%
\begin{eqnarray*}
\tilde{\sigma}_{sr}\left( s\right) &=&\frac{1}{s^{1-\beta }}\frac{%
b_{1}+b_{2}s^{\alpha }}{1+a_{1}s^{\alpha }+a_{2}s^{\beta }+a_{3}s^{\alpha
+\beta }} \\
\tilde{\sigma}_{sr}\left( s\right) &=&\frac{1}{s^{1-\beta }}\left(
b_{1}+\left( b_{2}-a_{1}b_{1}\right) s^{\alpha }+O\left( s^{\rho }\right)
\right) ,\;\;\text{as}\;\;s\rightarrow 0,
\end{eqnarray*}%
where the binomial formula (\ref{wke}) is used.

Note that the term $\frac{b_{2}}{b_{1}}-\frac{a_{3}}{a_{2}}$ is non-negative
due to the thermodynamical requirements (\ref{Model 6 - tdr}).

\subsection{Model VII \label{mdl-7}}

Model VII, given by (\ref{Model 7}) and subject to thermodynamical
restrictions (\ref{Model 7 - tdr}), is obtained from the unified model (\ref%
{UCE}) for $\eta =\beta .$

\subsubsection{Restrictions on range of model parameters}

The requirement that the creep compliance (\ref{CR-ILT}) is a Bernstein
function, while the relaxation modulus (\ref{SR-ILT}) is completely
monotonic narrows down the thermodynamical restriction (\ref{Model 7 - tdr})
to%
\begin{equation}
\frac{a_{3}}{a_{2}}\leq \frac{a_{2}}{2\cos ^{2}\left( \beta \pi \right) }%
\left( 1-\sqrt{1-\frac{4a_{3}\cos ^{2}\left( \beta \pi \right) }{a_{2}^{2}}}%
\right) \leq \frac{b_{2}}{b_{1}}\leq \frac{a_{2}}{\left\vert \cos \left(
\beta \pi \right) \right\vert },  \label{Model 7 - nerov}
\end{equation}%
provided that 
\begin{equation*}
2\sqrt{a_{3}}\leq \frac{a_{2}}{\left\vert \cos \left( \beta \pi \right)
\right\vert },
\end{equation*}%
guaranteeing the non-negativity of term under the square root. The
requirement (\ref{Model 7 - nerov}) is obtained by insuring the
non-negativity of function $Q,$ given by (\ref{Q}) and having the form%
\begin{eqnarray*}
Q\left( \rho \right) &=&\frac{a_{1}}{a_{3}}\frac{b_{1}}{b_{2}}\rho ^{\alpha
}\sin \left( \left( \beta -\alpha \right) \pi \right) +\frac{a_{1}}{a_{3}}%
\rho ^{\alpha +\beta }\sin \left( \left( 2\beta -\alpha \right) \pi \right)
\\
&&+\frac{a_{2}}{a_{3}}\frac{b_{1}}{b_{2}}\left( \left( \frac{b_{2}}{b_{1}}-%
\frac{a_{3}}{a_{2}}\right) \rho ^{2\beta }-2\rho ^{\beta }\frac{\left\vert
\cos \left( \beta \pi \right) \right\vert }{a_{2}}\frac{b_{2}}{b_{1}}+\frac{1%
}{a_{2}}\right) \sin \left( \beta \pi \right)
\end{eqnarray*}%
in the case of Model VII. Due to thermodynamical requirement (\ref{Model 7 -
tdr}), the first two terms in function $Q$ are non-negative, as well as $%
\frac{b_{2}}{b_{1}}-\frac{a_{3}}{a_{2}}\geq 0,$ so the non-negativity of $Q$
is guaranteed if the quadratic function in $\rho ^{\beta }$ is non-negative,
i.e., if its discriminant is non-positive, yielding%
\begin{equation*}
\left( \frac{\left\vert \cos \left( \beta \pi \right) \right\vert }{a_{2}}%
\right) ^{2}\left( \frac{b_{2}}{b_{1}}\right) ^{2}-\frac{1}{a_{2}}\frac{b_{2}%
}{b_{1}}+\frac{a_{3}}{a_{2}^{2}}\leq 0,
\end{equation*}%
that solved with respect to $\frac{b_{2}}{b_{1}}$ gives%
\begin{equation}
\frac{a_{2}}{2\cos ^{2}\left( \beta \pi \right) }\left( 1-\sqrt{1-\frac{%
4a_{3}\cos ^{2}\left( \beta \pi \right) }{a_{2}^{2}}}\right) \leq \frac{b_{2}%
}{b_{1}}\leq \frac{a_{2}}{2\cos ^{2}\left( \beta \pi \right) }\left( 1+\sqrt{%
1-\frac{4a_{3}\cos ^{2}\left( \beta \pi \right) }{a_{2}^{2}}}\right) .
\label{pos-K-MD7}
\end{equation}%
In order to prove the left-hand-side of (\ref{Model 7 - nerov}), one uses
the binomial formula (\ref{bin-form}) in (\ref{pos-K-MD7}) and obtains%
\begin{equation*}
\frac{a_{2}}{2\cos ^{2}\left( \beta \pi \right) }\left( 1-\sqrt{1-\frac{%
4a_{3}\cos ^{2}\left( \beta \pi \right) }{a_{2}^{2}}}\right) =\frac{a_{3}}{%
a_{2}}+\frac{a_{2}}{2\cos ^{2}\left( \beta \pi \right) }\sum_{k=2}^{\infty }%
\frac{\left( 2k-3\right) !!}{2^{k}k!}\left( \frac{4a_{3}\cos ^{2}\left(
\beta \pi \right) }{a_{2}^{2}}\right) ^{k}\leq \frac{a_{3}}{a_{2}}.
\end{equation*}%
On the other hand, from (\ref{pos-K-MD7}) one has%
\begin{equation*}
\frac{b_{2}}{b_{1}}\leq \frac{a_{2}}{2\cos ^{2}\left( \beta \pi \right) }%
\left( 1+\sqrt{1-\frac{4a_{3}\cos ^{2}\left( \beta \pi \right) }{a_{2}^{2}}}%
\right) \leq \frac{a_{2}}{\cos ^{2}\left( \beta \pi \right) }
\end{equation*}%
and since $\frac{1}{\left\vert \cos \left( \beta \pi \right) \right\vert }%
\leq \frac{1}{\cos ^{2}\left( \beta \pi \right) },$ the thermodynamical
requirement remains on the right-hand-side of (\ref{Model 7 - nerov}).

\subsubsection{Asymptotic expansions}

The asymptotic expansion of creep compliance (\ref{CR-ILT}) for Model VII
near initial time-instant is obtained in the form%
\begin{eqnarray}
\varepsilon _{cr}\left( t\right) &=&\frac{a_{3}}{b_{2}}+\frac{a_{2}b_{1}}{%
b_{2}^{2}}\left( \frac{b_{2}}{b_{1}}-\frac{a_{3}}{a_{2}}\right) \frac{%
t^{\beta }}{\Gamma \left( 1+\beta \right) }+\frac{a_{1}}{b_{2}}\frac{%
t^{2\beta -\alpha }}{\Gamma \left( 1+2\beta -\alpha \right) }  \notag \\
&&+\frac{1}{b_{2}}\left( 1-\frac{a_{2}b_{1}}{b_{2}}+\frac{a_{3}b_{1}^{2}}{%
b_{2}^{2}}\right) \frac{t^{2\beta }}{\Gamma \left( 1+2\beta \right) }%
+O\left( t^{3\beta }\right) ,\;\;\text{as}\;\;t%
\begin{tabular}{l}
$\rightarrow $%
\end{tabular}%
0,  \label{MD7-krip-0}
\end{eqnarray}%
as the inverse Laplace transform of the creep compliance in complex domain (%
\ref{cr-LT-1}) rewritten as 
\begin{eqnarray*}
\tilde{\varepsilon}_{cr}\left( s\right) &=&\frac{1}{b_{2}s^{1+2\beta }}\frac{%
1+a_{1}s^{\alpha }+a_{2}s^{\beta }+a_{3}s^{2\beta }}{1+\frac{b_{1}}{b_{2}}%
\frac{1}{s^{\beta }}} \\
\tilde{\varepsilon}_{cr}\left( s\right) &=&\frac{1}{b_{2}s^{1+2\beta }}%
\left( a_{3}s^{2\beta }+\left( a_{2}-\frac{a_{3}b_{1}}{b_{2}}\right)
s^{\beta }+a_{1}s^{\alpha }+1-\frac{a_{2}b_{1}}{b_{2}}+\frac{a_{3}b_{1}^{2}}{%
b_{2}^{2}}+O\left( s^{-\beta }\right) \right) ,\;\;\text{as}\;\;s\rightarrow
\infty ,
\end{eqnarray*}%
using the binomial formula (\ref{wke}), while the asymptotic expansion of
creep compliance (\ref{CR-ILT}) for large time takes the form 
\begin{equation}
\varepsilon _{cr}\left( t\right) =\frac{1}{b_{1}}\frac{t^{\beta }}{\Gamma
\left( 1+\beta \right) }+\frac{a_{1}}{b_{1}}\frac{t^{\beta -\alpha }}{\Gamma
\left( 1+\beta -\alpha \right) }+\frac{1}{b_{1}}\left( a_{2}-\frac{b_{2}}{%
b_{1}}\right) -\frac{a_{1}b_{2}}{b_{1}^{2}}\frac{t^{-\alpha }}{\Gamma \left(
1-\alpha \right) }+O\left( t^{-\beta }\right) ,\;\;\text{as}\;\;t\rightarrow
\infty ,  \label{MD7-krip-besk}
\end{equation}%
and it is obtained as the inverse Laplace transform of (\ref{cr-LT-1}):%
\begin{eqnarray*}
\tilde{\varepsilon}_{cr}\left( s\right) &=&\frac{1}{b_{1}s^{1+\beta }}\frac{%
1+a_{1}s^{\alpha }+a_{2}s^{\beta }+a_{3}s^{2\beta }}{1+\frac{b_{2}}{b_{1}}%
s^{\beta }} \\
\tilde{\varepsilon}_{cr}\left( s\right) &=&\frac{1}{b_{1}s^{1+\beta }}\left(
1+a_{1}s^{\alpha }+\left( a_{2}-\frac{b_{2}}{b_{1}}\right) s^{\beta }-\frac{%
a_{1}b_{2}}{b_{1}}s^{\alpha +\beta }+O\left( s^{2\beta }\right) \right) ,\;\;%
\text{as}\;\;s\rightarrow 0,
\end{eqnarray*}%
where the binomial formula (\ref{wke}) is used.

The asymptotic expansion of relaxation modulus (\ref{SR-ILT}) for Model VII
near initial time-instant is obtained in the form%
\begin{equation}
\sigma _{sr}\left( t\right) =\frac{b_{2}}{a_{3}}-\frac{a_{2}b_{1}}{a_{3}^{2}}%
\left( \frac{b_{2}}{b_{1}}-\frac{a_{3}}{a_{2}}\right) \frac{t^{\beta }}{%
\Gamma \left( 1+\beta \right) }-\frac{a_{1}b_{2}}{a_{3}^{2}}\frac{t^{2\beta
-\alpha }}{\Gamma \left( 1+2\beta -\alpha \right) }+O\left( t^{2\beta
}\right) ,\;\;\text{as}\;\;t\rightarrow 0,  \label{MD7-sr-0}
\end{equation}%
as the inverse Laplace transform of the relaxation modulus in complex domain
(\ref{sr-lap-1}) rewritten as 
\begin{eqnarray*}
\tilde{\sigma}_{sr}\left( s\right) &=&\frac{1}{a_{3}s^{1+\beta }}\frac{%
b_{1}+b_{2}s^{\beta }}{1+\frac{a_{1}}{a_{3}}\frac{1}{s^{2\beta -\alpha }}+%
\frac{a_{2}}{a_{3}}\frac{1}{s^{\beta }}+\frac{1}{a_{3}}\frac{1}{s^{2\beta }}}
\\
\tilde{\sigma}_{sr}\left( s\right) &=&\frac{1}{a_{3}s^{1+\beta }}\left(
b_{2}s^{\beta }+\frac{a_{3}b_{1}-a_{2}b_{2}}{a_{3}}-\frac{a_{1}b_{2}}{a_{3}}%
s^{\alpha }+O\left( 1\right) \right) ,\;\;\text{as}\;\;s\rightarrow \infty ,
\end{eqnarray*}%
using the binomial formula (\ref{wke}), while the asymptotic expansion of
the relaxation modulus (\ref{SR-ILT}) for large time takes the form%
\begin{eqnarray}
\sigma _{sr}\left( t\right) &=&b_{1}\frac{t^{-\beta }}{\Gamma \left( 1-\beta
\right) }-a_{1}b_{1}\frac{t^{-\beta -\alpha }}{\Gamma \left( 1-\beta -\alpha
\right) }-b_{1}\left( a_{2}-\frac{b_{2}}{b_{1}}\right) \frac{t^{-2\beta }}{%
\Gamma \left( 1-2\beta \right) }  \notag \\
&&+a_{1}^{2}b_{1}\frac{t^{-\beta -2\alpha }}{\Gamma \left( 1-\beta -2\alpha
\right) }+O\left( t^{-2\beta -\alpha }\right) ,\;\;\text{as}\;\;t%
\begin{tabular}{l}
$\rightarrow $%
\end{tabular}%
\infty ,  \label{MD7-sr-besk}
\end{eqnarray}%
and it is obtained as the inverse Laplace transform of (\ref{sr-lap-1}):%
\begin{eqnarray*}
\tilde{\sigma}_{sr}\left( s\right) &=&\frac{1}{s^{1-\beta }}\frac{%
b_{1}+b_{2}s^{\beta }}{1+a_{1}s^{\alpha }+a_{2}s^{\beta }+a_{3}s^{2\beta }}
\\
\tilde{\sigma}_{sr}\left( s\right) &=&\frac{1}{s^{1-\beta }}\left(
b_{1}-a_{1}b_{1}s^{\alpha }+\left( b_{2}-a_{2}b_{1}\right) s^{\beta
}+a_{1}^{2}b_{1}s^{2\alpha }+O\left( s^{\alpha +\beta }\right) \right) ,\;\;%
\text{as}\;\;s\rightarrow 0,
\end{eqnarray*}%
where the binomial formula (\ref{wke}) is used.

Note that the term $\frac{b_{2}}{b_{1}}-\frac{a_{3}}{a_{2}}$ is non-negative
due to the thermodynamical requirements (\ref{Model 7 - tdr}).

\subsection{Model VIII \label{mdl-8}}

Model VIII, given by (\ref{Model 8}) and subject to thermodynamical
restrictions (\ref{Model 8 - tdr}), is obtained from the unified model (\ref%
{UCE}) for $\eta =\beta =\alpha ,$ $\bar{a}_{1}=a_{1}+a_{2}$ and $\bar{a}%
_{2}=a_{3}.$

\subsubsection{Restrictions on range of model parameters}

The requirement that the creep compliance (\ref{CR-ILT-1}) is a Bernstein
function, while the relaxation modulus (\ref{SR-ILT-1}) is completely
monotonic narrows down the thermodynamical restriction (\ref{Model 8 - tdr})
to%
\begin{equation}
\frac{\bar{a}_{2}}{\bar{a}_{1}}\leq \frac{\bar{a}_{1}}{2\cos ^{2}\left(
\alpha \pi \right) }\left( 1-\sqrt{1-\frac{4\bar{a}_{2}\cos ^{2}\left(
\alpha \pi \right) }{\bar{a}_{1}^{2}}}\right) \leq \frac{b_{2}}{b_{1}}\leq 
\frac{\bar{a}_{1}}{\left\vert \cos \left( \alpha \pi \right) \right\vert },
\label{Model 8 - nerov}
\end{equation}%
provided that 
\begin{equation*}
2\sqrt{\bar{a}_{2}}\leq \frac{\bar{a}_{1}}{\left\vert \cos \left( \alpha \pi
\right) \right\vert },
\end{equation*}%
guaranteeing the non-negativity of term under the square root. The
requirement (\ref{Model 8 - nerov}) is obtained by insuring the
non-negativity of function $\bar{Q},$ given by (\ref{Q-bar}) and having the
form%
\begin{equation*}
\bar{Q}\left( \rho \right) =\frac{\bar{a}_{1}}{\bar{a}_{2}}\frac{b_{1}}{b_{2}%
}\left( \left( \frac{b_{2}}{b_{1}}-\frac{\bar{a}_{2}}{\bar{a}_{1}}\right)
\rho ^{2\alpha }-2\frac{b_{2}}{b_{1}}\frac{\left\vert \cos \left( \alpha \pi
\right) \right\vert }{\bar{a}_{1}}\rho ^{\alpha }+\frac{1}{\bar{a}_{1}}%
\right) \sin \left( \alpha \pi \right)
\end{equation*}%
in the case of Model VIII. Due to thermodynamical requirement (\ref{Model 8
- tdr}), one has $\frac{b_{2}}{b_{1}}-\frac{\bar{a}_{2}}{\bar{a}_{1}}\geq 0,$
while the non-negativity of $\bar{Q}$ is guaranteed if the quadratic
function in $\rho ^{\alpha }$ is non-negative, i.e., if its discriminant is
non-positive, yielding%
\begin{equation*}
\left( \frac{\left\vert \cos \left( \alpha \pi \right) \right\vert }{\bar{a}%
_{1}}\right) ^{2}\left( \frac{b_{2}}{b_{1}}\right) ^{2}-\frac{1}{\bar{a}_{1}}%
\frac{b_{2}}{b_{1}}+\frac{\bar{a}_{2}}{\bar{a}_{1}^{2}}\leq 0,
\end{equation*}%
that solved with respect to $\frac{b_{2}}{b_{1}}$ gives%
\begin{equation}
\frac{\bar{a}_{1}}{2\cos ^{2}\left( \alpha \pi \right) }\left( 1-\sqrt{1-%
\frac{4\bar{a}_{2}\cos ^{2}\left( \alpha \pi \right) }{\bar{a}_{1}^{2}}}%
\right) \leq \frac{b_{2}}{b_{1}}\leq \frac{\bar{a}_{1}}{2\cos ^{2}\left(
\alpha \pi \right) }\left( 1+\sqrt{1-\frac{4\bar{a}_{2}\cos ^{2}\left(
\alpha \pi \right) }{\bar{a}_{1}^{2}}}\right) .  \label{pos-K-MD8}
\end{equation}%
In order to prove the left-hand-side of (\ref{Model 8 - nerov}), one uses
the binomial formula (\ref{bin-form}) in (\ref{pos-K-MD8}) and obtains%
\begin{equation*}
\frac{\bar{a}_{1}}{2\cos ^{2}\left( \alpha \pi \right) }\left( 1-\sqrt{1-%
\frac{4\bar{a}_{2}\cos ^{2}\left( \alpha \pi \right) }{\bar{a}_{1}^{2}}}%
\right) =\frac{\bar{a}_{2}}{\bar{a}_{1}}+\frac{\bar{a}_{1}}{2\cos ^{2}\left(
\alpha \pi \right) }\sum_{k=2}^{\infty }\frac{\left( 2k-3\right) !!}{2^{k}k!}%
\left( \frac{4\bar{a}_{2}\cos ^{2}\left( \alpha \pi \right) }{\bar{a}_{1}^{2}%
}\right) ^{k}\leq \frac{\bar{a}_{2}}{\bar{a}_{1}}.
\end{equation*}%
On the other hand, from (\ref{pos-K-MD8}) one has%
\begin{equation*}
\frac{b_{2}}{b_{1}}\leq \frac{\bar{a}_{1}}{2\cos ^{2}\left( \alpha \pi
\right) }\left( 1+\sqrt{1-\frac{4\bar{a}_{2}\cos ^{2}\left( \alpha \pi
\right) }{\bar{a}_{1}^{2}}}\right) \leq \frac{\bar{a}_{1}}{\cos ^{2}\left(
\alpha \pi \right) }
\end{equation*}%
and since $\frac{1}{\left\vert \cos \left( \alpha \pi \right) \right\vert }%
\leq \frac{1}{\cos ^{2}\left( \alpha \pi \right) },$ the thermodynamical
requirement remains on the right-hand-side of (\ref{Model 8 - nerov}).

\subsubsection{Asymptotic expansions}

The asymptotic expansion of creep compliance (\ref{CR-ILT-1}) for Model VIII
near initial time-instant is obtained in the form%
\begin{equation}
\varepsilon _{cr}\left( t\right) =\frac{\bar{a}_{2}}{b_{2}}+\frac{\bar{a}%
_{1}b_{1}}{b_{2}^{2}}\left( \frac{b_{2}}{b_{1}}-\frac{\bar{a}_{2}}{\bar{a}%
_{1}}\right) \frac{t^{\alpha }}{\Gamma \left( 1+\alpha \right) }+O\left(
t^{2\alpha }\right) ,\;\;\text{as}\;\;t\rightarrow 0,  \label{MD8-krip-0}
\end{equation}%
as the inverse Laplace transform the creep compliance in complex domain (\ref%
{cr-LT-2}) rewritten as 
\begin{eqnarray*}
\tilde{\varepsilon}_{cr}\left( s\right) &=&\frac{1}{b_{2}s^{1+2\alpha }}%
\frac{1+\bar{a}_{1}s^{\alpha }+\bar{a}_{2}s^{2\alpha }}{1+\frac{b_{1}}{b_{2}}%
\frac{1}{s^{\alpha }}} \\
\tilde{\varepsilon}_{cr}\left( s\right) &=&\frac{1}{b_{2}s^{1+2\alpha }}%
\left( \bar{a}_{2}s^{2\alpha }+\frac{\bar{a}_{1}b_{2}-\bar{a}_{2}b_{1}}{b_{2}%
}s^{\alpha }+O\left( 1\right) \right) ,\;\;\text{as}\;\;s\rightarrow \infty ,
\end{eqnarray*}%
using the binomial formula (\ref{wke}), while the asymptotic expansion of
creep compliance (\ref{CR-ILT-1}) for large time takes the form 
\begin{equation}
\varepsilon _{cr}\left( t\right) =\frac{1}{b_{1}}\frac{t^{\alpha }}{\Gamma
\left( 1+\alpha \right) }+\bar{a}_{1}-\frac{b_{2}}{b_{1}}+O\left( t^{-\alpha
}\right) ,\;\;\text{as}\;\;t\rightarrow \infty ,  \label{MD8-krip-besk}
\end{equation}%
and it is obtained as the inverse Laplace transform of (\ref{cr-LT-2}):%
\begin{eqnarray*}
\tilde{\varepsilon}_{cr}\left( s\right) &=&\frac{1}{b_{1}s^{1+\alpha }}\frac{%
1+\bar{a}_{1}s^{\alpha }+\bar{a}_{2}s^{2\alpha }}{1+\frac{b_{2}}{b_{1}}%
s^{\alpha }} \\
\tilde{\varepsilon}_{cr}\left( s\right) &=&\frac{1}{b_{1}s^{1+\alpha }}%
\left( 1+\frac{\bar{a}_{1}b_{1}-b_{2}}{b_{1}}s^{\alpha }+O\left( s^{2\alpha
}\right) \right) ,\;\;\text{as}\;\;s\rightarrow 0,
\end{eqnarray*}%
where the binomial formula (\ref{wke}) is used.

The asymptotic expansion of relaxation modulus (\ref{SR-ILT-1}) for Model
VIII near initial time-instant is obtained in the form%
\begin{equation}
\sigma _{sr}\left( t\right) =\frac{b_{2}}{\bar{a}_{2}}-\frac{\bar{a}_{1}b_{1}%
}{\bar{a}_{2}^{2}}\left( \frac{b_{2}}{b_{1}}-\frac{\bar{a}_{2}}{\bar{a}_{1}}%
\right) \frac{t^{\alpha }}{\Gamma \left( 1+\alpha \right) }+O\left(
t^{2\alpha }\right) ,\;\;\text{as}\;\;t\rightarrow 0,  \label{MD8-sr-0}
\end{equation}%
as the inverse Laplace transform of the relaxation modulus in complex domain
(\ref{sr-lap-2}) rewritten as 
\begin{eqnarray*}
\tilde{\sigma}_{sr}\left( s\right) &=&\frac{1}{\bar{a}_{2}s^{1+\alpha }}%
\frac{b_{1}+b_{2}s^{\alpha }}{1+\frac{\bar{a}_{1}}{\bar{a}_{2}}\frac{1}{%
s^{\alpha }}+\frac{1}{\bar{a}_{2}}\frac{1}{s^{2\alpha }}} \\
\tilde{\sigma}_{sr}\left( s\right) &=&\frac{1}{\bar{a}_{2}s^{1+\alpha }}%
\left( b_{2}s^{\alpha }+\frac{\bar{a}_{2}b_{1}-\bar{a}_{1}b_{2}}{\bar{a}_{2}}%
+O\left( \frac{1}{s^{\alpha }}\right) \right) ,\;\;\text{as}\;\;s\rightarrow
\infty ,
\end{eqnarray*}%
using the binomial formula (\ref{wke}), while the asymptotic expansion of
relaxation modulus (\ref{SR-ILT-1}) for large time takes the form%
\begin{equation}
\sigma _{sr}\left( t\right) =b_{1}\frac{t^{-\alpha }}{\Gamma \left( 1-\alpha
\right) }-b_{1}\left( \bar{a}_{1}-\frac{b_{2}}{b_{1}}\right) \frac{%
t^{-2\alpha }}{\Gamma \left( 1-2\alpha \right) }+O\left( t^{-3\alpha
}\right) ,\;\;\text{as}\;\;t\rightarrow \infty ,  \label{MD8-sr-besk}
\end{equation}%
and it is obtained as the inverse Laplace transform of (\ref{sr-lap-2}):%
\begin{eqnarray*}
\tilde{\sigma}_{sr}\left( s\right) &=&\frac{1}{s^{1-\alpha }}\frac{%
b_{1}+b_{2}s^{\alpha }}{1+\bar{a}_{1}s^{\alpha }+\bar{a}_{2}s^{2\alpha }} \\
\tilde{\sigma}_{sr}\left( s\right) &=&\frac{1}{s^{1-\alpha }}\left(
b_{1}+\left( b_{2}-\bar{a}_{1}b_{1}\right) s^{\alpha }+O\left( s^{2\alpha
}\right) \right) ,\;\;\text{as}\;\;s\rightarrow 0,
\end{eqnarray*}%
where the binomial formula (\ref{wke}) is used.

Note that the term $\frac{b_{2}}{b_{1}}-\frac{\bar{a}_{2}}{\bar{a}_{1}}$ is
non-negative due to the thermodynamical requirements (\ref{Model 8 - tdr}).

\section{Numerical examples}

Models I - VIII display similar behavior in creep and stress relation tests,
except for the behavior near the initial time-instant when the creep
compliance either starts at zero deformation (Models I - V), or has a jump
(Models VI - VIII), while the relaxation modulus decreases either from
infinite (Models I - V), or from a finite value of stress (Models VI -
VIII), see (\ref{gg}) and (\ref{jg}). The growth of creep compliances to
infinity is governed by (\ref{CR-1-5}) or (\ref{CR-ILT-1-5}) in the case of
Models I - V and by (\ref{CR}) or (\ref{CR-ILT}) in the case of Models VI -
VII, i.e., by (\ref{CR-1}) or (\ref{CR-ILT-1}) for Model VIII, while the
relaxation modulus either monotonically, or non-monotonically tends to zero,
governed by (\ref{SR}) for Models I - V, or by (\ref{SR-ILT}) for Models VI
- VII, i.e., by (\ref{SR-ILT-1}) in the case of Model VII. The material
responses in creep and stress relaxation tests will be illustrated taking
the example of Model VII.

\begin{figure}[tbph]
\begin{center}
\includegraphics[width=0.6\columnwidth]{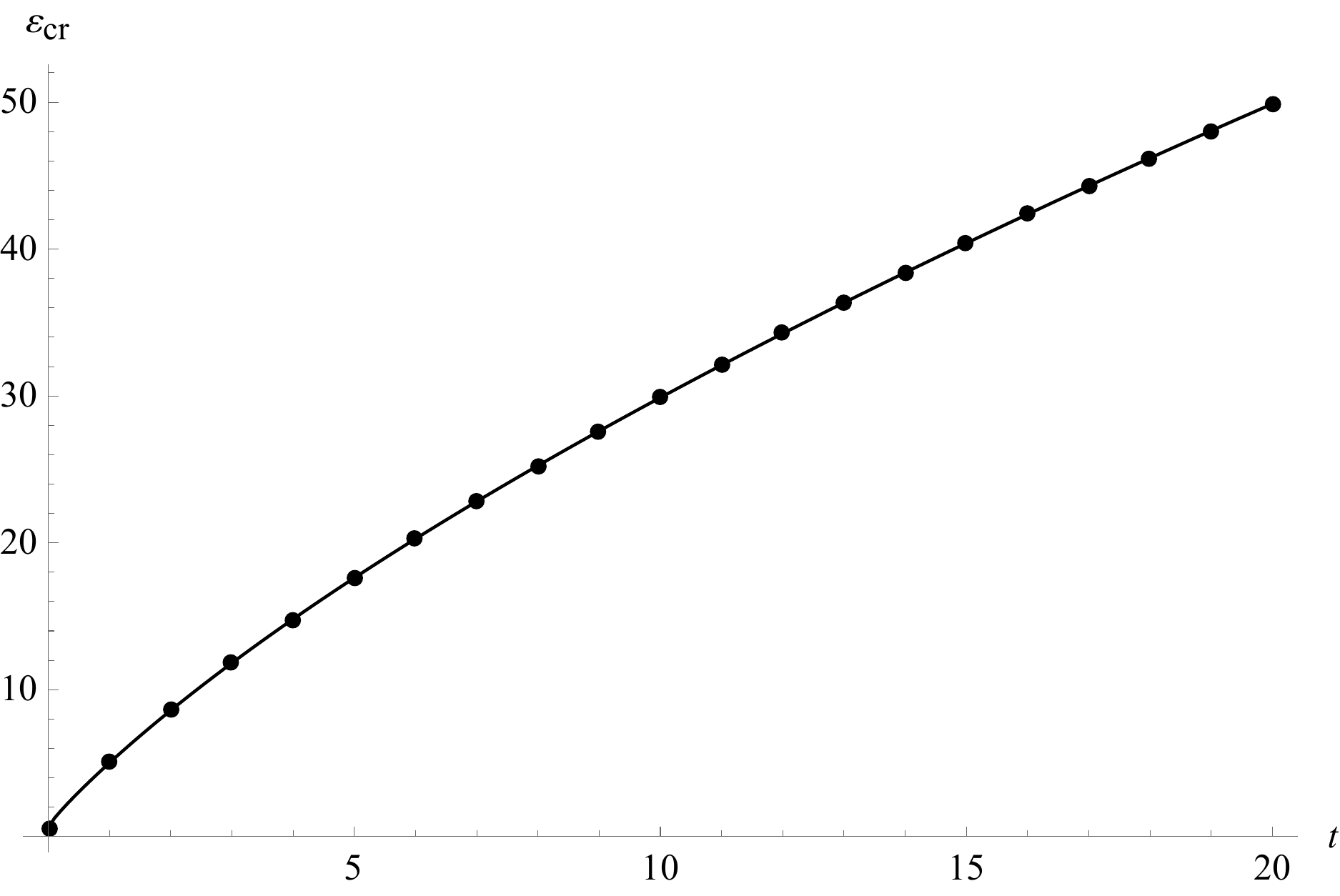}
\end{center}
\caption{Creep compliance as Bernstein function expressed through: integral
form - solid line, two-parameter Mittag-Leffler function - dots.}
\label{krip-nema-nula}
\end{figure}

\begin{figure}[tbph]
\begin{center}
\begin{minipage}{0.48\columnwidth}
  \subfloat[Asymptotics for small time.]{
   \includegraphics[width=\columnwidth]{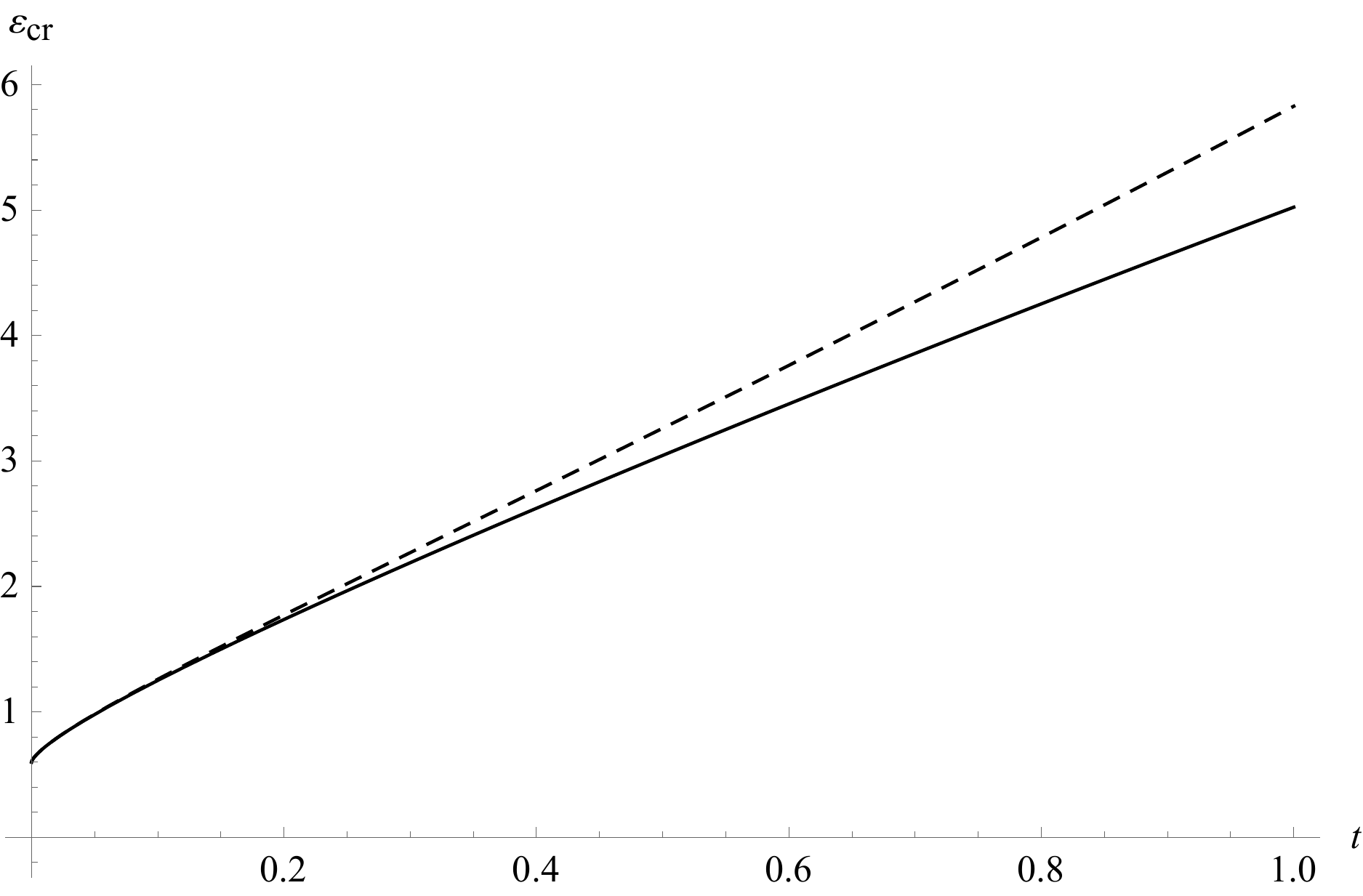}
   \label{krip-asimpt-nula}}
  \end{minipage}
\hfil
\begin{minipage}{0.48\columnwidth}
  \subfloat[Asymptotics for large time.]{
   \includegraphics[width=\columnwidth]{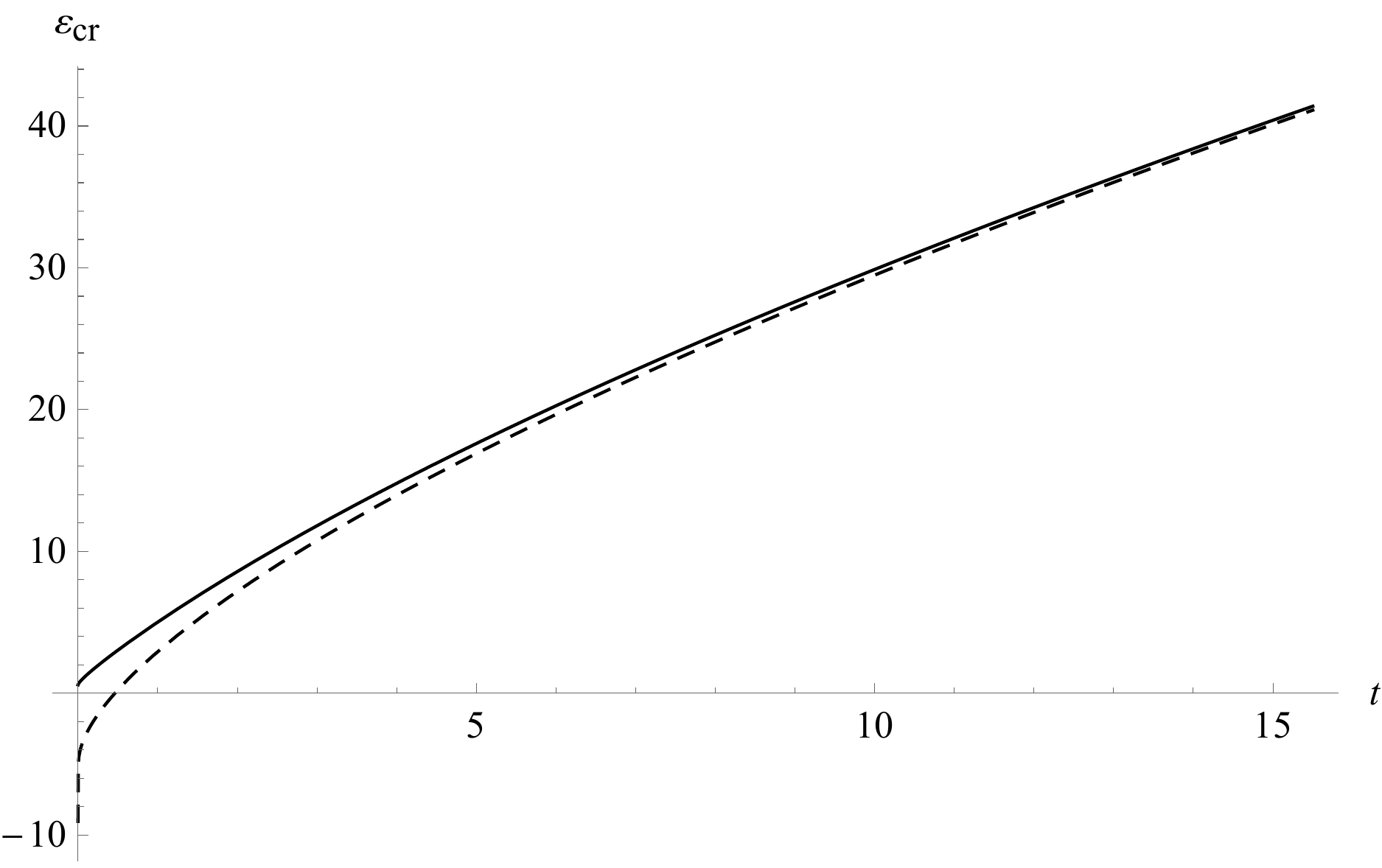}
   \label{krip-asimpt-inf}}
  \end{minipage}
\end{center}
\caption{Comparison of creep compliance (solid line) and its asymptotic
expansions (dashed line).}
\end{figure}

\begin{figure}[tbph]
\begin{center}
\begin{minipage}{0.48\columnwidth}
  \subfloat[Case 1: completely monotonic function.]{
   \includegraphics[width=\columnwidth]{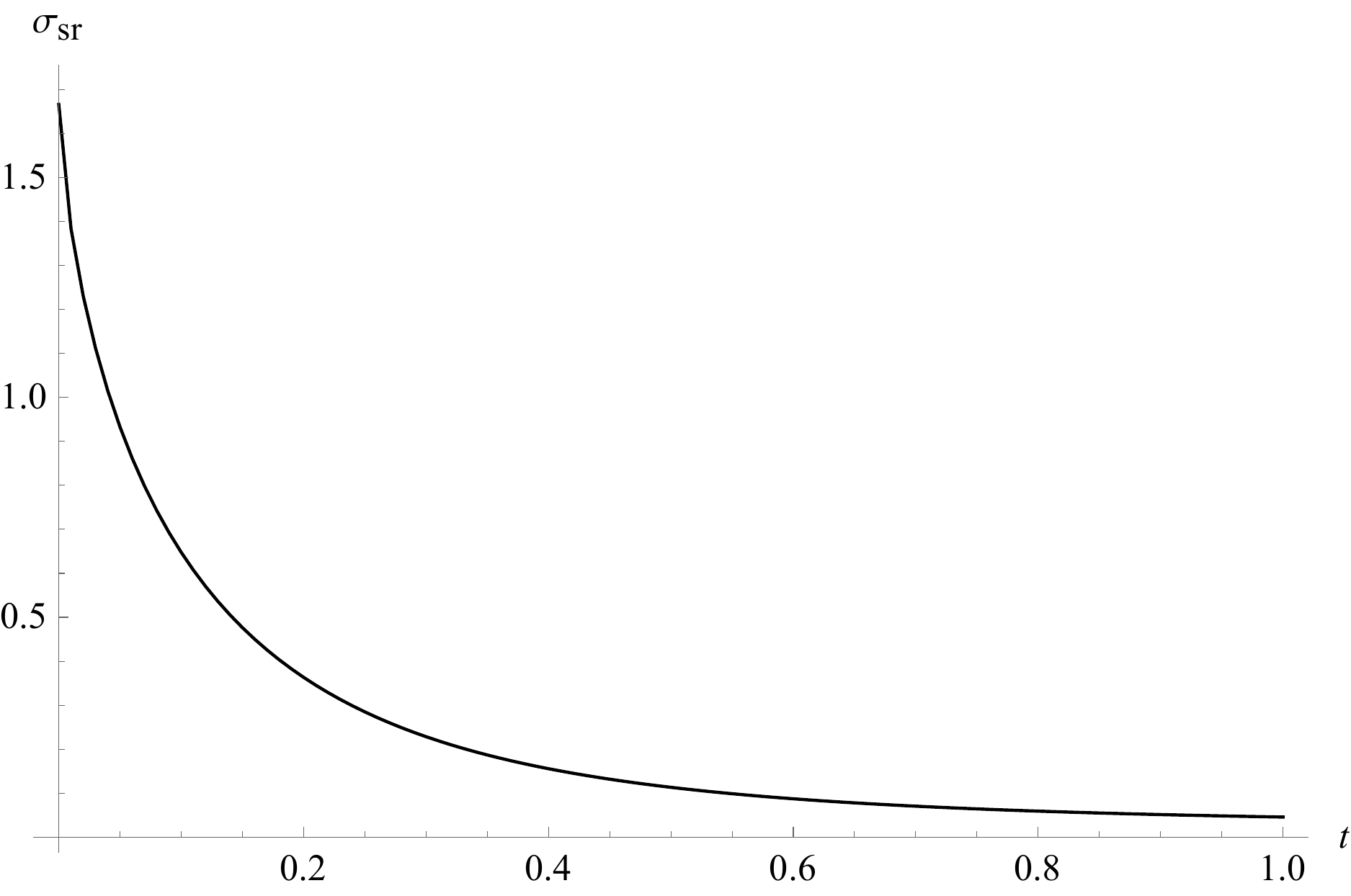}
   \label{sr-nema-nula}}
  \end{minipage}
\hfil
\begin{minipage}{0.48\columnwidth}
  \subfloat[Case 2.]{
   \includegraphics[width=\columnwidth]{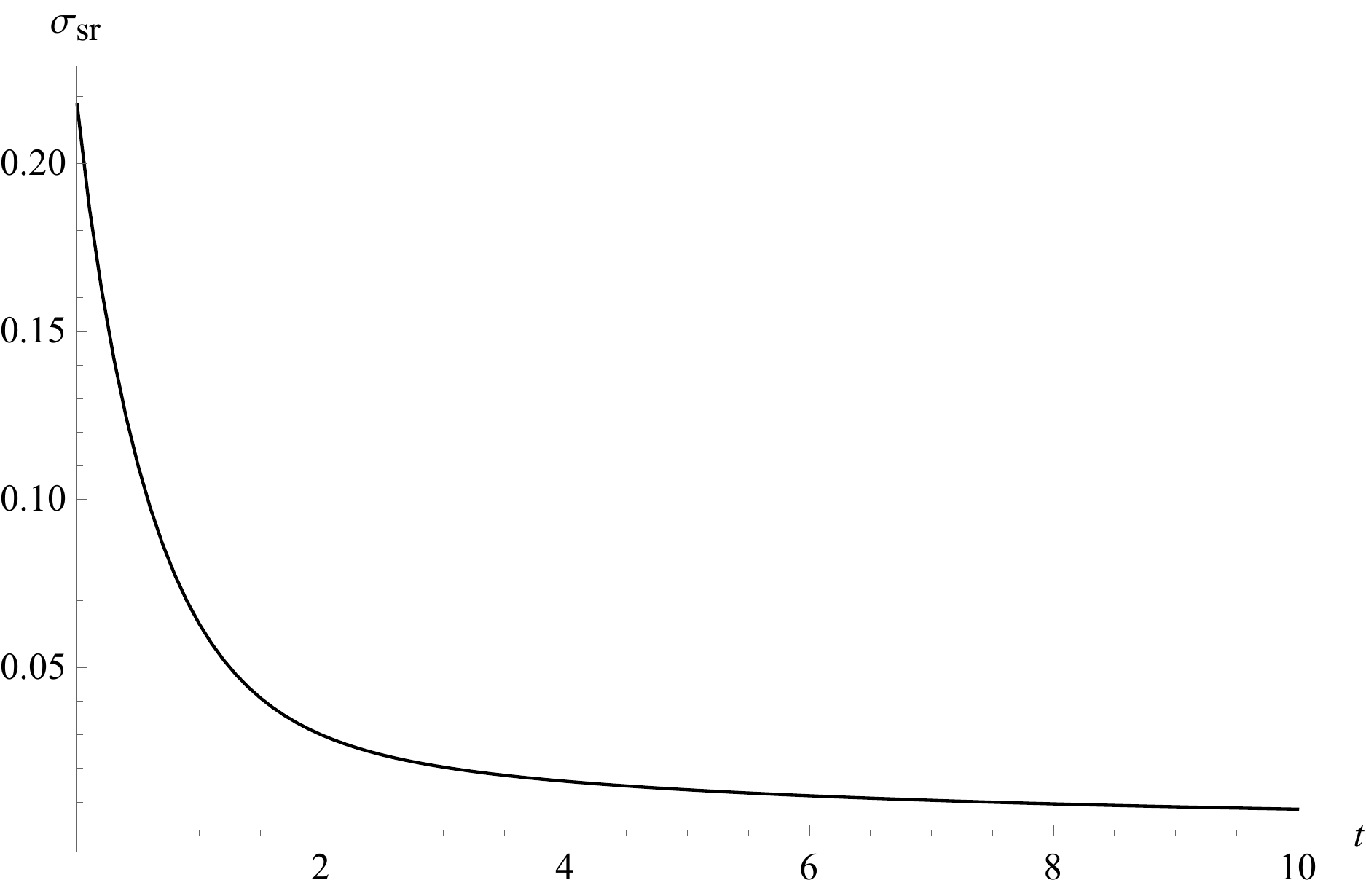}
   \label{sr-ima-real-nulu}}
  \end{minipage}
\end{center}
\caption{Relaxation modulus in Cases 1 and 2.}
\end{figure}

Figure \ref{krip-nema-nula} displays the comparison of creep compliances
obtained through the representation via two-parameter Mittag-Leffler
function (\ref{CR}), presented by dots and the integral representation (\ref%
{SR-ILT}), presented by the solid line. The agreement between two different
representations of the creep compliance is evidently very good. The model
parameters: $\alpha =0.1,$ $\beta =0.7,$ $a_{1}=0.25,$ $a_{2}=0.75,$ $%
a_{3}=0.15,$ $b_{1}=0.2,$ and $b_{2}=0.25,$ taken in order to produce plots
from Figure \ref{krip-nema-nula} guarantee not only that the thermodynamical
restrictions (\ref{Model 7 - tdr}) are fulfilled, but also that the
requirement (\ref{Model 7 - nerov}) is met, implying that the creep
compliance is a Bernstein function. The same set of model parameters is used
for plots shown in Figures \ref{krip-asimpt-nula}, \ref{krip-asimpt-inf},
and \ref{sr-nema-nula}. 

The plots of asymptotic expansions of the creep compliance near initial
time-instant (\ref{MD7-krip-0}) and for large time (\ref{MD7-krip-besk}) are
presented by the dashed lines in Figures \ref{krip-asimpt-nula} and \ref%
{krip-asimpt-inf} respectively, along with the creep compliance calculated
by (\ref{SR-ILT}) and presented by the solid line. One notices the good
agreement between creep compliance curves and its asymptotic expansions. 

Figures \ref{sr-nema-nula} and \ref{sr-ima-real-nulu} display the relaxation
modulus, calculated according to (\ref{SR-ILT}) in Case 1 and 2,
respectively. Model parameters used to produce plots from Figure \ref%
{sr-nema-nula} are the same as for the previous plots guaranteeing that the
function $\Psi $ (\ref{psi}) has no zeros in the complex plane (thus
corresponding to Case 1) and moreover that the relaxation modulus is a
completely monotonic function. Curve from Figure \ref{sr-ima-real-nulu} is
produced for the model parameters: $\alpha =0.55,$ $\beta =0.8192,$ $%
a_{1}=1, $ $a_{2}=1.5,$ $a_{3}=1.15,$ $b_{1}=0.2,$ and $b_{2}=0.25,$
guaranteeing that the function $\Psi $ has one negative real zero. 

\begin{figure}[p]
\begin{center}
\includegraphics[width=0.6\columnwidth]{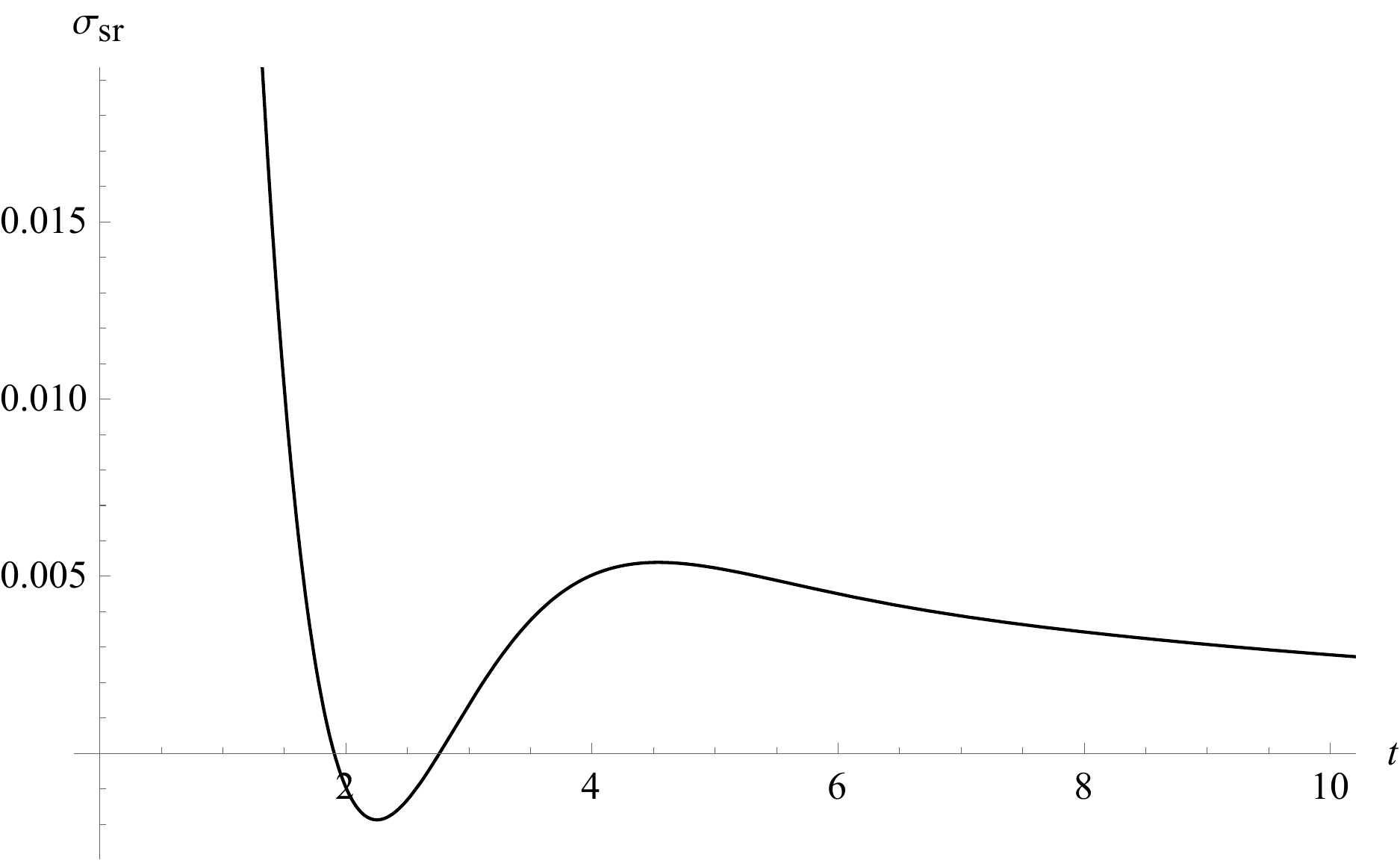}
\end{center}
\caption{Relaxation modulus in Case 3: oscillatory function having
decreasing amplitude.}
\label{sr-ima-nula}
\end{figure}

Figure \ref{sr-ima-nula} displays the relaxation modulus, calculated for
model parameters: $\alpha =0.4,$ $\beta =0.95,$ $a_{1}=1.25,$ $a_{2}=1.5,$ $%
a_{3}=1.15,$ $b_{1}=0.2,$ and $b_{2}=0.25$ according to (\ref{SR-ILT}) in
Case 3, i.e., in the case when function $\Psi $ (\ref{psi}) has a pair of
complex conjugated having negative real part. One notices the non-monotonic
behavior of the relaxation modulus.

\begin{figure}[p]
\begin{center}
\begin{minipage}{0.48\columnwidth}
  \subfloat[Asymptotics for small time.]{
   \includegraphics[width=\columnwidth]{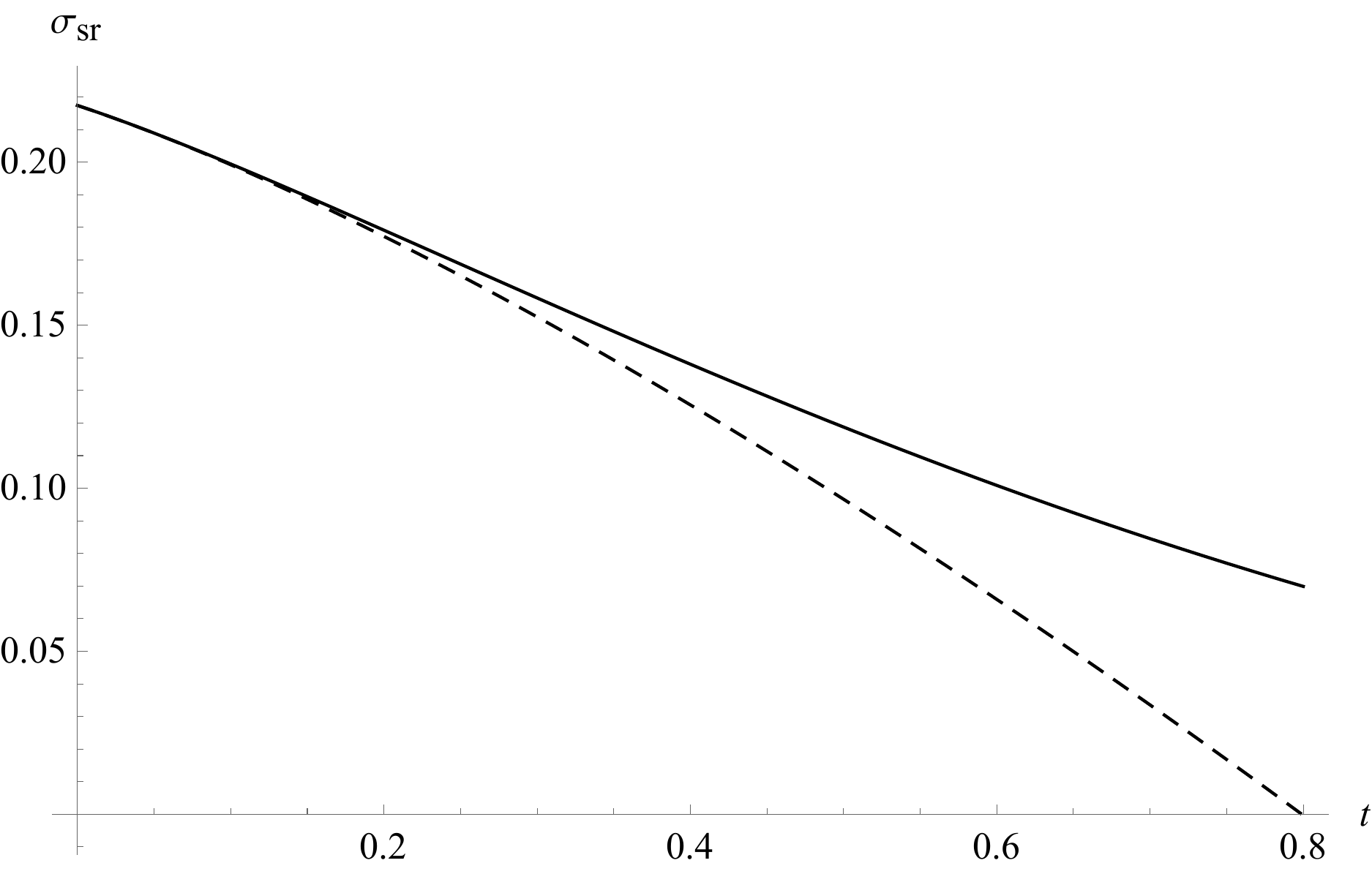}
   \label{sr-asimpt-nula}}
  \end{minipage}
\hfil
\begin{minipage}{0.48\columnwidth}
  \subfloat[Asymptotics for large time.]{
   \includegraphics[width=\columnwidth]{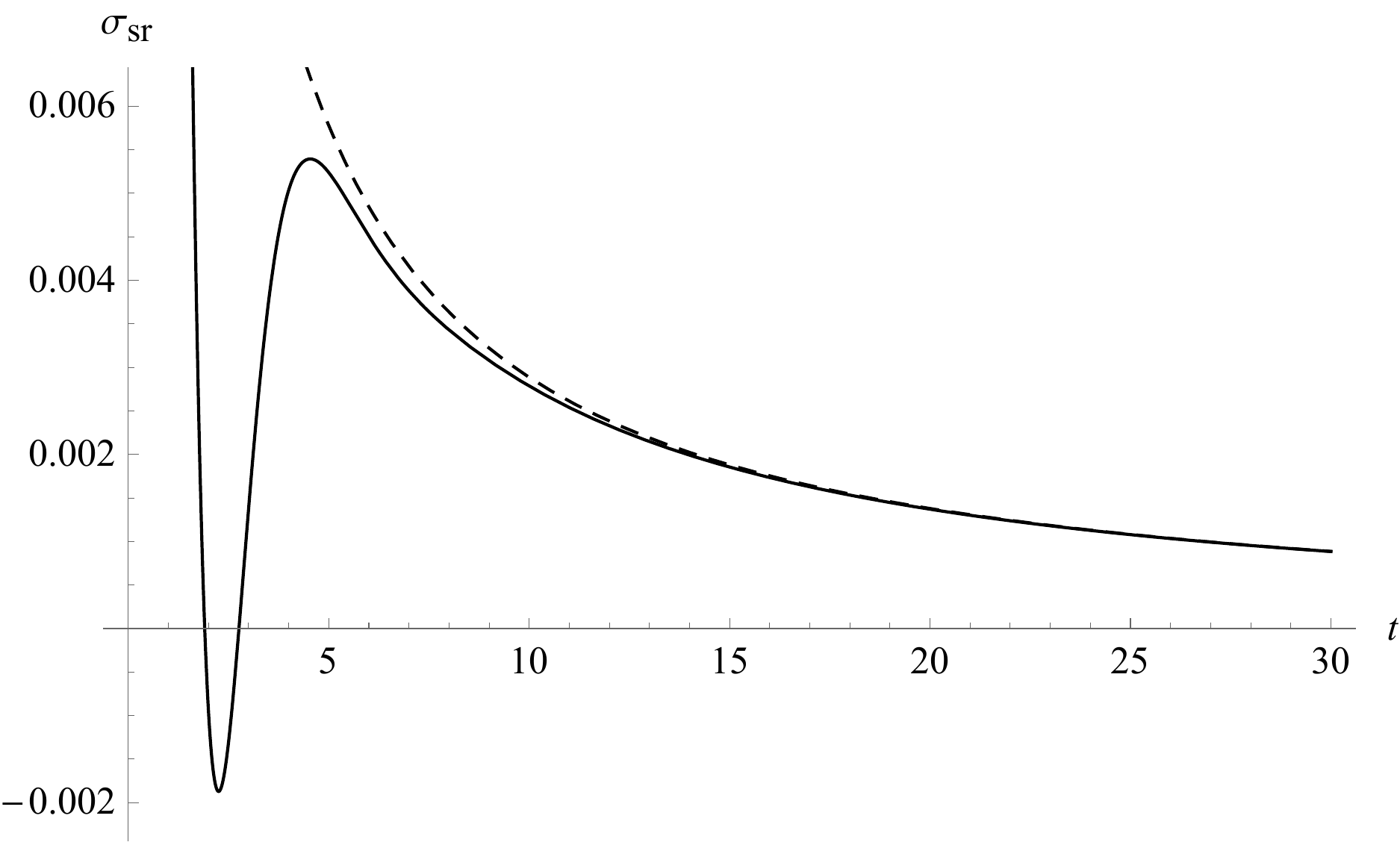}
   \label{sr-asimpt-inf}}
  \end{minipage}
\end{center}
\caption{Comparison of relaxation modulus (solid line) and its asymptotic
expansions (dashed line).}
\end{figure}

Figures \ref{sr-asimpt-nula} and \ref{sr-asimpt-inf} display the comparison
of relaxation modulus calculated according to (\ref{SR-ILT}) in Case 3,
presented by the solid line, and its asymptotic expansions near initial
time-instant (\ref{MD7-sr-0}) and for large time (\ref{MD7-sr-besk}),
presented by the dashed line, for the previously mentioned model parameters.
Regardless of the non-monotonic behavior of relaxation modulus, the
agreement between curves is good. 
\begin{table}[p]
\caption{Summary of the asymptotic analysis.}
\label{tbl}
\begin{center}
\begin{tabular}{c|c|c|c|c}
 & \multicolumn{2}{c|}{Asymptotics as $t\rightarrow 0$.} &  \multicolumn{2}{c}{Asymptotics as $t\rightarrow \infty$.} \\ \hline
 Model & creep compliance & relaxation modulus & creep compliance & relaxation modulus \\ \hline
 I & $\sim t^{\mu -\gamma +\varkappa }$ & $\sim t^{-\left( \mu -\gamma\right) -\varkappa }$ & \multirow{5}{*}{$\sim t^{\mu }$} & \multirow{5}{*}{$\sim t^{-\mu }$} \\ 
 II & $\sim t^{\mu -\alpha }$ & $\sim t^{-\left( \mu -\alpha \right) }$ &  & \\ 
 III & $\sim t^{\mu -\beta }$ & $\sim t^{-\left( \mu -\beta \right) }$ &  & \\ 
 IV & $\sim t^{\mu -\alpha }$ & $\sim t^{-\left( \mu -\alpha \right) }$ &  & \\ 
 V & $\sim t^{\mu -\beta }$ & $\sim t^{-\left( \mu -\beta \right) }$ &  &  \\ \hline 
 VI & incr. $\sim t^{\alpha }$ from $\frac{a_{3}}{b_{2}}$ & decr. $\sim t^{\alpha }$ from $\frac{b_{2}}{a_{3}}$ & \multirow{2}{*}{$\sim t^{\beta }$} & \multirow{2}{*}{$\sim t^{-\beta }$} \\  
 VII & incr. $\sim t^{\beta }$ from $\frac{a_{3}}{b_{2}}$ & decr. $\sim t^{\beta }$ from $\frac{b_{2}}{a_{3}}$ &  &  \\ \hline
 VIII & incr. $\sim t^{\alpha }$ from $\frac{\bar{a}_{2}}{b_{2}}$ & decr. $\sim t^{\alpha }$ from $\frac{b_{2}}{\bar{a}_{2}}$ & $\sim t^{\alpha }$ & $ \sim t^{-\alpha }$ \\ \hline
\end{tabular}
\end{center}
\end{table}

\section{Conclusion}

Thermodynamically consistent classical and fractional Burgers models I -
VIII are examined in creep and stress relaxation tests. Using the Laplace
transform method, explicit forms of creep compliance in representation via
Mittag-Leffler function and in integral representation, as well as the
integral representation of relaxation modulus are obtained taking into
account the material behavior at initial time-instant, since Models I - V
have zero glass compliance (infinite glass modulus), while classical model
and Models VI - VIII, have non-zero glass compliance (glass modulus). For
all Burgers models equilibrium compliance is infinite implying the zero
equilibrium modulus.

By requiring kernels' non-negativity, the integral forms of the creep
compliance and relaxation modulus proved useful in showing that the
thermodynamical requirements allow wider range of model parameters than the
range in which the creep compliance is a Bernstein function and the
relaxation modulus is a completely monotonic function. For the model
parameters outside this restrictive interval the creep compliance and
relaxation modulus do not have to be monotonic functions. In addition, the
relaxation modulus may even be oscillatory function having decreasing
amplitude, which is due to the possible zeros of denominator of the
relaxation modulus in complex domain, yielding the damped oscillatory term
in the relaxation modulus. The conditions for appearance of zeros is
independent of the conditions insuring non-negativity of kernel in the
integral representation.

The asymptotic analysis of both creep compliance and relaxation modulus
conducted in the case of all thermodynamically consistent fractional Burgers
models in the vicinity of initial time-instant and for large time as well is
summarized in Table \ref{tbl}. 

\appendix

\section{Determination of position and number of zeros of functions $\Psi $
and $\Phi $ \label{DPNZ}}

\subsection{Case of function $\Psi $ \label{CFpsi}}

Introducing the substitution $s=\rho \mathrm{e}^{\mathrm{i}\varphi }$ into
function $\Psi ,$ given by (\ref{psi}), and by separating real and imaginary
parts in $\Psi $ one obtains 
\begin{eqnarray}
\func{Re}\Psi \left( \rho ,\varphi \right) &=&1+a_{1}\rho ^{\alpha }\cos
\left( \alpha \varphi \right) +a_{2}\rho ^{\beta }\cos \left( \beta \varphi
\right) +a_{3}\rho ^{\gamma }\cos \left( \gamma \varphi \right) ,
\label{repsi} \\
\func{Im}\Psi \left( \rho ,\varphi \right) &=&a_{1}\rho ^{\alpha }\sin
\left( \alpha \varphi \right) +a_{2}\rho ^{\beta }\sin \left( \beta \varphi
\right) +a_{3}\rho ^{\gamma }\sin \left( \gamma \varphi \right) ,
\label{impsi}
\end{eqnarray}%
where $0<\alpha \leq \beta <1,$ $\gamma \in \left( 0,2\right) ,$ and $%
a_{1},a_{2},a_{3}>0.$ Note that if $s_{0}=\rho _{0}\mathrm{e}^{\mathrm{i}%
\varphi _{0}},$ $\varphi _{0}\in \left( 0,\pi \right) ,$ is zero of function 
$\Psi ,$ then its complex conjugate, $\bar{s}_{0}=\rho _{0}\mathrm{e}^{-%
\mathrm{i}\varphi _{0}},$ is also a zero of $\Psi ,$ since $\func{Im}\Psi
\left( \rho ,-\varphi \right) =-\func{Im}\Psi \left( \rho ,\varphi \right) .$
Therefore, it is sufficient to consider the upper complex half-plane only.

Considering the imaginary part of $\Psi ,$ (\ref{impsi}), one concludes that
function $\Psi $ does not have zeroes in the right complex half-plane.
Namely, function $\Psi $ cannot have positive real zeroes, since for $%
\varphi =0,$ although $\func{Im}\Psi =0$, it holds that $\func{Re}\Psi >0$.
Also, since $\alpha ,\beta \in \left( 0,1\right) $ and $\gamma \in \left(
0,2\right) ,$ for $\varphi \in \left( 0,\frac{\pi }{2}\right] $ all sine
functions in (\ref{impsi}) are positive, implying that $\func{Im}\Psi >0$.
Moreover, in the case when $\gamma \in \left( 0,1\right) ,$ function $\Psi $
does not have zeroes in the whole complex plane, since for $\varphi \in
\left( 0,\pi \right] $ all sine functions in (\ref{impsi}) are positive
implying that $\func{Im}\Psi >0.$ Therefore, zeros of function $\Psi $ may
exist only if $\gamma \in \left( 1,2\right) ,$ and, if zeros exist, they are
complex conjugated and located in the left complex half-plane.

\begin{figure}[htbp]
\centering
\includegraphics[scale=0.7]{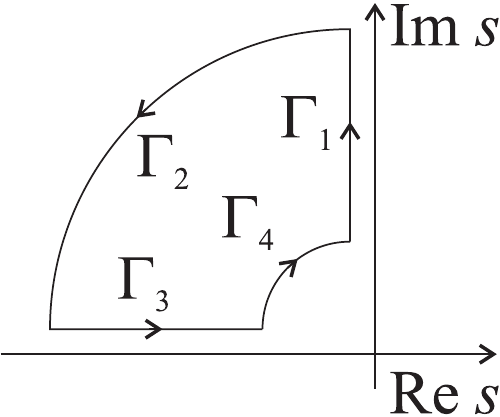}
\caption{Contour $\Gamma =\Gamma _{1}\cup \Gamma _{2}\cup \Gamma _{3}\cup
\Gamma _{4}$.}
\label{gama}
\end{figure}

In the case when $\gamma \in \left( 1,2\right) $, the zeroes of function $%
\Psi $ will be sought using the argument principle and contour $\Gamma
=\Gamma _{1}\cup \Gamma _{2}\cup \Gamma _{3}\cup \Gamma _{4}$ placed in the
upper left complex quarter-plane, shown in Figure \ref{gama}. 

Parametrizing the contour $\Gamma _{1}$ by $s=\rho \mathrm{e}^{\mathrm{i}%
\frac{\pi }{2}},$ $\rho \in \left( r,R\right) ,$ with $r\rightarrow 0$ and $%
R\rightarrow \infty ,$ the real and imaginary parts of $\Psi ,$ (\ref{repsi}%
) and (\ref{impsi}), become 
\begin{eqnarray*}
\func{Re}\Psi \left( \rho \right) &=&1+a_{1}\rho ^{\alpha }\cos \frac{\alpha
\pi }{2}+a_{2}\rho ^{\beta }\cos \frac{\beta \pi }{2}+a_{3}\rho ^{\gamma
}\cos \frac{\gamma \pi }{2}, \\
\func{Im}\Psi \left( \rho \right) &=&a_{1}\rho ^{\alpha }\sin \frac{\alpha
\pi }{2}+a_{2}\rho ^{\beta }\sin \frac{\beta \pi }{2}+a_{3}\rho ^{\gamma
}\sin \frac{\gamma \pi }{2}>0,
\end{eqnarray*}%
which in the limiting cases yield%
\begin{gather*}
\func{Re}\Psi \left( \rho \right) \sim 1,\;\;\func{Im}\Psi \left( \rho
\right) \sim a_{1}\rho ^{\alpha }\sin \frac{\alpha \pi }{2}\rightarrow
0^{+},\;\;\text{as}\;\;\rho =r\rightarrow 0, \\
\func{Re}\Psi \left( \rho \right) \sim a_{3}\rho ^{\gamma }\cos \frac{\gamma
\pi }{2}\rightarrow -\infty ,\;\;\func{Im}\Psi \left( \rho \right) \sim
a_{3}\rho ^{\gamma }\sin \frac{\gamma \pi }{2}\rightarrow \infty ,\;\;\text{%
as}\;\;\rho =R\rightarrow \infty .
\end{gather*}

The real and imaginary parts of $\Psi ,$ (\ref{repsi}) and (\ref{impsi}),
along the contour $\Gamma _{2}$, parametrized by $s=R\mathrm{e}^{\mathrm{i}%
\varphi },$ $\varphi \in \left[ \frac{\pi }{2},\pi \right] ,$ with $%
R\rightarrow \infty ,$ take the form%
\begin{equation}
\func{Re}\Psi \left( \varphi \right) \sim a_{3}R^{\gamma }\cos (\gamma
\varphi ),\;\;\func{Im}\Psi \left( \varphi \right) \sim a_{3}R^{\gamma }\sin
(\gamma \varphi ),\;\;\text{as}\;\;R\rightarrow \infty .  \label{impsig2}
\end{equation}%
Note, when $\gamma \in \left( 1,\frac{3}{2}\right) $ it holds that $\func{Re}%
\Psi <0,$ since $\cos (\gamma \varphi )<0.$ If $\varphi =\frac{\pi }{2},$
then (\ref{impsig2}) becomes%
\begin{equation*}
\func{Re}\Psi \left( \frac{\pi }{2}\right) \rightarrow -\infty ,\;\;\func{Im}%
\Psi \left( \frac{\pi }{2}\right) \rightarrow \infty ,\;\;\text{as}%
\;\;R\rightarrow \infty ,
\end{equation*}%
while if $\varphi =\pi ,$ then, as $R\rightarrow \infty ,$ (\ref{impsig2})
becomes either 
\begin{gather*}
\func{Re}\Psi \left( \pi \right) \rightarrow -\infty ,\;\;\func{Im}\Psi
\left( \pi \right) \rightarrow -\infty ,\;\;\text{for}\;\;\gamma \in \left(
1,\frac{3}{2}\right) ,\;\;\text{or} \\
\func{Re}\Psi \left( \pi \right) \rightarrow \infty ,\;\;\func{Im}\Psi
\left( \pi \right) \rightarrow -\infty ,\;\;\text{for}\;\;\gamma \in \left[ 
\frac{3}{2},2\right) .
\end{gather*}

Using the parametrization $s=\rho \mathrm{e}^{\mathrm{i}\pi },$ $\rho \in
\left( r,R\right) ,$ with $r\rightarrow 0$ and $R\rightarrow \infty ,$ of
contour $\Gamma _{3},$ the real and imaginary parts of $\Psi ,$ (\ref{repsi}%
) and (\ref{impsi}), become%
\begin{eqnarray}
\func{Re}\Psi \left( \rho \right) &=&1+a_{1}\rho ^{\alpha }\cos \left(
\alpha \pi \right) +a_{2}\rho ^{\beta }\cos \left( \beta \pi \right)
+a_{3}\rho ^{\gamma }\cos \left( \gamma \pi \right) ,  \label{repsig3} \\
\func{Im}\Psi \left( \rho \right) &=&a_{1}\rho ^{\alpha }\sin \left( \alpha
\pi \right) +a_{2}\rho ^{\beta }\sin \left( \beta \pi \right) +a_{3}\rho
^{\gamma }\sin \left( \gamma \pi \right) ,  \label{impsig3}
\end{eqnarray}%
which, as $\rho =r\rightarrow 0,$ yield%
\begin{equation*}
\func{Re}\Psi \left( \rho \right) \sim 1,\;\;\func{Im}\Psi \left( \rho
\right) \sim a_{1}\rho ^{\alpha }\sin \left( \alpha \pi \right) \rightarrow
0^{+},
\end{equation*}%
and, as $\rho =R\rightarrow \infty ,$ either 
\begin{gather*}
\func{Re}\Psi \left( \rho \right) \sim a_{3}\rho ^{\gamma }\cos \left(
\gamma \pi \right) \rightarrow -\infty ,\;\;\func{Im}\Psi \left( \rho
\right) \sim a_{3}\rho ^{\gamma }\sin \left( \gamma \pi \right) \rightarrow
\infty ,\;\;\text{for}\;\;\gamma \in \left( 1,\frac{3}{2}\right) ,\;\;\text{%
or} \\
\func{Re}\Psi \left( \rho \right) \sim a_{3}\rho ^{\gamma }\cos \left(
\gamma \pi \right) \rightarrow \infty ,\;\;\func{Im}\Psi \left( \rho \right)
\sim a_{3}\rho ^{\gamma }\sin \left( \gamma \pi \right) \rightarrow \infty
,\;\;\text{for}\;\;\gamma \in \left[ \frac{3}{2},2\right) ,
\end{gather*}%
Since $\alpha ,\beta \in \left( 0,1\right) $ and $\gamma \in \left(
1,2\right) ,$ the first two terms in (\ref{impsig3}) are positive, while the
third term is negative, so that there exists at least one $\rho ^{\ast }\neq
0$ such that 
\begin{equation*}
\func{Im}\Psi \left( \rho ^{\ast }\right) =a_{3}\left( \rho ^{\ast }\right)
^{\alpha }\left\vert \sin \left( \gamma \pi \right) \right\vert \left( \frac{%
a_{1}\sin \left( \alpha \pi \right) }{a_{3}\left\vert \sin \left( \gamma \pi
\right) \right\vert }+\frac{a_{2}\sin \left( \beta \pi \right) }{%
a_{3}\left\vert \sin \left( \gamma \pi \right) \right\vert }\left( \rho
^{\ast }\right) ^{\beta -\alpha }-\left( \rho ^{\ast }\right) ^{\gamma
-\alpha }\right) =0.
\end{equation*}%
The equation%
\begin{equation}
\frac{a_{1}\sin \left( \alpha \pi \right) }{a_{3}\left\vert \sin \left(
\gamma \pi \right) \right\vert }+\frac{a_{2}\sin \left( \beta \pi \right) }{%
a_{3}\left\vert \sin \left( \gamma \pi \right) \right\vert }\left( \rho
^{\ast }\right) ^{\beta -\alpha }=\left( \rho ^{\ast }\right) ^{\gamma
-\alpha }  \label{ro-zvezda}
\end{equation}%
has a single solution $\rho ^{\ast }.$ Namely, since $0<\alpha \leq \beta <1$
and $\gamma \in \left( 1,2\right) ,$ the function $\left( \rho ^{\ast
}\right) ^{\beta -\alpha }$ is concave, due to $\beta -\alpha \in \left(
0,1\right) $ (or even constant if $\beta =\alpha $), and $\left( \rho ^{\ast
}\right) ^{\gamma -\alpha }$ is either convex, if $\gamma -\alpha \in \left(
1,2\right) $, or concave, if $\gamma -\alpha \in \left( 0,1\right) $ (or
even linear if $\gamma -\alpha =1$), but always increases faster than $%
\left( \rho ^{\ast }\right) ^{\beta -\alpha }.$ Depending on the value of $%
\rho ^{\ast },$ the real part of $\Psi ,$ (\ref{repsig3}), can be either
negative, zero, or positive, which will determine the change of argument of
function $\Psi $ along the contour $\Gamma .$

Along the contour $\Gamma _{4},$ parametrized by $s=r\mathrm{e}^{\mathrm{i}%
\varphi },$ $\varphi \in \left[ \frac{\pi }{2},\pi \right] ,$ with $%
r\rightarrow 0,$ the real and imaginary parts of $\Psi ,$ (\ref{repsi}) and (%
\ref{impsi}), take the form%
\begin{equation*}
\func{Re}\Psi \left( \varphi \right) \sim 1,\;\;\func{Im}\Psi \left( \varphi
\right) \sim a_{1}r^{\gamma }\sin (\alpha \varphi )>0,\;\;\text{as}%
\;\;r\rightarrow 0,
\end{equation*}%
so that $\func{Im}\Psi \left( \varphi \right) \rightarrow 0^{+},$ for $%
\varphi \in \left[ \frac{\pi }{2},\pi \right] ,$ as $r\rightarrow 0.$

By the argument principle, function $\Psi $ has no zeros (has one complex
zero) in the upper left complex quarter-plane if $\func{Re}\Psi \left( \rho
^{\ast }\right) <0$ ($\func{Re}\Psi \left( \rho ^{\ast }\right) >0$) in (\ref%
{repsig3}), since the change of argument of function $\Psi $ is $\Delta \arg
\Psi =0$ ($\Delta \arg \Psi =2\pi $) along the contour $\Gamma ,$ while if $%
\func{Re}\Psi \left( \rho ^{\ast }\right) =0,$ then function $\Psi $ has a
negative real zero $-\rho ^{\ast },$ with $\rho ^{\ast }$ determined from (%
\ref{ro-zvezda}). This holds true for the whole complex plane as well if $%
\gamma \in \left( 1,2\right) ,$ while if $\gamma \in \left( 0,1\right) ,$
then function $\Psi $ does not have zeros in the complex plane.

\subsection{Case of function $\Phi $ \label{CFfi}}

Function $\Phi ,$ given by (\ref{fi}), being a quadratic function in terms
of $s^{\alpha },$ is decomposed as%
\begin{equation*}
\Phi \left( s\right) =\bar{a}_{2}\left( s^{\alpha }+a_{1}\right) \left(
s^{\alpha }+a_{2}\right) ,\;\;\text{with}\;\;a_{1,2}=\frac{\bar{a}_{1}}{2%
\bar{a}_{2}}\pm \sqrt{\left( \frac{\bar{a}_{1}}{2\bar{a}_{2}}\right) ^{2}-%
\frac{1}{\bar{a}_{2}}}>0,
\end{equation*}%
for $\left( \frac{\bar{a}_{1}}{2\bar{a}_{2}}\right) ^{2}\geq \frac{1}{\bar{a}%
_{2}},$ implying that function $\Phi $ has no zeros in the principal Riemann
branch, i.e., for $\arg s\in \left( -\pi ,\pi \right) ,$ since it is
well-known that equation 
\begin{equation*}
s^{\alpha }+a_{1,2}=0,\;\;s\in 
\mathbb{C}
,
\end{equation*}%
has no solutions for $\arg s\in \left( -\pi ,\pi \right) .$

For $\left( \frac{\bar{a}_{1}}{2\bar{a}_{2}}\right) ^{2}<\frac{1}{\bar{a}_{2}%
},$ function $\Phi $ is decomposed as%
\begin{equation*}
\Phi \left( s\right) =\bar{a}_{2}\left( s^{\alpha }+a-\mathrm{i}b\right)
\left( s^{\alpha }+a+\mathrm{i}b\right) ,\;\;\text{with}\;\;a=\frac{\bar{a}%
_{1}}{2\bar{a}_{2}}\;\;\text{and}\;\;b=\sqrt{\frac{1}{\bar{a}_{2}}-\left( 
\frac{\bar{a}_{1}}{2\bar{a}_{2}}\right) ^{2}}.
\end{equation*}%
Define a function $\phi $ by%
\begin{equation}
\phi \left( s\right) =s^{\alpha }+a-\mathrm{i}b,\;\;s\in 
\mathbb{C}
,  \label{s^alfa+a+ib}
\end{equation}%
where $\alpha \in \left( 0,1\right) ,$ The real and imaginary parts of
function $\phi $ are obtained as%
\begin{eqnarray}
\func{Re}\phi \left( \rho ,\varphi \right) &=&\rho ^{\alpha }\cos \left(
\alpha \varphi \right) +a,  \label{refi} \\
\func{Im}\phi \left( \rho ,\varphi \right) &=&\rho ^{\alpha }\sin \left(
\alpha \varphi \right) -b,  \label{imfi}
\end{eqnarray}%
using $s=\rho \mathrm{e}^{\mathrm{i}\varphi }$ in (\ref{s^alfa+a+ib}). If
function $\phi $ has zero $s_{0}=\rho _{0}\mathrm{e}^{\mathrm{i}\varphi
_{0}} $ in the upper complex half-plane (since $\sin \left( \alpha \varphi
\right) >0$ for $\varphi \in \left( 0,\pi \right) $), then the complex
conjugated zero $\bar{s}_{0}=\rho _{0}\mathrm{e}^{-\mathrm{i}\varphi _{0}}$
is obtained as a solution of equation $s^{\alpha }+a+\mathrm{i}b=0.$ Thus, $%
s_{0}$ and $\bar{s}_{0}$ are complex conjugated zeros of function $\Phi $ as
well. Function $\phi $ does not have zeroes in the upper right complex
quarter-plane, since for $\varphi \in \left[ 0,\frac{\pi }{2}\right] $ its
real part (\ref{refi}) is positive.

Therefore, the zeros of $\phi $ will be sought for in the upper left complex
quarter-plane within the contour $\Gamma ,$ shown in Figure \ref{gama}, by
using the argument principle. Parametrizing the contour $\Gamma =\Gamma
_{1}\cup \Gamma _{2}\cup \Gamma _{3}\cup \Gamma _{4}$ as in Section \ref%
{CFpsi}, along $\Gamma _{1},$ by (\ref{refi}) and (\ref{imfi}), one obtains%
\begin{gather*}
\func{Re}\phi \left( \rho \right) =\rho ^{\alpha }\cos \frac{\alpha \pi }{2}%
+a>0,\;\;\func{Im}\phi \left( \rho \right) =\rho ^{\alpha }\sin \frac{\alpha
\pi }{2}-b, \\
\func{Re}\phi \left( \rho \right) \sim a,\;\;\func{Im}\phi \left( \rho
\right) \sim -b,\;\;\text{as}\;\;\rho =r\rightarrow 0, \\
\func{Re}\phi \left( \rho \right) \rightarrow \infty ,\;\;\func{Im}\phi
\left( \rho \right) \rightarrow \infty ,\;\;\text{as}\;\;\rho =R\rightarrow
\infty ,
\end{gather*}%
as real and imaginary parts of $\phi $, and their limiting cases. Along $%
\Gamma _{2},$ as $R\rightarrow \infty ,$ for arbitrary value of $\varphi $,
as well as for $\varphi =\frac{\pi }{2}$ and $\varphi =\pi ,$ (\ref{refi})
and (\ref{imfi}) take the following forms%
\begin{gather*}
\func{Re}\phi \left( \varphi \right) \sim R^{\alpha }\cos (\alpha \varphi
),\;\;\func{Im}\phi \left( \varphi \right) \sim R^{\alpha }\sin (\alpha
\varphi )>0, \\
\func{Re}\phi \left( \frac{\pi }{2}\right) \rightarrow \infty ,\;\;\func{Im}%
\phi \left( \frac{\pi }{2}\right) \rightarrow \infty , \\
\func{Re}\phi \left( \pi \right) \rightarrow -\infty ,\;\;\func{Im}\phi
\left( \pi \right) \rightarrow \infty .
\end{gather*}%
In the case of contour $\Gamma _{3},$ (\ref{refi}), (\ref{imfi}) and their
limiting cases are%
\begin{gather}
\func{Re}\phi \left( \rho \right) =-\rho ^{\alpha }\left\vert \cos \left(
\alpha \pi \right) \right\vert +a,\;\;\func{Im}\phi \left( \rho \right)
=\rho ^{\alpha }\sin \left( \alpha \pi \right) -b,  \label{refiinfi} \\
\func{Re}\phi \left( \rho \right) \sim a,\;\;\func{Im}\phi \left( \rho
\right) \sim -b,\;\;\text{as}\;\;\rho =r\rightarrow 0,  \notag \\
\func{Re}\phi \left( \rho \right) \rightarrow -\infty ,\;\;\func{Im}\phi
\left( \rho \right) \rightarrow \infty ,\;\;\text{as}\;\;\rho =R\rightarrow
\infty .  \notag
\end{gather}%
Depending on the value of solution $\rho ^{\ast }=\left( \frac{b}{\sin
\left( \alpha \pi \right) }\right) ^{\frac{1}{\alpha }}$ of equation $\func{%
Im}\phi \left( \rho ^{\ast }\right) =0,$ (\ref{refiinfi})$_{2},$ the real
part of $\phi ,$ (\ref{refiinfi})$_{1}$, can be either negative, zero, or
positive, which will determine the change of argument of function $\phi $
along the whole contour $\Gamma .$ Along the contour $\Gamma _{4},$ one has%
\begin{equation*}
\func{Re}\phi \left( \varphi \right) \sim a,\;\;\func{Im}\phi \left( \varphi
\right) \sim -b^{+},\;\;\text{as}\;\;r\rightarrow 0.
\end{equation*}

By the argument principle, function $\phi $ has no zeros (has one complex
zero) in the upper left complex quarter-plane and therefore in the whole
upper complex half-plane as well, if $\func{Re}\phi \left( \rho ^{\ast
}\right) =-b\frac{\left\vert \cos \left( \alpha \pi \right) \right\vert }{%
\sin \left( \alpha \pi \right) }+a<0$ ($\func{Re}\phi \left( \rho ^{\ast
}\right) =-b\frac{\left\vert \cos \left( \alpha \pi \right) \right\vert }{%
\sin \left( \alpha \pi \right) }+a>0$) in (\ref{refiinfi}), since the change
of argument of function $\phi $ is $\Delta \arg \phi =0$ ($\Delta \arg \phi
=2\pi $) along the contour $\Gamma $. If $\func{Re}\phi \left( \rho ^{\ast
}\right) =0,$ then function $\phi $ has a negative real zero $-\rho ^{\ast
}. $

\section{Calculation of creep compliance and relaxation modulus}

The creep compliance and relaxation modulus corresponding to Models I - V,
respectively to Models VI and VII, will be obtained in the integral forms (%
\ref{CR-ILT-1-5}) and (\ref{SR}), respectively as (\ref{CR-ILT}) and (\ref%
{SR-ILT}), by inverting the creep compliance and relaxation modulus in
complex domain, given by (\ref{cr-LT}) and (\ref{sr-lap}), respectively by (%
\ref{cr-LT-1}) and (\ref{sr-lap-1}), using the definition of inverse Laplace
transform 
\begin{equation}
f\left( t\right) =\mathcal{L}^{-1}\left[ \tilde{f}\left( s\right) \right]
\left( t\right) =\frac{1}{2\pi \mathrm{i}}\int_{p_{0}-\mathrm{i}\infty
}^{p_{0}+\mathrm{i}\infty }\tilde{f}\left( s\right) \mathrm{e}^{st}\mathrm{d}%
s.  \label{inv-laplas}
\end{equation}%
In the case of creep compliance in complex domain, as well as for the
relaxation modulus in complex domain when function $\Psi ,$ given by (\ref%
{psi}), has either no zeros in complex plane, or has one negative real zero,
the Laplace transform inversion will be performed by using the Cauchy
integral theorem $\oint_{\Gamma }f(z)\mathrm{d}z=0.$ In the case of
relaxation modulus in complex domain when function $\Psi $ (\ref{psi}) has a
pair of complex conjugated zeros, the Cauchy residue theorem $\oint_{\Gamma
}f(z)\mathrm{d}z=2\pi \mathrm{i}\sum_{k}\func{Res}\left( f\left( z\right)
,z_{k}\right) $ will be employed.

In the case of Model VIII, the creep compliance and relaxation modulus,
given by (\ref{CR-ILT-1}) and (\ref{SR-ILT-1}), are also calculated by the
Laplace transform inversion of the creep compliance and relaxation modulus
in complex domain, given by (\ref{cr-LT-2}) and (\ref{sr-lap-2}). Being
analogous to the case of Models VI and VII the calculation is omitted.

\subsection{Calculation of creep compliance \label{Cr-Calc}}

\subsubsection{Creep compliance in the case of models having zero glass
compliance}

The rate of creep is obtained as%
\begin{equation}
\dot{\varepsilon}_{cr}\left( t\right) =\mathcal{L}^{-1}\left[ s\tilde{%
\varepsilon}_{cr}\left( s\right) \right] \left( t\right) =\frac{1}{2\pi 
\mathrm{i}}\lim_{\substack{ R\rightarrow \infty ,  \\ r\rightarrow 0}}%
\int_{\Gamma _{0}}s\tilde{\varepsilon}_{cr}\left( s\right) \mathrm{e}^{st}%
\mathrm{d}s,  \label{roc-ilt}
\end{equation}%
using the inverse Laplace transform (\ref{inv-laplas}) of the creep
compliance in complex domain (\ref{cr-LT}) and zero value of the glass
compliance (\ref{gg}). The Cauchy integral theorem%
\begin{equation}
\oint_{\Gamma ^{\left( \mathrm{I}\right) }}s\tilde{\varepsilon}_{cr}\left(
s\right) \mathrm{e}^{st}\mathrm{d}s=0,  \label{kif}
\end{equation}%
where the contour $\Gamma ^{\left( \mathrm{I}\right) }=\Gamma _{0}\cup
\Gamma _{1}\cup \Gamma _{2}\cup \Gamma _{3}\cup \Gamma _{4}\cup \Gamma
_{5}\cup \Gamma _{6}\cup \Gamma _{7}$ is chosen as in Figure \ref{fig-nema
nula}, 
\begin{figure}[tbph]
\centering
\includegraphics[scale=0.7]{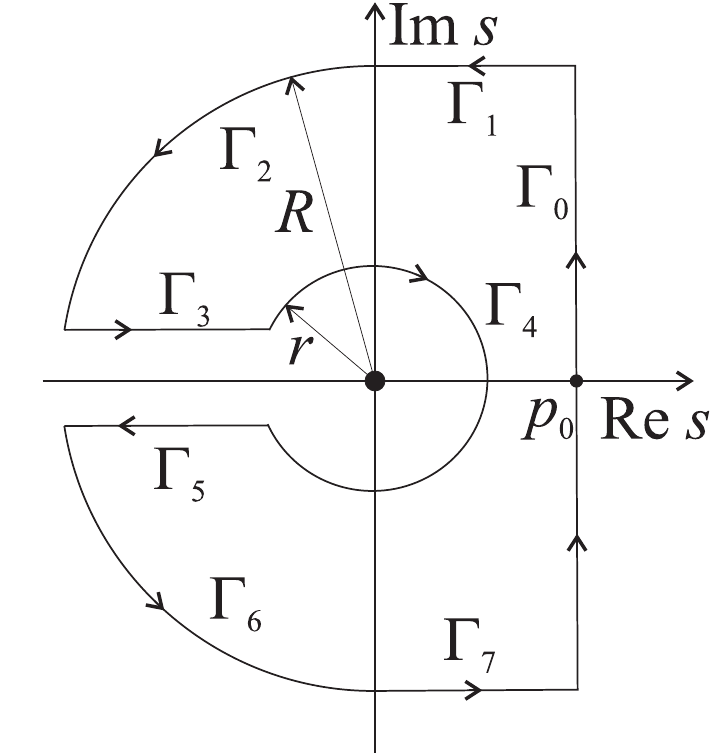}
\caption{Contour $\Gamma ^{\left( \mathrm{I}\right) }.$}
\label{fig-nema nula}
\end{figure}
yields the rate of creep in the form 
\begin{eqnarray}
\dot{\varepsilon}_{cr}\left( t\right) &=&\frac{1}{2\pi \mathrm{i}}%
\int_{0}^{\infty }\left( \frac{\Psi \left( \rho \mathrm{e}^{-\mathrm{i}\pi
}\right) }{b_{1}+b_{2}\rho ^{\eta }\mathrm{e}^{-\mathrm{i}\eta \pi }}\mathrm{%
e}^{\mathrm{i}\mu \pi }-\frac{\Psi \left( \rho \mathrm{e}^{\mathrm{i}\pi
}\right) }{b_{1}+b_{2}\rho ^{\eta }\mathrm{e}^{\mathrm{i}\eta \pi }}\mathrm{e%
}^{-\mathrm{i}\mu \pi }\right) \frac{\mathrm{e}^{-\rho t}}{\rho ^{\mu }}%
\mathrm{d}\rho  \label{roc-0} \\
&=&\frac{1}{\pi }\int_{0}^{\infty }\frac{K\left( \rho \right) }{\left\vert
b_{1}+b_{2}\rho ^{\eta }\mathrm{e}^{\mathrm{i}\eta \pi }\right\vert ^{2}}%
\frac{\mathrm{e}^{-\rho t}}{\rho ^{\mu }}\mathrm{d}\rho ,  \label{roc}
\end{eqnarray}%
and thus the creep compliance becomes%
\begin{equation*}
\varepsilon _{cr}\left( t\right) =\frac{1}{\pi }\int_{0}^{\infty }\frac{%
K\left( \rho \right) }{\left\vert b_{1}+b_{2}\rho ^{\eta }\mathrm{e}^{%
\mathrm{i}\eta \pi }\right\vert ^{2}}\frac{1-\mathrm{e}^{-\rho t}}{\rho
^{1+\mu }}\mathrm{d}\rho ,
\end{equation*}%
where,%
\begin{eqnarray*}
K\left( \rho \right) &=&b_{1}K_{1}\left( \rho \right) +b_{2}\rho ^{\eta
}K_{2}\left( \rho \right) ,\;\;\text{with} \\
K_{1}\left( \rho \right) &=&\sin \left( \mu \pi \right) \func{Re}\Psi \left(
\rho \mathrm{e}^{\mathrm{i}\pi }\right) -\cos \left( \mu \pi \right) \func{Im%
}\Psi \left( \rho \mathrm{e}^{\mathrm{i}\pi }\right) \\
&=&\sin \left( \mu \pi \right) +a_{1}\rho ^{\alpha }\sin \left( \left( \mu
-\alpha \right) \pi \right) +a_{2}\rho ^{\beta }\sin \left( \left( \mu
-\beta \right) \pi \right) +a_{3}\rho ^{\gamma }\sin \left( \left( \mu
-\gamma \right) \pi \right) , \\
K_{2}\left( \rho \right) &=&\sin \left( \left( \mu +\eta \right) \pi \right) 
\func{Re}\Psi \left( \rho \mathrm{e}^{\mathrm{i}\pi }\right) -\cos \left(
\left( \mu +\eta \right) \pi \right) \func{Im}\Psi \left( \rho \mathrm{e}^{%
\mathrm{i}\pi }\right) \\
&=&\sin \left( \left( \mu +\eta \right) \pi \right) +a_{1}\rho ^{\alpha
}\sin \left( \left( \mu +\eta -\alpha \right) \pi \right) +a_{2}\rho ^{\beta
}\sin \left( \left( \mu +\eta -\beta \right) \pi \right) +a_{3}\rho ^{\gamma
}\sin \left( \left( \mu +\eta -\gamma \right) \pi \right) ,
\end{eqnarray*}%
since $\Psi \left( \bar{s}\right) =-\bar{\Psi}\left( s\right) .$

The integrals along contours $\Gamma _{3}$ (parametrized by $s=\rho \mathrm{e%
}^{\mathrm{i}\pi },$ $\rho \in \left( r,R\right) $) and $\Gamma _{5}$
(parametrized by $s=\rho \mathrm{e}^{-\mathrm{i}\pi },$ $\rho \in \left(
r,R\right) $) read%
\begin{eqnarray*}
\lim_{\substack{ R\rightarrow \infty ,  \\ r\rightarrow 0}}\int_{\Gamma
_{3}}s\tilde{\varepsilon}_{cr}\left( s\right) \mathrm{e}^{st}\mathrm{d}s
&=&\int_{\infty }^{0}\frac{1}{\rho ^{\mu }\mathrm{e}^{\mathrm{i}\mu \pi }}%
\frac{\Psi \left( \rho \mathrm{e}^{\mathrm{i}\pi }\right) }{b_{1}+b_{2}\rho
^{\eta }\mathrm{e}^{\mathrm{i}\eta \pi }}\mathrm{e}^{-\rho t}\mathrm{e}^{%
\mathrm{i}\pi }\mathrm{d}\rho =\int_{0}^{\infty }\frac{\Psi \left( \rho 
\mathrm{e}^{\mathrm{i}\pi }\right) }{b_{1}+b_{2}\rho ^{\eta }\mathrm{e}^{%
\mathrm{i}\eta \pi }}\mathrm{e}^{-\mathrm{i}\mu \pi }\frac{\mathrm{e}^{-\rho
t}}{\rho ^{\mu }}\mathrm{d}\rho , \\
\lim_{\substack{ R\rightarrow \infty ,  \\ r\rightarrow 0}}\int_{\Gamma
_{5}}s\tilde{\varepsilon}_{cr}\left( s\right) \mathrm{e}^{st}\mathrm{d}s
&=&\int_{0}^{\infty }\frac{1}{\rho ^{\mu }\mathrm{e}^{-\mathrm{i}\mu \pi }}%
\frac{\Psi \left( \rho \mathrm{e}^{-\mathrm{i}\pi }\right) }{b_{1}+b_{2}\rho
^{\eta }\mathrm{e}^{-\mathrm{i}\eta \pi }}\mathrm{e}^{-\rho t}\mathrm{e}^{-%
\mathrm{i}\pi }\mathrm{d}\rho =-\int_{0}^{\infty }\frac{\Psi \left( \rho 
\mathrm{e}^{-\mathrm{i}\pi }\right) }{b_{1}+b_{2}\rho ^{\eta }\mathrm{e}^{-%
\mathrm{i}\eta \pi }}\mathrm{e}^{\mathrm{i}\mu \pi }\frac{\mathrm{e}^{-\rho
t}}{\rho ^{\mu }}\mathrm{d}\rho ,
\end{eqnarray*}%
yielding the rate of creep (\ref{roc-0}) according to the Cauchy integral
theorem (\ref{kif}), since the inverse Laplace transform of the rate of
creep in complex domain is (\ref{roc-ilt}) and integrals along $\Gamma _{1},$
$\Gamma _{2},$ $\Gamma _{4},$ $\Gamma _{6},$ $\Gamma _{7}$ tend to zero as $%
R\rightarrow \infty $ and $r\rightarrow 0.$

The contour $\Gamma _{1}$ is parametrized by $s=p+\mathrm{i}R,$ $p\in \left(
0,p_{0}\right) ,$ with $R\rightarrow \infty ,$ so that the integral%
\begin{equation*}
I_{\Gamma _{1}}=\int_{\Gamma _{1}}s\tilde{\varepsilon}_{cr}\left( s\right) 
\mathrm{e}^{st}\mathrm{d}s=\int_{0}^{p_{0}}\frac{1}{(p+\mathrm{i}R)^{\mu }}%
\frac{\Psi \left( p+\mathrm{i}R\right) }{b_{1}+b_{2}(p+\mathrm{i}R)^{\eta }}%
\mathrm{e}^{\left( p+\mathrm{i}R\right) t}\mathrm{d}p
\end{equation*}%
is estimated as%
\begin{equation}
\left\vert I_{\Gamma _{1}}\right\vert \leq \int_{0}^{p_{0}}\frac{1}{%
\left\vert p+\mathrm{i}R\right\vert ^{\mu }}\frac{\left\vert \Psi \left( p+%
\mathrm{i}R\right) \right\vert }{\left\vert b_{1}+b_{2}(p+\mathrm{i}R)^{\eta
}\right\vert }\mathrm{e}^{pt}\mathrm{d}p.  \label{int-gama1-cr}
\end{equation}%
Assuming $s=\rho \mathrm{e}^{\mathrm{i}\varphi },$ since $R\rightarrow
\infty ,$ one obtains $\rho =\sqrt{p^{2}+R^{2}}\sim R$ and $\varphi =\arctan 
\frac{R}{p}\sim \frac{\pi }{2},$ so that (\ref{int-gama1-cr}) becomes%
\begin{equation*}
\lim_{R\rightarrow \infty }\left\vert I_{\Gamma _{1}}\right\vert \leq
\lim_{R\rightarrow \infty }\int_{0}^{p_{0}}\frac{1}{R^{\mu }}\frac{%
\left\vert \Psi \left( R\mathrm{e}^{\mathrm{i}\frac{\pi }{2}}\right)
\right\vert }{\left\vert b_{1}+b_{2}R^{\eta }\mathrm{e}^{\mathrm{i}\frac{%
\eta \pi }{2}}\right\vert }\mathrm{e}^{pt}\mathrm{d}p\leq \frac{a_{3}}{b_{2}}%
\lim_{R\rightarrow \infty }\int_{0}^{p_{0}}\frac{1}{R^{\mu +\eta -\gamma }}%
\mathrm{e}^{pt}\mathrm{d}p=0,
\end{equation*}%
since, by (\ref{psi}) one has%
\begin{equation}
\left\vert \Psi \left( R\mathrm{e}^{\mathrm{i}\frac{\pi }{2}}\right)
\right\vert \sim a_{3}R^{\gamma },\;\;\text{as}\;\;R\rightarrow \infty ,
\label{est-k}
\end{equation}%
as well as%
\begin{equation}
\mu +\eta -\gamma =\left\{ 
\begin{tabular}{ll}
$\mu ,$ or $\mu -\left( \gamma -\beta \right) ,$ or $\mu -\left( \gamma
-\alpha \right) ,$ & for Model I, \\ 
$\mu -\alpha ,$ & for Models II and IV, \\ 
$\mu -\beta ,$ & for Models III and V,%
\end{tabular}%
\right. \in \left( 0,1\right) ,  \label{mi-eta-gama}
\end{equation}%
due to $\eta =\kappa \in \left\{ \alpha ,\beta ,\gamma \right\} $, with
thermodynamical restriction (\ref{Model 1 - tdr}) for Model I, and due to $%
\left( \eta ,\gamma \right) \in \{\left( \alpha ,2\alpha \right) ,$ $\left(
\alpha ,\alpha +\beta \right) ,$ $\left( \beta ,\alpha +\beta \right) ,$ $%
\left( \beta ,2\beta \right) \},$ with thermodynamical restrictions (\ref%
{Model 2 - tdr}), (\ref{Model 3 - tdr}), (\ref{Model 4 - tdr}), and (\ref%
{Model 5 - tdr}) for Models II - V. Analogously, it can be proved that $%
\lim_{R\rightarrow \infty }\left\vert I_{\Gamma _{7}}\right\vert =0.$

The integral along contour $\Gamma _{2},$ parametrized by $s=R\mathrm{e}^{%
\mathrm{i}\varphi },$ $\varphi \in \left( \frac{\pi }{2},\pi \right) ,$ with 
$R\rightarrow \infty ,$ is%
\begin{equation*}
I_{\Gamma _{2}}=\int_{\Gamma _{2}}s\tilde{\varepsilon}_{cr}\left( s\right) 
\mathrm{e}^{st}\mathrm{d}s=\int_{\frac{\pi }{2}}^{\pi }\frac{1}{R^{\mu }%
\mathrm{e}^{\mathrm{i}\mu \varphi }}\frac{\Psi \left( R\mathrm{e}^{\mathrm{i}%
\varphi }\right) }{b_{1}+b_{2}R^{\eta }\mathrm{e}^{\mathrm{i}\eta \varphi }}%
\mathrm{e}^{Rt\mathrm{e}^{\mathrm{i}\varphi }}\mathrm{i}R\mathrm{e}^{\mathrm{%
i}\varphi }\mathrm{d}\varphi ,
\end{equation*}%
yielding the estimate%
\begin{equation}
\left\vert I_{\Gamma _{2}}\right\vert \leq \int_{\frac{\pi }{2}}^{\pi
}R^{1-\mu }\frac{\left\vert \Psi \left( R\mathrm{e}^{\mathrm{i}\varphi
}\right) \right\vert }{\left\vert b_{1}+b_{2}R^{\eta }\mathrm{e}^{\mathrm{i}%
\eta \varphi }\right\vert }\mathrm{e}^{Rt\cos \varphi }\mathrm{d}\varphi .
\label{int-gama2-cr}
\end{equation}%
Similarly as in (\ref{est-k}), one has $\left\vert \Psi \left( R\mathrm{e}^{%
\mathrm{i}\varphi }\right) \right\vert \sim a_{3}R^{\gamma },$ as $%
R\rightarrow \infty ,$ which, along with $\cos \varphi <0$ for $\varphi \in
\left( \frac{\pi }{2},\pi \right) ,$ implies that (\ref{int-gama2-cr}) in
the limit when $R\rightarrow \infty $ becomes%
\begin{equation*}
\lim_{R\rightarrow \infty }\left\vert I_{\Gamma _{2}}\right\vert \leq \frac{%
a_{3}}{b_{2}}\lim_{R\rightarrow \infty }\int_{\frac{\pi }{2}}^{\pi
}R^{1-\left( \mu +\eta -\gamma \right) }\mathrm{e}^{Rt\cos \varphi }\mathrm{d%
}\varphi =0,
\end{equation*}%
although, by (\ref{mi-eta-gama}), $1-\left( \mu +\eta -\gamma \right) >0.$
By the similar arguments, $\lim_{R\rightarrow \infty }\left\vert I_{\Gamma
_{6}}\right\vert =0,$ as well.

Parametrization of the contour $\Gamma _{4}$ is $s=r\mathrm{e}^{\mathrm{i}%
\varphi },$ $\varphi \in \left( -\pi ,\pi \right) ,$ with $r\rightarrow 0,$
so that%
\begin{equation*}
I_{\Gamma _{4}}=\int_{\Gamma _{4}}s\tilde{\varepsilon}_{cr}\left( s\right) 
\mathrm{e}^{st}\mathrm{d}s=\int_{-\pi }^{\pi }\frac{1}{r^{\mu }\mathrm{e}^{%
\mathrm{i}\mu \varphi }}\frac{\Psi \left( r\mathrm{e}^{\mathrm{i}\varphi
}\right) }{b_{1}+b_{2}r^{\eta }\mathrm{e}^{\mathrm{i}\eta \varphi }}\mathrm{e%
}^{rt\mathrm{e}^{\mathrm{i}\varphi }}\mathrm{i}r\mathrm{e}^{\mathrm{i}%
\varphi }\mathrm{d}\varphi
\end{equation*}%
can be estimated by 
\begin{equation*}
\lim_{r\rightarrow 0}\left\vert I_{\Gamma _{4}}\right\vert \leq
\lim_{r\rightarrow 0}\int_{-\pi }^{\pi }r^{1-\mu }\frac{\left\vert \Psi
\left( r\mathrm{e}^{\mathrm{i}\varphi }\right) \right\vert }{\left\vert
b_{1}+b_{2}r^{\eta }\mathrm{e}^{\mathrm{i}\eta \varphi }\right\vert }\mathrm{%
e}^{rt\cos \varphi }\mathrm{d}\varphi \leq \frac{2\pi }{b_{1}}%
\lim_{r\rightarrow 0}r^{1-\mu }=0,
\end{equation*}%
since $\left\vert \Psi \left( r\mathrm{e}^{\mathrm{i}\varphi }\right)
\right\vert \sim 1$ as $r\rightarrow 0.$

\subsubsection{Creep compliance in the case of models having non-zero glass
compliance}

Inverting the Laplace transform in the creep compliance in complex domain (%
\ref{cr-LT-1}), the creep compliance is obtained in the form%
\begin{equation*}
\varepsilon _{cr}\left( t\right) =\frac{a_{3}}{b_{2}}+\frac{a_{3}}{b_{2}}%
\int_{0}^{t}f_{cr}\left( \tau \right) \mathrm{d}\tau ,
\end{equation*}%
where $\varepsilon _{cr}^{\left( g\right) }=\frac{a_{3}}{b_{2}}$ is the
glass compliance (\ref{gg}) and function $f_{cr}$ is defined by its Laplace
transform as%
\begin{equation*}
\tilde{f}_{cr}\left( s\right) =\mathcal{L}\left[ f_{cr}\left( t\right) %
\right] \left( s\right) =\frac{1}{s^{\beta }}\frac{\psi \left( s\right) }{%
\frac{b_{1}}{b_{2}}+s^{\eta }},
\end{equation*}%
with function $\psi $ defined by (\ref{psi1}). Function $f_{cr},$ having the
form%
\begin{eqnarray*}
f_{cr}\left( t\right) &=&\frac{1}{2\pi \mathrm{i}}\int_{0}^{\infty }\left( 
\frac{\psi \left( \rho \mathrm{e}^{-\mathrm{i}\pi }\right) }{\frac{b_{1}}{%
b_{2}}+\rho ^{\eta }\mathrm{e}^{-\mathrm{i}\eta \pi }}\mathrm{e}^{\mathrm{i}%
\beta \pi }-\frac{\psi \left( \rho \mathrm{e}^{\mathrm{i}\pi }\right) }{%
\frac{b_{1}}{b_{2}}+\rho ^{\eta }\mathrm{e}^{\mathrm{i}\eta \pi }}\mathrm{e}%
^{-\mathrm{i}\beta \pi }\right) \frac{\mathrm{e}^{-\rho t}}{\rho ^{\beta }}%
\mathrm{d}\rho \\
&=&\frac{1}{\pi }\int_{0}^{\infty }\frac{Q\left( \rho \right) }{\left\vert 
\frac{b_{1}}{b_{2}}+\rho ^{\eta }\mathrm{e}^{\mathrm{i}\eta \pi }\right\vert
^{2}}\frac{\mathrm{e}^{-\rho t}}{\rho ^{\beta }}\mathrm{d}\rho ,
\end{eqnarray*}%
where, due to $\psi \left( \bar{s}\right) =-\bar{\psi}\left( s\right) ,$%
\begin{eqnarray*}
Q\left( \rho \right) &=&\frac{b_{1}}{b_{2}}Q_{1}\left( \rho \right) +\rho
^{\eta }Q_{2}\left( \rho \right) ,\;\;\text{with} \\
Q_{1}\left( \rho \right) &=&\sin \left( \beta \pi \right) \func{Re}\psi
\left( \rho \mathrm{e}^{\mathrm{i}\pi }\right) -\cos \left( \beta \pi
\right) \func{Im}\psi \left( \rho \mathrm{e}^{\mathrm{i}\pi }\right) \\
&=&\frac{1}{a_{3}}\sin \left( \beta \pi \right) +\frac{a_{1}}{a_{3}}\rho
^{\alpha }\sin \left( \left( \beta -\alpha \right) \pi \right) , \\
Q_{2}\left( \rho \right) &=&\sin \left( \left( \beta +\eta \right) \pi
\right) \func{Re}\psi \left( \rho \mathrm{e}^{\mathrm{i}\pi }\right) -\cos
\left( \left( \beta +\eta \right) \pi \right) \func{Im}\psi \left( \rho 
\mathrm{e}^{\mathrm{i}\pi }\right) \\
&=&\frac{1}{a_{3}}\sin \left( \left( \beta +\eta \right) \pi \right) +\frac{%
a_{1}}{a_{3}}\rho ^{\alpha }\sin \left( \left( \beta +\eta -\alpha \right)
\pi \right) +\left( \frac{a_{2}}{a_{3}}-\frac{b_{1}}{b_{2}}\right) \rho
^{\beta }\sin \left( \eta \pi \right) ,
\end{eqnarray*}%
is calculated by the inverse Laplace transform%
\begin{equation*}
f_{cr}\left( t\right) =\frac{1}{2\pi \mathrm{i}}\lim_{\substack{ %
R\rightarrow \infty ,  \\ r\rightarrow 0}}\int_{\Gamma _{0}}\tilde{f}%
_{cr}\left( s\right) \mathrm{e}^{st}\mathrm{d}s,
\end{equation*}%
see (\ref{inv-laplas}), using the Cauchy integral theorem%
\begin{equation*}
\oint_{\Gamma ^{\left( \mathrm{I}\right) }}\tilde{f}_{cr}\left( s\right) 
\mathrm{e}^{st}\mathrm{d}s=0,
\end{equation*}%
where the contour $\Gamma ^{\left( \mathrm{I}\right) }=\Gamma _{0}\cup
\Gamma _{1}\cup \Gamma _{2}\cup \Gamma _{3}\cup \Gamma _{4}\cup \Gamma
_{5}\cup \Gamma _{6}\cup \Gamma _{7}$ is chosen as in Figure \ref{fig-nema
nula}, since the integrals along contours $\Gamma _{3}$ (parametrized by $%
s=\rho \mathrm{e}^{\mathrm{i}\pi },$ $\rho \in \left( r,R\right) $) and $%
\Gamma _{5}$ (parametrized by $s=\rho \mathrm{e}^{-\mathrm{i}\pi },$ $\rho
\in \left( r,R\right) $) read 
\begin{eqnarray*}
\lim_{\substack{ R\rightarrow \infty ,  \\ r\rightarrow 0}}\int_{\Gamma _{3}}%
\tilde{f}_{cr}\left( s\right) \mathrm{e}^{st}\mathrm{d}s &=&\int_{\infty
}^{0}\frac{1}{\rho ^{\beta }\mathrm{e}^{\mathrm{i}\beta \pi }}\frac{\psi
\left( \rho \mathrm{e}^{\mathrm{i}\pi }\right) }{\frac{b_{1}}{b_{2}}+\rho
^{\eta }\mathrm{e}^{\mathrm{i}\eta \pi }}\mathrm{e}^{-\rho t}\mathrm{e}^{%
\mathrm{i}\pi }\mathrm{d}\rho =\int_{0}^{\infty }\frac{\psi \left( \rho 
\mathrm{e}^{\mathrm{i}\pi }\right) }{\frac{b_{1}}{b_{2}}+\rho ^{\eta }%
\mathrm{e}^{\mathrm{i}\eta \pi }}\mathrm{e}^{-\mathrm{i}\beta \pi }\frac{%
\mathrm{e}^{-\rho t}}{\rho ^{\beta }}\mathrm{d}\rho , \\
\lim_{\substack{ R\rightarrow \infty ,  \\ r\rightarrow 0}}\int_{\Gamma _{5}}%
\tilde{f}_{cr}\left( s\right) \mathrm{e}^{st}\mathrm{d}s &=&\int_{0}^{\infty
}\frac{1}{\rho ^{\beta }\mathrm{e}^{-\mathrm{i}\beta \pi }}\frac{\psi \left(
\rho \mathrm{e}^{-\mathrm{i}\pi }\right) }{\frac{b_{1}}{b_{2}}+\rho ^{\eta }%
\mathrm{e}^{-\mathrm{i}\eta \pi }}\mathrm{e}^{-\rho t}\mathrm{e}^{-\mathrm{i}%
\pi }\mathrm{d}\rho =-\int_{0}^{\infty }\frac{\psi \left( \rho \mathrm{e}^{-%
\mathrm{i}\pi }\right) }{\frac{b_{1}}{b_{2}}+\rho ^{\eta }\mathrm{e}^{-%
\mathrm{i}\eta \pi }}\mathrm{e}^{\mathrm{i}\beta \pi }\frac{\mathrm{e}%
^{-\rho t}}{\rho ^{\beta }}\mathrm{d}\rho ,
\end{eqnarray*}%
respectively, while the integrals along $\Gamma _{1},$ $\Gamma _{2},$ $%
\Gamma _{4},$ $\Gamma _{6},$ $\Gamma _{7}$ tend to zero as $R\rightarrow
\infty $ and $r\rightarrow 0.$

The contour $\Gamma _{1}$ is parametrized by $s=p+\mathrm{i}R,$ $p\in \left(
0,p_{0}\right) ,$ with $R\rightarrow \infty ,$ so that the integral%
\begin{equation*}
I_{\Gamma _{1}}=\int_{\Gamma _{1}}\tilde{f}_{cr}\left( s\right) \mathrm{e}%
^{st}\mathrm{d}s=\int_{0}^{p_{0}}\frac{1}{(p+\mathrm{i}R)^{\beta }}\frac{%
\psi \left( p+\mathrm{i}R\right) }{\frac{b_{1}}{b_{2}}+(p+\mathrm{i}R)^{\eta
}}\mathrm{e}^{\left( p+\mathrm{i}R\right) t}\mathrm{d}p
\end{equation*}%
is estimated as%
\begin{equation}
\left\vert I_{\Gamma _{1}}\right\vert \leq \int_{0}^{p_{0}}\frac{1}{%
\left\vert p+\mathrm{i}R\right\vert ^{\beta }}\frac{\left\vert \psi \left( p+%
\mathrm{i}R\right) \right\vert }{\left\vert \frac{b_{1}}{b_{2}}+(p+\mathrm{i}%
R)^{\eta }\right\vert }\mathrm{e}^{pt}\mathrm{d}p.  \label{int-gama1-cr1}
\end{equation}%
Assuming $s=\rho \mathrm{e}^{\mathrm{i}\varphi },$ since $R\rightarrow
\infty ,$ one obtains $\rho =\sqrt{p^{2}+R^{2}}\sim R$ and $\varphi =\arctan 
\frac{R}{p}\sim \frac{\pi }{2},$ so that (\ref{int-gama1-cr1}) becomes%
\begin{equation*}
\lim_{R\rightarrow \infty }\left\vert I_{\Gamma _{1}}\right\vert \leq
\lim_{R\rightarrow \infty }\int_{0}^{p_{0}}\frac{1}{R^{\beta }}\frac{%
\left\vert \psi \left( R\mathrm{e}^{\mathrm{i}\frac{\pi }{2}}\right)
\right\vert }{\left\vert \frac{b_{1}}{b_{2}}+R^{\eta }\mathrm{e}^{\mathrm{i}%
\frac{\eta \pi }{2}}\right\vert }\mathrm{e}^{pt}\mathrm{d}p\leq \left( \frac{%
a_{2}}{a_{3}}-\frac{b_{1}}{b_{2}}\right) \lim_{R\rightarrow \infty
}\int_{0}^{p_{0}}\frac{1}{R^{\eta }}\mathrm{e}^{pt}\mathrm{d}p=0,
\end{equation*}%
since, by (\ref{psi1}), one has%
\begin{equation}
\left\vert \psi \left( R\mathrm{e}^{\mathrm{i}\frac{\pi }{2}}\right)
\right\vert \sim \left( \frac{a_{2}}{a_{3}}-\frac{b_{1}}{b_{2}}\right)
R^{\beta },\;\;\text{as}\;\;R\rightarrow \infty .  \label{est-k1}
\end{equation}%
Analogously, it can be proved that $\lim_{R\rightarrow \infty }\left\vert
I_{\Gamma _{7}}\right\vert =0.$

The integral along contour $\Gamma _{2},$ parametrized by $s=R\mathrm{e}^{%
\mathrm{i}\varphi },$ $\varphi \in \left( \frac{\pi }{2},\pi \right) ,$ with 
$R\rightarrow \infty ,$ is%
\begin{equation*}
I_{\Gamma _{2}}=\int_{\Gamma _{2}}\tilde{f}_{cr}\left( s\right) \mathrm{e}%
^{st}\mathrm{d}s=\int_{\frac{\pi }{2}}^{\pi }\frac{1}{R^{\beta }\mathrm{e}^{%
\mathrm{i}\beta \varphi }}\frac{\psi \left( R\mathrm{e}^{\mathrm{i}\varphi
}\right) }{\frac{b_{1}}{b_{2}}+R^{\eta }\mathrm{e}^{\mathrm{i}\eta \varphi }}%
\mathrm{e}^{Rt\mathrm{e}^{\mathrm{i}\varphi }}\mathrm{i}R\mathrm{e}^{\mathrm{%
i}\varphi }\mathrm{d}\varphi ,
\end{equation*}%
yielding the estimate%
\begin{equation}
\left\vert I_{\Gamma _{2}}\right\vert \leq \int_{\frac{\pi }{2}}^{\pi
}R^{1-\beta }\frac{\left\vert \psi \left( R\mathrm{e}^{\mathrm{i}\varphi
}\right) \right\vert }{\left\vert \frac{b_{1}}{b_{2}}+R^{\eta }\mathrm{e}^{%
\mathrm{i}\eta \varphi }\right\vert }\mathrm{e}^{Rt\cos \varphi }\mathrm{d}%
\varphi .  \label{int-gama2-cr1}
\end{equation}%
Similarly as in (\ref{est-k1}), $\left\vert \psi \left( R\mathrm{e}^{\mathrm{%
i}\varphi }\right) \right\vert \sim \left( \frac{a_{2}}{a_{3}}-\frac{b_{1}}{%
b_{2}}\right) R^{\beta },$ as $R\rightarrow \infty ,$ which along with $\cos
\varphi <0$ for $\varphi \in \left( \frac{\pi }{2},\pi \right) $ implies
that (\ref{int-gama2-cr1}) in the limit when $R\rightarrow \infty $ becomes%
\begin{equation*}
\lim_{R\rightarrow \infty }\left\vert I_{\Gamma _{2}}\right\vert \leq \left( 
\frac{a_{2}}{a_{3}}-\frac{b_{1}}{b_{2}}\right) \lim_{R\rightarrow \infty
}\int_{\frac{\pi }{2}}^{\pi }R^{1-\eta }\mathrm{e}^{Rt\cos \varphi }\mathrm{d%
}\varphi =0.
\end{equation*}%
By the similar arguments, $\lim_{R\rightarrow \infty }\left\vert I_{\Gamma
_{6}}\right\vert =0,$ as well.

Parametrization of the contour $\Gamma _{4}$ is $s=r\mathrm{e}^{\mathrm{i}%
\varphi },$ $\varphi \in \left( -\pi ,\pi \right) ,$ with $r\rightarrow 0,$
so that%
\begin{equation*}
I_{\Gamma _{4}}=\int_{\Gamma _{4}}\tilde{f}_{cr}\left( s\right) \mathrm{e}%
^{st}\mathrm{d}s=\int_{-\pi }^{\pi }\frac{1}{r^{\beta }\mathrm{e}^{\mathrm{i}%
\beta \varphi }}\frac{\psi \left( r\mathrm{e}^{\mathrm{i}\varphi }\right) }{%
\frac{b_{1}}{b_{2}}+r^{\eta }\mathrm{e}^{\mathrm{i}\eta \varphi }}\mathrm{e}%
^{rt\mathrm{e}^{\mathrm{i}\varphi }}\mathrm{i}r\mathrm{e}^{\mathrm{i}\varphi
}\mathrm{d}\varphi
\end{equation*}%
can be estimated by 
\begin{equation*}
\lim_{r\rightarrow 0}\left\vert I_{\Gamma _{4}}\right\vert \leq
\lim_{r\rightarrow 0}\int_{-\pi }^{\pi }r^{1-\beta }\frac{\left\vert \psi
\left( r\mathrm{e}^{\mathrm{i}\varphi }\right) \right\vert }{\left\vert 
\frac{b_{1}}{b_{2}}+r^{\eta }\mathrm{e}^{\mathrm{i}\eta \varphi }\right\vert 
}\mathrm{e}^{rt\cos \varphi }\mathrm{d}\varphi \leq 2\pi \frac{b_{2}}{%
a_{3}b_{1}}\lim_{r\rightarrow 0}r^{1-\beta }=0,
\end{equation*}%
since $\left\vert \psi \left( r\mathrm{e}^{\mathrm{i}\varphi }\right)
\right\vert \sim \frac{1}{a_{3}}$ as $r\rightarrow 0.$

\subsection{Calculation of relaxation modulus \label{SR-Calc}}

\subsubsection{Case when function $\Psi $ has no zeros in complex plane 
\label{Nemanula}}

\textbf{Relaxation modulus in the case of models having zero glass
compliance. }The Cauchy integral theorem with the relaxation modulus in
complex domain (\ref{sr-lap}) as an integrand becomes 
\begin{equation}
\oint_{\Gamma ^{\left( \mathrm{I}\right) }}\tilde{\sigma}_{sr}\left(
s\right) \mathrm{e}^{st}\mathrm{d}s=0,  \label{kit-sr}
\end{equation}%
where the contour $\Gamma ^{\left( \mathrm{I}\right) }=\Gamma _{0}\cup
\Gamma _{1}\cup \Gamma _{2}\cup \Gamma _{3}\cup \Gamma _{4}\cup \Gamma
_{5}\cup \Gamma _{6}\cup \Gamma _{7}$ is chosen as in Figure \ref{fig-nema
nula}. The relaxation modulus is obtained in the form%
\begin{eqnarray}
\sigma _{sr}\left( t\right) &=&\frac{1}{2\pi \mathrm{i}}\int_{0}^{\infty
}\left( \frac{b_{1}+b_{2}\rho ^{\eta }\mathrm{e}^{\mathrm{i}\eta \pi }}{\Psi
\left( \rho \mathrm{e}^{\mathrm{i}\pi }\right) }\mathrm{e}^{\mathrm{i}\mu
\pi }-\frac{b_{1}+b_{2}\rho ^{\eta }\mathrm{e}^{-\mathrm{i}\eta \pi }}{\Psi
\left( \rho \mathrm{e}^{-\mathrm{i}\pi }\right) }\mathrm{e}^{-\mathrm{i}\mu
\pi }\right) \frac{\mathrm{e}^{-\rho t}}{\rho ^{1-\mu }}\mathrm{d}\rho 
\notag \\
&=&\frac{1}{\pi }\int_{0}^{\infty }\frac{K\left( \rho \right) }{\left\vert
\Psi \left( \rho \mathrm{e}^{\mathrm{i}\pi }\right) \right\vert ^{2}}\frac{%
\mathrm{e}^{-\rho t}}{\rho ^{1-\mu }}\mathrm{d}\rho ,  \label{PrvoResenje}
\end{eqnarray}%
with function $K$ given by (\ref{funct-K}), as a consequence of the Cauchy
integral theorem (\ref{kit-sr}), since the integrals along contours $\Gamma
_{3}$ (parametrized by $s=\rho \mathrm{e}^{\mathrm{i}\pi },$ $\rho \in
\left( r,R\right) $) and $\Gamma _{5}$ (parametrized by $s=\rho \mathrm{e}^{-%
\mathrm{i}\pi },$ $\rho \in \left( r,R\right) $) read%
\begin{eqnarray}
\lim_{\substack{ R\rightarrow \infty ,  \\ r\rightarrow 0}}\int_{\Gamma _{3}}%
\tilde{\sigma}_{sr}\left( s\right) \mathrm{e}^{st}\mathrm{d}s
&=&\int_{\infty }^{0}\frac{1}{\left( \rho \mathrm{e}^{\mathrm{i}\pi }\right)
^{1-\mu }}\frac{b_{1}+b_{2}\rho ^{\eta }\mathrm{e}^{\mathrm{i}\eta \pi }}{%
\Psi \left( \rho \mathrm{e}^{\mathrm{i}\pi }\right) }\mathrm{e}^{-\rho t}%
\mathrm{e}^{\mathrm{i}\pi }\mathrm{d}\rho =-\int_{0}^{\infty }\frac{%
b_{1}+b_{2}\rho ^{\eta }\mathrm{e}^{\mathrm{i}\eta \pi }}{\Psi \left( \rho 
\mathrm{e}^{\mathrm{i}\pi }\right) }\mathrm{e}^{\mathrm{i}\mu \pi }\frac{%
\mathrm{e}^{-\rho t}}{\rho ^{1-\mu }}\mathrm{d}\rho ,  \label{int-gama3-1-5}
\\
\lim_{\substack{ R\rightarrow \infty ,  \\ r\rightarrow 0}}\int_{\Gamma _{5}}%
\tilde{\sigma}_{sr}\left( s\right) \mathrm{e}^{st}\mathrm{d}s
&=&\int_{0}^{\infty }\frac{1}{\left( \rho \mathrm{e}^{-\mathrm{i}\pi
}\right) ^{1-\mu }}\frac{b_{1}+b_{2}\rho ^{\eta }\mathrm{e}^{-\mathrm{i}\eta
\pi }}{\Psi \left( \rho \mathrm{e}^{-\mathrm{i}\pi }\right) }\mathrm{e}%
^{-\rho t}\mathrm{e}^{-\mathrm{i}\pi }\mathrm{d}\rho =\int_{0}^{\infty }%
\frac{b_{1}+b_{2}\rho ^{\eta }\mathrm{e}^{-\mathrm{i}\eta \pi }}{\Psi \left(
\rho \mathrm{e}^{-\mathrm{i}\pi }\right) }\mathrm{e}^{-\mathrm{i}\mu \pi }%
\frac{\mathrm{e}^{-\rho t}}{\rho ^{1-\mu }}\mathrm{d}\rho ,  \notag \\
&&  \label{int-gama5-1-5}
\end{eqnarray}%
respectively, with%
\begin{equation}
\sigma _{sr}\left( t\right) =\frac{1}{2\pi \mathrm{i}}\lim_{\substack{ %
R\rightarrow \infty ,  \\ r\rightarrow 0}}\int_{\Gamma _{0}}\tilde{\sigma}%
_{sr}\left( s\right) \mathrm{e}^{st}\mathrm{d}s,  \label{int-gama0-sr}
\end{equation}%
as the inverse Laplace transform, given by (\ref{inv-laplas}), while the
integrals along $\Gamma _{1},$ $\Gamma _{2},$ $\Gamma _{4},$ $\Gamma _{6},$ $%
\Gamma _{7}$ tend to zero as $R\rightarrow \infty $ and $r\rightarrow 0.$

The contour $\Gamma _{1}$ is parametrized by $s=p+\mathrm{i}R,$ $p\in \left(
0,p_{0}\right) ,$ with $R\rightarrow \infty ,$ so that the integral 
\begin{equation*}
I_{\Gamma _{1}}=\int_{\Gamma _{1}}\tilde{\sigma}_{sr}\left( s\right) \mathrm{%
e}^{st}\mathrm{d}s=\int_{0}^{p_{0}}\frac{1}{(p+\mathrm{i}R)^{1-\mu }}\frac{%
b_{1}+b_{2}(p+\mathrm{i}R)^{\eta }}{\Psi \left( p+\mathrm{i}R\right) }%
\mathrm{e}^{\left( p+\mathrm{i}R\right) t}\mathrm{d}p
\end{equation*}%
is estimated as%
\begin{equation}
\left\vert I_{\Gamma _{1}}\right\vert \leq \int_{0}^{p_{0}}\frac{1}{%
\left\vert p+\mathrm{i}R\right\vert ^{1-\mu }}\frac{\left\vert b_{1}+b_{2}(p+%
\mathrm{i}R)^{\eta }\right\vert }{\left\vert \Psi \left( p+\mathrm{i}%
R\right) \right\vert }\mathrm{e}^{pt}\mathrm{d}p.  \label{int-gama1}
\end{equation}%
Assuming $s=\rho \mathrm{e}^{\mathrm{i}\varphi },$ since $R\rightarrow
\infty ,$ one obtains $\rho =\sqrt{p^{2}+R^{2}}\sim R$ and $\varphi =\arctan 
\frac{R}{p}\sim \frac{\pi }{2},$ so that (\ref{int-gama1}) becomes%
\begin{equation*}
\lim_{R\rightarrow \infty }\left\vert I_{\Gamma _{1}}\right\vert \leq
\lim_{R\rightarrow \infty }\int_{0}^{p_{0}}\frac{1}{R^{1-\mu }}\frac{%
\left\vert b_{1}+b_{2}R^{\eta }\mathrm{e}^{\mathrm{i}\frac{\eta \pi }{2}%
}\right\vert }{\left\vert \Psi \left( R\mathrm{e}^{\mathrm{i}\frac{\pi }{2}%
}\right) \right\vert }\mathrm{e}^{pt}\mathrm{d}p\leq \frac{b_{2}}{a_{3}}%
\lim_{R\rightarrow \infty }\int_{0}^{p_{0}}\frac{1}{R^{1-\mu -\eta +\gamma }}%
\mathrm{e}^{pt}\mathrm{d}p=0,
\end{equation*}%
due to (\ref{est-k}) and inequality $1-\left( \mu +\eta -\gamma \right) >0$
following from (\ref{mi-eta-gama}). Analogously, it can be proved that $%
\lim_{R\rightarrow \infty }\left\vert I_{\Gamma _{7}}\right\vert =0.$

The integral along contour $\Gamma _{2},$ parametrized by $s=R\mathrm{e}^{%
\mathrm{i}\varphi },$ $\varphi \in \left( \frac{\pi }{2},\pi \right) ,$ with 
$R\rightarrow \infty ,$ is%
\begin{equation*}
I_{\Gamma _{2}}=\int_{\Gamma _{2}}\tilde{\sigma}_{sr}\left( s\right) \mathrm{%
e}^{st}\mathrm{d}s=\int_{\frac{\pi }{2}}^{\pi }\frac{1}{R^{1-\mu }\mathrm{e}%
^{\mathrm{i}\left( 1-\mu \right) \varphi }}\frac{b_{1}+b_{2}R^{\eta }\mathrm{%
e}^{\mathrm{i}\eta \varphi }}{\Psi \left( R\mathrm{e}^{\mathrm{i}\varphi
}\right) }\mathrm{e}^{Rt\mathrm{e}^{\mathrm{i}\varphi }}\mathrm{i}R\mathrm{e}%
^{\mathrm{i}\varphi }\mathrm{d}\varphi ,
\end{equation*}%
yielding the estimate%
\begin{equation}
\left\vert I_{\Gamma _{2}}\right\vert \leq \int_{\frac{\pi }{2}}^{\pi
}R^{\mu }\frac{\left\vert b_{1}+b_{2}R^{\eta }\mathrm{e}^{\mathrm{i}\eta
\varphi }\right\vert }{\left\vert \Psi \left( R\mathrm{e}^{\mathrm{i}\varphi
}\right) \right\vert }\mathrm{e}^{Rt\cos \varphi }\mathrm{d}\varphi .
\label{int-gama2}
\end{equation}%
Similarly as in (\ref{est-k}), $\left\vert \Psi \left( R\mathrm{e}^{\mathrm{i%
}\varphi }\right) \right\vert \sim a_{3}R^{\gamma },$ as $R\rightarrow
\infty ,$ which along with $\cos \varphi <0$ for $\varphi \in \left( \frac{%
\pi }{2},\pi \right) $ implies that (\ref{int-gama2}) in the limit when $%
R\rightarrow \infty $ becomes%
\begin{equation*}
\lim_{R\rightarrow \infty }\left\vert I_{\Gamma _{2}}\right\vert \leq \frac{%
b_{2}}{a_{3}}\lim_{R\rightarrow \infty }\int_{\frac{\pi }{2}}^{\pi }R^{\mu
+\eta -\gamma }\mathrm{e}^{Rt\cos \varphi }\mathrm{d}\varphi =0,
\end{equation*}%
although, by (\ref{mi-eta-gama}) $\mu +\eta -\gamma >0.$ By the similar
arguments, $\lim_{R\rightarrow \infty }\left\vert I_{\Gamma _{6}}\right\vert
=0,$ as well.

Parametrization of the contour $\Gamma _{4}$ is $s=r\mathrm{e}^{\mathrm{i}%
\varphi },$ $\varphi \in \left( -\pi ,\pi \right) ,$ with $r\rightarrow 0,$
so that%
\begin{equation*}
I_{\Gamma _{4}}=\int_{\Gamma _{4}}\tilde{\sigma}_{sr}\left( s\right) \mathrm{%
e}^{st}\mathrm{d}s=\int_{-\pi }^{\pi }\frac{1}{r^{1-\mu }\mathrm{e}^{\mathrm{%
i}\left( 1-\mu \right) \varphi }}\frac{b_{1}+b_{2}r^{\eta }\mathrm{e}^{%
\mathrm{i}\eta \varphi }}{\Psi \left( r\mathrm{e}^{\mathrm{i}\varphi
}\right) }\mathrm{e}^{rt\mathrm{e}^{\mathrm{i}\varphi }}\mathrm{i}r\mathrm{e}%
^{\mathrm{i}\varphi }\mathrm{d}\varphi ,
\end{equation*}%
can be estimated by 
\begin{equation*}
\lim_{r\rightarrow 0}\left\vert I_{\Gamma _{4}}\right\vert \leq
\lim_{r\rightarrow 0}\int_{-\pi }^{\pi }r^{\mu }\frac{\left\vert
b_{1}+b_{2}r^{\eta }\mathrm{e}^{\mathrm{i}\eta \varphi }\right\vert }{%
\left\vert \Psi \left( r\mathrm{e}^{\mathrm{i}\varphi }\right) \right\vert }%
\mathrm{e}^{rt\cos \varphi }\mathrm{d}\varphi \leq 2\pi
b_{1}\lim_{r\rightarrow 0}r^{\mu }=0,
\end{equation*}%
since $\left\vert \Psi \left( r\mathrm{e}^{\mathrm{i}\varphi }\right)
\right\vert \sim 1$ as $r\rightarrow 0.$

\textbf{Relaxation modulus in the case of models having non-zero glass
compliance. }Inverting the Laplace transform in the relaxation modulus in
complex domain (\ref{sr-lap-1}), the relaxation modulus is obtained in the
form%
\begin{eqnarray*}
\sigma _{sr}\left( t\right) &=&\frac{b_{2}}{a_{3}}-\frac{b_{2}}{a_{3}}%
\int_{0}^{t}\mathcal{L}^{-1}\left[ \tilde{f}_{sr}\left( s\right) \right]
\left( \tau \right) \mathrm{d}\tau \\
&=&\frac{b_{2}}{a_{3}}-\frac{b_{2}}{a_{3}}\int_{0}^{t}f_{sr}\left( \tau
\right) \mathrm{d}\tau ,
\end{eqnarray*}%
where $\sigma _{sr}^{\left( g\right) }=\frac{a_{3}}{b_{2}}$ is the glass
modulus (\ref{jg}), and function $f_{sr}$ is defined by its Laplace
transform as%
\begin{equation}
\tilde{f}_{sr}\left( s\right) =\frac{\psi \left( s\right) }{\psi \left(
s\right) +s^{\beta }\left( \frac{b_{1}}{b_{2}}+s^{\eta }\right) },
\label{f-tilda-sr}
\end{equation}%
with function $\psi $ defined by (\ref{psi1}). Function $f_{sr},$ having the
form%
\begin{eqnarray*}
f_{sr}\left( t\right) &=&\frac{1}{2\pi \mathrm{i}}\int_{0}^{\infty }\left( 
\frac{\psi \left( \rho \mathrm{e}^{-\mathrm{i}\pi }\right) }{\psi \left(
\rho \mathrm{e}^{-\mathrm{i}\pi }\right) +\rho ^{\beta }\mathrm{e}^{-\mathrm{%
i}\beta \pi }\left( \frac{b_{1}}{b_{2}}+\rho ^{\eta }\mathrm{e}^{-\mathrm{i}%
\eta \pi }\right) }-\frac{\psi \left( \rho \mathrm{e}^{\mathrm{i}\pi
}\right) }{\psi \left( \rho \mathrm{e}^{\mathrm{i}\pi }\right) +\rho ^{\beta
}\mathrm{e}^{\mathrm{i}\beta \pi }\left( \frac{b_{1}}{b_{2}}+\rho ^{\eta }%
\mathrm{e}^{\mathrm{i}\eta \pi }\right) }\right) \mathrm{e}^{-\rho t}\mathrm{%
d}\rho \\
&=&\frac{1}{\pi }\int_{0}^{\infty }\frac{Q\left( \rho \right) }{\left\vert
\psi \left( \rho \mathrm{e}^{\mathrm{i}\pi }\right) +\rho ^{\beta }\mathrm{e}%
^{\mathrm{i}\beta \pi }\left( \frac{b_{1}}{b_{2}}+\rho ^{\eta }\mathrm{e}^{%
\mathrm{i}\eta \pi }\right) \right\vert ^{2}}\rho ^{\beta }\mathrm{e}^{-\rho
t}\mathrm{d}\rho ,
\end{eqnarray*}%
with function $Q$ given by (\ref{Q}), is calculated as the inverse Laplace
transform%
\begin{equation}
\mathcal{L}^{-1}\left[ \tilde{f}_{sr}\left( s\right) \right] =\frac{1}{2\pi 
\mathrm{i}}\lim_{\substack{ R\rightarrow \infty ,  \\ r\rightarrow 0}}%
\int_{\Gamma _{0}}\tilde{f}_{sr}\left( s\right) \mathrm{e}^{st}\mathrm{d}s,
\label{int-gama0}
\end{equation}%
see (\ref{inv-laplas}), using the Cauchy integral theorem%
\begin{equation*}
\oint_{\Gamma ^{\left( \mathrm{I}\right) }}\tilde{f}_{sr}\left( s\right) 
\mathrm{e}^{st}\mathrm{d}s=0,
\end{equation*}%
where the contour $\Gamma ^{\left( \mathrm{I}\right) }=\Gamma _{0}\cup
\Gamma _{1}\cup \Gamma _{2}\cup \Gamma _{3}\cup \Gamma _{4}\cup \Gamma
_{5}\cup \Gamma _{6}\cup \Gamma _{7}$ is chosen as in Figure \ref{fig-nema
nula}, since the integrals along contours $\Gamma _{3}$ (parametrized by $%
s=\rho \mathrm{e}^{\mathrm{i}\pi },$ $\rho \in \left( r,R\right) $) and $%
\Gamma _{5}$ (parametrized by $s=\rho \mathrm{e}^{-\mathrm{i}\pi },$ $\rho
\in \left( r,R\right) $) read%
\begin{eqnarray}
\lim_{\substack{ R\rightarrow \infty ,  \\ r\rightarrow 0}}\int_{\Gamma _{3}}%
\tilde{f}_{sr}\left( s\right) \mathrm{e}^{st}\mathrm{d}s &=&\int_{\infty
}^{0}\frac{\psi \left( \rho \mathrm{e}^{\mathrm{i}\pi }\right) }{\psi \left(
\rho \mathrm{e}^{\mathrm{i}\pi }\right) +\rho ^{\beta }\mathrm{e}^{\mathrm{i}%
\beta \pi }\left( \frac{b_{1}}{b_{2}}+\rho ^{\eta }\mathrm{e}^{\mathrm{i}%
\eta \pi }\right) }\mathrm{e}^{-\rho t}\mathrm{e}^{\mathrm{i}\pi }\mathrm{d}%
\rho  \notag \\
&=&\int_{0}^{\infty }\frac{\psi \left( \rho \mathrm{e}^{\mathrm{i}\pi
}\right) }{\psi \left( \rho \mathrm{e}^{\mathrm{i}\pi }\right) +\rho ^{\beta
}\mathrm{e}^{\mathrm{i}\beta \pi }\left( \frac{b_{1}}{b_{2}}+\rho ^{\eta }%
\mathrm{e}^{\mathrm{i}\eta \pi }\right) }\mathrm{e}^{-\rho t}\mathrm{d}\rho ,
\label{int-gama3} \\
\lim_{\substack{ R\rightarrow \infty ,  \\ r\rightarrow 0}}\int_{\Gamma _{5}}%
\tilde{f}_{sr}\left( s\right) \mathrm{e}^{st}\mathrm{d}s &=&\int_{0}^{\infty
}\frac{\psi \left( \rho \mathrm{e}^{-\mathrm{i}\pi }\right) }{\psi \left(
\rho \mathrm{e}^{-\mathrm{i}\pi }\right) +\rho ^{\beta }\mathrm{e}^{-\mathrm{%
i}\beta \pi }\left( \frac{b_{1}}{b_{2}}+\rho ^{\eta }\mathrm{e}^{-\mathrm{i}%
\eta \pi }\right) }\mathrm{e}^{-\rho t}\mathrm{e}^{-\mathrm{i}\pi }\mathrm{d}%
\rho  \notag \\
&=&-\int_{0}^{\infty }\frac{\psi \left( \rho \mathrm{e}^{-\mathrm{i}\pi
}\right) }{\psi \left( \rho \mathrm{e}^{-\mathrm{i}\pi }\right) +\rho
^{\beta }\mathrm{e}^{-\mathrm{i}\beta \pi }\left( \frac{b_{1}}{b_{2}}+\rho
^{\eta }\mathrm{e}^{-\mathrm{i}\eta \pi }\right) }\mathrm{e}^{-\rho t}%
\mathrm{d}\rho ,  \label{int-gama5}
\end{eqnarray}%
respectively, while the integrals along $\Gamma _{1},$ $\Gamma _{2},$ $%
\Gamma _{4},$ $\Gamma _{6},$ $\Gamma _{7}$ tend to zero as $R\rightarrow
\infty $ and $r\rightarrow 0.$

The contour $\Gamma _{1}$ is parametrized by $s=p+\mathrm{i}R,$ $p\in \left(
0,p_{0}\right) ,$ with $R\rightarrow \infty ,$ so that the integral%
\begin{equation*}
I_{\Gamma _{1}}=\int_{\Gamma _{1}}\tilde{f}_{sr}\left( s\right) \mathrm{e}%
^{st}\mathrm{d}s=\int_{0}^{p_{0}}\frac{\psi \left( p+\mathrm{i}R\right) }{%
\psi \left( p+\mathrm{i}R\right) +(p+\mathrm{i}R)^{\beta }\left( \frac{b_{1}%
}{b_{2}}+(p+\mathrm{i}R)^{\eta }\right) }\mathrm{e}^{\left( p+\mathrm{i}%
R\right) t}\mathrm{d}p
\end{equation*}%
is estimated as%
\begin{equation}
\left\vert I_{\Gamma _{1}}\right\vert \leq \int_{0}^{p_{0}}\frac{\left\vert
\psi \left( p+\mathrm{i}R\right) \right\vert }{\left\vert \psi \left( p+%
\mathrm{i}R\right) +(p+\mathrm{i}R)^{\beta }\left( \frac{b_{1}}{b_{2}}+(p+%
\mathrm{i}R)^{\eta }\right) \right\vert }\mathrm{e}^{pt}\mathrm{d}p.
\label{int-gama1-sr1}
\end{equation}%
Assuming $s=\rho \mathrm{e}^{\mathrm{i}\varphi },$ since $R\rightarrow
\infty ,$ one obtains $\rho =\sqrt{p^{2}+R^{2}}\sim R$ and $\varphi =\arctan 
\frac{R}{p}\sim \frac{\pi }{2},$ so that (\ref{int-gama1-sr1}) becomes%
\begin{equation*}
\lim_{R\rightarrow \infty }\left\vert I_{\Gamma _{1}}\right\vert \leq \left( 
\frac{a_{2}}{a_{3}}-\frac{b_{1}}{b_{2}}\right) \lim_{R\rightarrow \infty
}\int_{0}^{p_{0}}\frac{1}{R^{\eta }}\mathrm{e}^{pt}\mathrm{d}p=0,
\end{equation*}%
due to (\ref{est-k1}). Analogously, it can be proved that $%
\lim_{R\rightarrow \infty }\left\vert I_{\Gamma _{7}}\right\vert =0.$

The integral along contour $\Gamma _{2},$ parametrized by $s=R\mathrm{e}^{%
\mathrm{i}\varphi },$ $\varphi \in \left( \frac{\pi }{2},\pi \right) ,$ with 
$R\rightarrow \infty ,$ is%
\begin{equation*}
I_{\Gamma _{2}}=\int_{\Gamma _{2}}\tilde{f}_{sr}\left( s\right) \mathrm{e}%
^{st}\mathrm{d}s=\int_{\frac{\pi }{2}}^{\pi }\frac{\psi \left( R\mathrm{e}^{%
\mathrm{i}\varphi }\right) }{\psi \left( R\mathrm{e}^{\mathrm{i}\varphi
}\right) +R^{\beta }\mathrm{e}^{\mathrm{i}\beta \varphi }\left( \frac{b_{1}}{%
b_{2}}+R^{\eta }\mathrm{e}^{\mathrm{i}\eta \varphi }\right) }\mathrm{e}^{Rt%
\mathrm{e}^{\mathrm{i}\varphi }}\mathrm{i}R\mathrm{e}^{\mathrm{i}\varphi }%
\mathrm{d}\varphi ,
\end{equation*}%
yielding the estimate%
\begin{equation}
\left\vert I_{\Gamma _{2}}\right\vert \leq \int_{\frac{\pi }{2}}^{\pi }\frac{%
\left\vert \psi \left( R\mathrm{e}^{\mathrm{i}\varphi }\right) \right\vert }{%
\left\vert \psi \left( R\mathrm{e}^{\mathrm{i}\varphi }\right) +R^{\beta }%
\mathrm{e}^{\mathrm{i}\beta \varphi }\left( \frac{b_{1}}{b_{2}}+R^{\eta }%
\mathrm{e}^{\mathrm{i}\eta \varphi }\right) \right\vert }R\mathrm{e}^{Rt\cos
\varphi }\mathrm{d}\varphi .  \label{int-gama2-sr1}
\end{equation}%
Similarly as in (\ref{est-k1}), $\left\vert \psi \left( R\mathrm{e}^{\mathrm{%
i}\varphi }\right) \right\vert \sim \left( \frac{a_{2}}{a_{3}}-\frac{b_{1}}{%
b_{2}}\right) R^{\beta },$ as $R\rightarrow \infty ,$ which along with $\cos
\varphi <0$ for $\varphi \in \left( \frac{\pi }{2},\pi \right) $ implies
that (\ref{int-gama2-sr1}) in the limit when $R\rightarrow \infty $ becomes%
\begin{equation*}
\lim_{R\rightarrow \infty }\left\vert I_{\Gamma _{2}}\right\vert \leq \left( 
\frac{a_{2}}{a_{3}}-\frac{b_{1}}{b_{2}}\right) \lim_{R\rightarrow \infty
}\int_{\frac{\pi }{2}}^{\pi }R^{1-\eta }\mathrm{e}^{Rt\cos \varphi }\mathrm{d%
}\varphi =0.
\end{equation*}%
By the similar arguments, $\lim_{R\rightarrow \infty }\left\vert I_{\Gamma
_{6}}\right\vert =0,$ as well.

Parametrization of the contour $\Gamma _{4}$ is $s=r\mathrm{e}^{\mathrm{i}%
\varphi },$ $\varphi \in \left( -\pi ,\pi \right) ,$ with $r\rightarrow 0,$
so that%
\begin{equation*}
I_{\Gamma _{4}}=\int_{\Gamma _{4}}\tilde{f}_{sr}\left( s\right) \mathrm{e}%
^{st}\mathrm{d}s=\int_{-\pi }^{\pi }\frac{\psi \left( r\mathrm{e}^{\mathrm{i}%
\varphi }\right) }{\psi \left( r\mathrm{e}^{\mathrm{i}\varphi }\right)
+r^{\beta }\mathrm{e}^{\mathrm{i}\beta \varphi }\left( \frac{b_{1}}{b_{2}}%
+r^{\eta }\mathrm{e}^{\mathrm{i}\eta \varphi }\right) }\mathrm{e}^{rt\mathrm{%
e}^{\mathrm{i}\varphi }}\mathrm{i}r\mathrm{e}^{\mathrm{i}\varphi }\mathrm{d}%
\varphi
\end{equation*}%
can be estimated by 
\begin{equation*}
\lim_{r\rightarrow 0}\left\vert I_{\Gamma _{4}}\right\vert \leq
\lim_{r\rightarrow 0}\int_{-\pi }^{\pi }\frac{\left\vert \psi \left( r%
\mathrm{e}^{\mathrm{i}\varphi }\right) \right\vert }{\left\vert \psi \left( r%
\mathrm{e}^{\mathrm{i}\varphi }\right) +r^{\beta }\mathrm{e}^{\mathrm{i}%
\beta \varphi }\left( \frac{b_{1}}{b_{2}}+r^{\eta }\mathrm{e}^{\mathrm{i}%
\eta \varphi }\right) \right\vert }r\mathrm{e}^{rt\cos \varphi }\mathrm{d}%
\varphi \leq 2\pi \lim_{r\rightarrow 0}r=0,
\end{equation*}%
since $\left\vert \psi \left( r\mathrm{e}^{\mathrm{i}\varphi }\right)
\right\vert \sim \frac{1}{a_{3}}$ as $r\rightarrow 0.$

\subsubsection{Case when function $\Psi $ has a negative real zero}

\textbf{Relaxation modulus in the case of models having zero glass
compliance. }The Cauchy integral theorem with the relaxation modulus in
complex domain (\ref{sr-lap}) as an integrand becomes 
\begin{equation}
\oint_{\Gamma ^{\left( \mathrm{II}\right) }}\tilde{\sigma}_{sr}\left(
s\right) \mathrm{e}^{st}\mathrm{d}s=0,  \label{Kit1}
\end{equation}%
where the contour $\Gamma ^{\left( \mathrm{II}\right) }=\Gamma _{0}\cup
\Gamma _{1}\cup \Gamma _{2}\cup \Gamma _{3a}\cup \Gamma ^{\ast }\cup \Gamma
_{3b}\cup \Gamma _{4}\cup \Gamma _{5a}\cup \Gamma _{\ast }\cup \Gamma
_{5b}\cup \Gamma _{6}\cup \Gamma _{7}$ is chosen as in Figure \ref{fig-neg
real nula} due to the existence of negative real zero $-\rho ^{\ast },$ with 
$\rho ^{\ast }$ determined from (\ref{rho-zvezda-1-5}), of function $\Psi ,$
given by (\ref{psi}). 
\begin{figure}[tbph]
\centering
\includegraphics[scale=0.7]{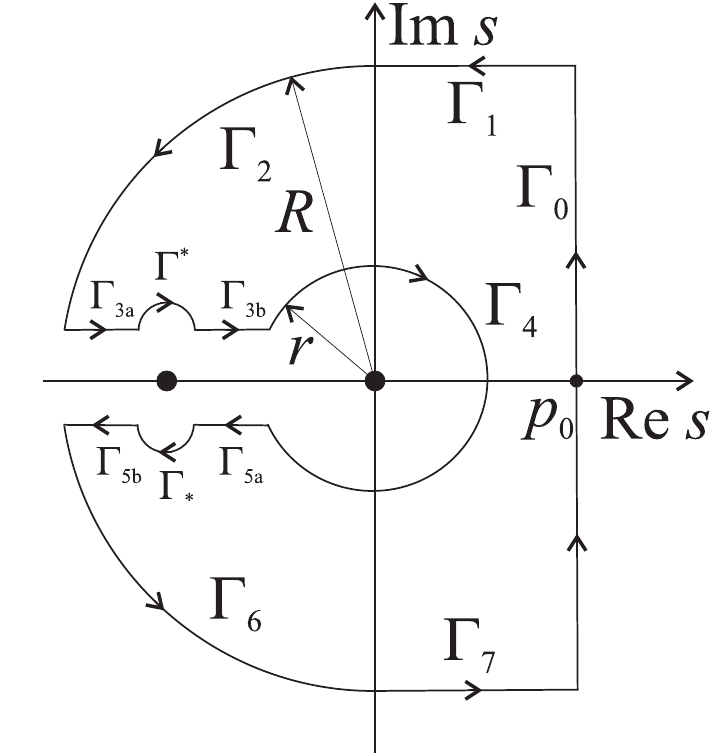}
\caption{Contour $\Gamma ^{\left( \mathrm{II}\right) }.$}
\label{fig-neg real nula}
\end{figure}

The relaxation modulus is obtained in the form%
\begin{equation}
\sigma _{sr}\left( t\right) =\frac{1}{\pi }\int_{0}^{\infty }\frac{K\left(
\rho \right) }{\left\vert \Psi \left( \rho \mathrm{e}^{\mathrm{i}\pi
}\right) \right\vert ^{2}}\frac{\mathrm{e}^{-\rho t}}{\rho ^{1-\mu }}\mathrm{%
d}\rho +f_{sr}^{\ast }\left( \rho ^{\ast }\right) \mathrm{e}^{-\rho ^{\ast
}t},  \label{DrugoResenje}
\end{equation}%
with functions $K$ and $f_{sr}^{\ast }$ given by (\ref{funct-K}) and (\ref%
{f-sr-zvezda-1-5}), using the Cauchy integral theorem (\ref{Kit1}). Namely,
the contour $\Gamma _{3a}$ is parametrized by $s=\rho \mathrm{e}^{\mathrm{i}%
\pi },$ $\rho \in \left( \rho ^{\ast }+r^{\ast },R\right) ,$ while $\Gamma
_{3b}$ has the same parametrization with $\rho \in \left( \rho ^{\ast
}-r^{\ast },r\right) $, so that in the limit when $R\rightarrow \infty ,$ $%
r\rightarrow 0,$ and $r^{\ast }\rightarrow 0$ the integral along contour $%
\Gamma _{3a}\cup \Gamma _{3b},$ as in (\ref{int-gama3-1-5}), reads%
\begin{equation}
\lim_{\substack{ R\rightarrow \infty ,  \\ r\rightarrow 0,  \\ r^{\ast
}\rightarrow 0}}\int_{\Gamma _{3a}\cup \Gamma _{3b}}\tilde{\sigma}%
_{sr}\left( s\right) \mathrm{e}^{st}\mathrm{d}s=-\int_{0}^{\infty }\frac{%
b_{1}+b_{2}\rho ^{\eta }\mathrm{e}^{\mathrm{i}\eta \pi }}{\Psi \left( \rho 
\mathrm{e}^{\mathrm{i}\pi }\right) }\mathrm{e}^{\mathrm{i}\mu \pi }\frac{%
\mathrm{e}^{-\rho t}}{\rho ^{1-\mu }}\mathrm{d}\rho ,  \label{Intgama3ab}
\end{equation}%
and similarly the integral along contour $\Gamma _{5a}\cup \Gamma _{5b}$
(parametrized by $s=\rho \mathrm{e}^{-\mathrm{i}\pi },$ $\rho \in \left(
\rho ^{\ast }-r^{\ast },r\right) $ for $\Gamma _{5a}$ and $\rho \in \left(
\rho ^{\ast }+r^{\ast },R\right) $ for $\Gamma _{5b}$), as in (\ref%
{int-gama5-1-5}), is 
\begin{equation}
\lim_{\substack{ R\rightarrow \infty ,  \\ r\rightarrow 0,  \\ r^{\ast
}\rightarrow 0}}\int_{\Gamma _{5a}\cup \Gamma _{5b}}\tilde{\sigma}%
_{sr}\left( s\right) \mathrm{e}^{st}\mathrm{d}s=\int_{0}^{\infty }\frac{%
b_{1}+b_{2}\rho ^{\eta }\mathrm{e}^{-\mathrm{i}\eta \pi }}{\Psi \left( \rho 
\mathrm{e}^{-\mathrm{i}\pi }\right) }\mathrm{e}^{-\mathrm{i}\mu \pi }\frac{%
\mathrm{e}^{-\rho t}}{\rho ^{1-\mu }}\mathrm{d}\rho ,  \label{Intgama5ab}
\end{equation}%
so that, along with the inverse Laplace transform (\ref{int-gama0-sr}), the
integrals in (\ref{Intgama3ab}) and (\ref{Intgama5ab}) yield the first term
in the relaxation modulus (\ref{DrugoResenje}), while, as it will be
calculated, the integrals along contours $\Gamma ^{\ast }$ and $\Gamma
_{\ast }$ yield the second term in (\ref{DrugoResenje}), as the integrals
along $\Gamma _{1},$ $\Gamma _{2},$ $\Gamma _{4},$ $\Gamma _{6},$ $\Gamma
_{7}$ tend to zero as $R\rightarrow \infty $ and $r\rightarrow 0,$ as
already proved in Section \ref{Nemanula}.

The contour $\Gamma ^{\ast }$ is parametrized by $s-\rho ^{\ast }\mathrm{e}^{%
\mathrm{i}\pi }=r^{\ast }\mathrm{e}^{\mathrm{i}\varphi },$ $\varphi \in
(0,\pi ),$ with $r^{\ast }\rightarrow 0,$ and the corresponding integral
reads%
\begin{equation*}
I_{\Gamma ^{\ast }}=\int_{\Gamma ^{\ast }}\tilde{\sigma}_{sr}\left( s\right) 
\mathrm{e}^{st}\mathrm{d}s=\int_{\pi }^{0}\frac{1}{\left( \rho ^{\ast }%
\mathrm{e}^{\mathrm{i}\pi }+r^{\ast }\mathrm{e}^{\mathrm{i}\varphi }\right)
^{1-\mu }}\frac{b_{1}+b_{2}\left( \rho ^{\ast }\mathrm{e}^{\mathrm{i}\pi
}+r^{\ast }\mathrm{e}^{\mathrm{i}\varphi }\right) ^{\eta }}{\Psi \left( \rho
^{\ast }\mathrm{e}^{\mathrm{i}\pi }+r^{\ast }\mathrm{e}^{\mathrm{i}\varphi
}\right) }\mathrm{e}^{\left( \rho ^{\ast }\mathrm{e}^{\mathrm{i}\pi
}+r^{\ast }\mathrm{e}^{\mathrm{i}\varphi }\right) t}\mathrm{i}r^{\ast }%
\mathrm{e}^{\mathrm{i}\varphi }\mathrm{d}\varphi ,
\end{equation*}%
so that by letting $r^{\ast }\rightarrow 0$ in the previous expression one
obtains%
\begin{eqnarray}
\lim_{_{\substack{ r^{\ast }\rightarrow 0}}}I_{\Gamma ^{\ast }} &=&-\mathrm{i%
}\frac{b_{1}+b_{2}\left( \rho ^{\ast }\right) ^{\eta }\mathrm{e}^{\mathrm{i}%
\eta \pi }}{\left( \rho ^{\ast }\right) ^{1-\mu }\mathrm{e}^{\mathrm{i}%
\left( 1-\mu \right) \pi }}\mathrm{e}^{-\rho ^{\ast }t}\lim_{_{\substack{ %
r^{\ast }\rightarrow 0}}}\int_{0}^{\pi }\frac{r^{\ast }\mathrm{e}^{\mathrm{i}%
\varphi }}{\Psi \left( \rho ^{\ast }\mathrm{e}^{\mathrm{i}\pi }+r^{\ast }%
\mathrm{e}^{\mathrm{i}\varphi }\right) }\mathrm{d}\varphi ,  \notag \\
&=&-\mathrm{i}\pi \left( \rho ^{\ast }\right) ^{\mu }\mathrm{e}^{\mathrm{i}%
\mu \pi }\frac{b_{1}+b_{2}\left( \rho ^{\ast }\right) ^{\eta }\mathrm{e}^{%
\mathrm{i}\eta \pi }}{\alpha a_{1}\left( \rho ^{\ast }\right) ^{\alpha }%
\mathrm{e}^{\mathrm{i}\alpha \pi }+\beta a_{2}\left( \rho ^{\ast }\right)
^{\beta }\mathrm{e}^{\mathrm{i}\beta \pi }+\gamma a_{3}\left( \rho ^{\ast
}\right) ^{\gamma }\mathrm{e}^{\mathrm{i}\gamma \pi }}\mathrm{e}^{-\rho
^{\ast }t}.  \label{intgama*}
\end{eqnarray}%
In calculating (\ref{intgama*}), function $\Psi $ (\ref{psi}) for $r^{\ast
}\rightarrow 0$ is written as%
\begin{eqnarray}
\Psi \left( \rho ^{\ast }\mathrm{e}^{\mathrm{i}\pi }+r^{\ast }\mathrm{e}^{%
\mathrm{i}\varphi }\right) \!\!\!\! &=&\!\!\!\!1+a_{1}\left( \rho ^{\ast
}\right) ^{\alpha }\mathrm{e}^{\mathrm{i}\alpha \pi }\left( 1-\frac{r^{\ast }%
\mathrm{e}^{\mathrm{i}\varphi }}{\rho ^{\ast }}\right) ^{\alpha
}+a_{2}\left( \rho ^{\ast }\right) ^{\beta }\mathrm{e}^{\mathrm{i}\beta \pi
}\left( 1-\frac{r^{\ast }\mathrm{e}^{\mathrm{i}\varphi }}{\rho ^{\ast }}%
\right) ^{\beta }+a_{3}\left( \rho ^{\ast }\right) ^{\gamma }\mathrm{e}^{%
\mathrm{i}\gamma \pi }\left( 1-\frac{r^{\ast }\mathrm{e}^{\mathrm{i}\varphi }%
}{\rho ^{\ast }}\right) ^{\gamma }  \notag \\
\!\!\!\! &\approx &\!\!\!\!1+a_{1}\left( \rho ^{\ast }\right) ^{\alpha }%
\mathrm{e}^{\mathrm{i}\alpha \pi }\left( 1-\alpha \frac{r^{\ast }\mathrm{e}^{%
\mathrm{i}\varphi }}{\rho ^{\ast }}\right) +a_{2}\left( \rho ^{\ast }\right)
^{\beta }\mathrm{e}^{\mathrm{i}\beta \pi }\left( 1-\beta \frac{r^{\ast }%
\mathrm{e}^{\mathrm{i}\varphi }}{\rho ^{\ast }}\right) +a_{3}\left( \rho
^{\ast }\right) ^{\gamma }\mathrm{e}^{\mathrm{i}\gamma \pi }\left( 1-\gamma 
\frac{r^{\ast }\mathrm{e}^{\mathrm{i}\varphi }}{\rho ^{\ast }}\right)  \notag
\\
\!\!\!\! &\approx &\!\!\!\!-\frac{r^{\ast }\mathrm{e}^{\mathrm{i}\varphi }}{%
\rho ^{\ast }}\left( \alpha a_{1}\left( \rho ^{\ast }\right) ^{\alpha }%
\mathrm{e}^{\mathrm{i}\alpha \pi }+\beta a_{2}\left( \rho ^{\ast }\right)
^{\beta }\mathrm{e}^{\mathrm{i}\beta \pi }+\gamma a_{3}\left( \rho ^{\ast
}\right) ^{\gamma }\mathrm{e}^{\mathrm{i}\gamma \pi }\right) ,
\label{psi-az-r-tezi-0}
\end{eqnarray}%
where the approximation $\left( 1+x\right) ^{\xi }\approx 1+\xi x,$ for $%
\left\vert x\right\vert \ll 1,$ and fact that $\Psi \left( \rho ^{\ast }%
\mathrm{e}^{\mathrm{i}\pi }\right) =0$ (since $\rho ^{\ast }$ is negative
real zero of $\Psi $) are used, implying 
\begin{equation}
\lim_{_{\substack{ r^{\ast }\rightarrow 0}}}\frac{r^{\ast }\mathrm{e}^{%
\mathrm{i}\varphi }}{\Psi \left( \rho ^{\ast }\mathrm{e}^{\mathrm{i}\pi
}+r^{\ast }\mathrm{e}^{\mathrm{i}\varphi }\right) }=-\frac{\rho ^{\ast }}{%
\alpha a_{1}\left( \rho ^{\ast }\right) ^{\alpha }\mathrm{e}^{\mathrm{i}%
\alpha \pi }+\beta a_{2}\left( \rho ^{\ast }\right) ^{\beta }\mathrm{e}^{%
\mathrm{i}\beta \pi }+\gamma a_{3}\left( \rho ^{\ast }\right) ^{\gamma }%
\mathrm{e}^{\mathrm{i}\gamma \pi }}.  \label{limint}
\end{equation}%
The integral corresponding to contour $\Gamma _{\ast }$, parametrized by $%
s-\rho ^{\ast }\mathrm{e}^{-\mathrm{i}\pi }=r^{\ast }\mathrm{e}^{\mathrm{i}%
\varphi },$ $\varphi \in (-\pi ,0),$ in the limit when $r^{\ast }\rightarrow
0$ is obtained as%
\begin{eqnarray*}
\lim_{_{\substack{ r^{\ast }\rightarrow 0}}}I_{\Gamma _{\ast }} &=&\lim_{ 
_{\substack{ r^{\ast }\rightarrow 0}}}\int_{\Gamma _{\ast }}\tilde{\sigma}%
_{sr}\left( s\right) \mathrm{e}^{st}\mathrm{d}s \\
&=&\mathrm{i}\frac{b_{1}+b_{2}\left( \rho ^{\ast }\right) ^{\eta }\mathrm{e}%
^{-\mathrm{i}\eta \pi }}{\left( \rho ^{\ast }\right) ^{1-\mu }\mathrm{e}^{-%
\mathrm{i}\left( 1-\mu \right) \pi }}\mathrm{e}^{-\rho ^{\ast }t}\lim_{ 
_{\substack{ r^{\ast }\rightarrow 0}}}\int_{0}^{-\pi }\frac{r^{\ast }\mathrm{%
e}^{\mathrm{i}\varphi }}{\Psi \left( \rho ^{\ast }\mathrm{e}^{-\mathrm{i}\pi
}+r^{\ast }\mathrm{e}^{\mathrm{i}\varphi }\right) }\mathrm{d}\varphi \\
&=&-\mathrm{i}\pi \left( \rho ^{\ast }\right) ^{\mu }\mathrm{e}^{-\mathrm{i}%
\mu \pi }\frac{b_{1}+b_{2}\left( \rho ^{\ast }\right) ^{\eta }\mathrm{e}^{-%
\mathrm{i}\eta \pi }}{\alpha a_{1}\left( \rho ^{\ast }\right) ^{\alpha }%
\mathrm{e}^{-\mathrm{i}\alpha \pi }+\beta a_{2}\left( \rho ^{\ast }\right)
^{\beta }\mathrm{e}^{-\mathrm{i}\beta \pi }+\gamma a_{3}\left( \rho ^{\ast
}\right) ^{\gamma }\mathrm{e}^{-\mathrm{i}\gamma \pi }}\mathrm{e}^{-\rho
^{\ast }t},
\end{eqnarray*}%
using the similar procedure as in calculating $\lim_{_{\substack{ r^{\ast
}\rightarrow 0}}}I_{\Gamma ^{\ast }}$.

Therefore, one has%
\begin{eqnarray*}
\lim_{_{\substack{ r^{\ast }\rightarrow 0}}}\left( I_{\Gamma ^{\ast
}}+I_{\Gamma _{\ast }}\right) &=&-\mathrm{i}\pi \left( \frac{%
b_{1}+b_{2}\left( \rho ^{\ast }\right) ^{\eta }\mathrm{e}^{\mathrm{i}\eta
\pi }}{\alpha a_{1}\left( \rho ^{\ast }\right) ^{\alpha }\mathrm{e}^{\mathrm{%
i}\alpha \pi }+\beta a_{2}\left( \rho ^{\ast }\right) ^{\beta }\mathrm{e}^{%
\mathrm{i}\beta \pi }+\gamma a_{3}\left( \rho ^{\ast }\right) ^{\gamma }%
\mathrm{e}^{\mathrm{i}\gamma \pi }}\mathrm{e}^{\mathrm{i}\mu \pi }\right. \\
&&+\left. \frac{b_{1}+b_{2}\left( \rho ^{\ast }\right) ^{\eta }\mathrm{e}^{-%
\mathrm{i}\eta \pi }}{\alpha a_{1}\left( \rho ^{\ast }\right) ^{\alpha }%
\mathrm{e}^{-\mathrm{i}\alpha \pi }+\beta a_{2}\left( \rho ^{\ast }\right)
^{\beta }\mathrm{e}^{-\mathrm{i}\beta \pi }+\gamma a_{3}\left( \rho ^{\ast
}\right) ^{\gamma }\mathrm{e}^{-\mathrm{i}\gamma \pi }}\mathrm{e}^{-\mathrm{i%
}\mu \pi }\right) \left( \rho ^{\ast }\right) ^{\mu }\mathrm{e}^{-\rho
^{\ast }t},
\end{eqnarray*}%
yielding function $f_{sr}^{\ast }$ in the form (\ref{f-sr-zvezda-1-5}) and
representing the second term in (\ref{DrugoResenje}).

\textbf{Relaxation modulus in the case of models having non-zero glass
compliance. }Inverting the Laplace transform in the relaxation modulus in
complex domain (\ref{sr-lap-1}), one obtains%
\begin{eqnarray}
\sigma _{sr}\left( t\right) &=&\frac{b_{2}}{a_{3}}-\frac{b_{2}}{a_{3}}%
\int_{0}^{t}\mathcal{L}^{-1}\left[ \tilde{f}_{sr}\left( s\right) \right]
\left( \tau \right) \mathrm{d}\tau  \notag \\
&=&\frac{b_{2}}{a_{3}}-\frac{b_{2}}{a_{3}}\int_{0}^{t}f_{sr}\left( \tau
\right) \mathrm{d}\tau +\frac{b_{2}}{a_{3}}f_{sr}^{\ast }\left( \rho ^{\ast
}\right) \left( 1-\mathrm{e}^{-\rho ^{\ast }t}\right) ,
\label{DrugoResenje-6-7}
\end{eqnarray}%
where functions $\tilde{f}_{sr},$ $f_{sr},$ and $f_{sr}^{\ast }$ are
respectively given by (\ref{f-tilda-sr}), (\ref{f-sr}), and (\ref%
{f-sr-zvezda}), due to the existence of negative real zero $-\rho ^{\ast }$
of function $\Psi $ (\ref{psi}), with $\rho ^{\ast }$ determined from (\ref%
{rho-zvezda-1-5}) with $\gamma =\beta +\eta .$ Namely, using function $%
\tilde{f}_{sr}$ in the form%
\begin{equation}
\tilde{f}_{sr}\left( s\right) =\frac{\psi \left( s\right) }{\psi \left(
s\right) +s^{\beta }\left( \frac{b_{1}}{b_{2}}+s^{\eta }\right) }=\frac{%
1+a_{1}s^{\alpha }+a_{3}\left( \frac{a_{2}}{a_{3}}-\frac{b_{1}}{b_{2}}%
\right) s^{\beta }}{\Psi \left( s\right) },  \label{f-tilda-sr-1}
\end{equation}%
see (\ref{psi1}) and (\ref{psi}), as an integrand in the Cauchy integral
theorem%
\begin{equation}
\oint_{\Gamma ^{\left( \mathrm{II}\right) }}\tilde{f}_{sr}\left( s\right) 
\mathrm{e}^{st}\mathrm{d}s=0,  \label{kit}
\end{equation}%
where the contour $\Gamma ^{\left( \mathrm{II}\right) }=\Gamma _{0}\cup
\Gamma _{1}\cup \Gamma _{2}\cup \Gamma _{3a}\cup \Gamma ^{\ast }\cup \Gamma
_{3b}\cup \Gamma _{4}\cup \Gamma _{5a}\cup \Gamma _{\ast }\cup \Gamma
_{5b}\cup \Gamma _{6}\cup \Gamma _{7}$ is chosen as in Figure \ref{fig-neg
real nula}, one obtains the second and third term in (\ref{DrugoResenje-6-7}%
).

The contour $\Gamma _{3a}$ is parametrized by $s=\rho \mathrm{e}^{\mathrm{i}%
\pi },$ $\rho \in \left( \rho ^{\ast }+r^{\ast },R\right) ,$ while $\Gamma
_{3b}$ has the same parametrization with $\rho \in \left( \rho ^{\ast
}-r^{\ast },r\right) $, so that in the limit when $R\rightarrow \infty ,$ $%
r\rightarrow 0$ and $r^{\ast }\rightarrow 0$ the integral along contour $%
\Gamma _{3a}\cup \Gamma _{3b},$ as in (\ref{int-gama3}), reads%
\begin{equation}
\lim_{\substack{ R\rightarrow \infty ,  \\ r\rightarrow 0,  \\ r^{\ast
}\rightarrow 0}}\int_{\Gamma _{3a}\cup \Gamma _{3b}}\tilde{f}_{sr}\left(
s\right) \mathrm{e}^{st}\mathrm{d}s=\int_{0}^{\infty }\frac{\psi \left( \rho 
\mathrm{e}^{\mathrm{i}\pi }\right) }{\psi \left( \rho \mathrm{e}^{\mathrm{i}%
\pi }\right) +\rho ^{\beta }\mathrm{e}^{\mathrm{i}\beta \pi }\left( \frac{%
b_{1}}{b_{2}}+\rho ^{\eta }\mathrm{e}^{\mathrm{i}\eta \pi }\right) }\mathrm{e%
}^{-\rho t}\mathrm{d}\rho .  \label{Intgama3ab-6-7}
\end{equation}%
Similarly, the integral along contour $\Gamma _{5a}\cup \Gamma _{5b}$
(parametrized by $s=\rho \mathrm{e}^{-\mathrm{i}\pi },$ $\rho \in \left(
\rho ^{\ast }-r^{\ast },r\right) $ for $\Gamma _{5a}$ and $\rho \in \left(
\rho ^{\ast }+r^{\ast },R\right) $ for $\Gamma _{5b}$), as in (\ref%
{int-gama5}), is 
\begin{equation}
\lim_{_{\substack{ R\rightarrow \infty ,  \\ r\rightarrow 0,  \\ r^{\ast
}\rightarrow 0}}}\int_{\Gamma _{5a}\cup \Gamma _{5b}}\tilde{f}_{sr}\left(
s\right) \mathrm{e}^{st}\mathrm{d}s=-\int_{0}^{\infty }\frac{\psi \left(
\rho \mathrm{e}^{-\mathrm{i}\pi }\right) }{\psi \left( \rho \mathrm{e}^{-%
\mathrm{i}\pi }\right) +\rho ^{\beta }\mathrm{e}^{-\mathrm{i}\beta \pi
}\left( \frac{b_{1}}{b_{2}}+\rho ^{\eta }\mathrm{e}^{-\mathrm{i}\eta \pi
}\right) }\mathrm{e}^{-\rho t}\mathrm{d}\rho .  \label{Intgama5ab-6-7}
\end{equation}

The contour $\Gamma ^{\ast }$ is parametrized by $s-\rho ^{\ast }\mathrm{e}^{%
\mathrm{i}\pi }=r^{\ast }\mathrm{e}^{\mathrm{i}\varphi },$ $\varphi \in
(0,\pi ),$ with $r^{\ast }\rightarrow 0,$ so that the corresponding integral
reads%
\begin{eqnarray*}
I_{\Gamma ^{\ast }} &=&\int_{\Gamma ^{\ast }}\tilde{f}_{sr}\left( s\right) 
\mathrm{e}^{st}\mathrm{d}s \\
&=&\int_{\pi }^{0}\frac{1+a_{1}\left( \rho ^{\ast }\mathrm{e}^{\mathrm{i}\pi
}+r^{\ast }\mathrm{e}^{\mathrm{i}\varphi }\right) ^{\alpha }+a_{3}\left( 
\frac{a_{2}}{a_{3}}-\frac{b_{1}}{b_{2}}\right) \left( \rho ^{\ast }\mathrm{e}%
^{\mathrm{i}\pi }+r^{\ast }\mathrm{e}^{\mathrm{i}\varphi }\right) ^{\beta }}{%
\Psi \left( \rho ^{\ast }\mathrm{e}^{\mathrm{i}\pi }+r^{\ast }\mathrm{e}^{%
\mathrm{i}\varphi }\right) }\mathrm{e}^{\left( \rho ^{\ast }\mathrm{e}^{%
\mathrm{i}\pi }+r^{\ast }\mathrm{e}^{\mathrm{i}\varphi }\right) t}\mathrm{i}%
r^{\ast }\mathrm{e}^{\mathrm{i}\varphi }\mathrm{d}\varphi ,
\end{eqnarray*}%
so that by letting $r^{\ast }\rightarrow 0$ in the previous expression one
obtains%
\begin{eqnarray}
\lim_{_{\substack{ r^{\ast }\rightarrow 0}}}I_{\Gamma ^{\ast }} &=&-\mathrm{i%
}\left( 1+a_{1}\left( \rho ^{\ast }\right) ^{\alpha }\mathrm{e}^{\mathrm{i}%
\alpha \pi }+a_{3}\left( \frac{a_{2}}{a_{3}}-\frac{b_{1}}{b_{2}}\right)
\left( \rho ^{\ast }\right) ^{\beta }\mathrm{e}^{\mathrm{i}\beta \pi
}\right) \mathrm{e}^{-\rho ^{\ast }t}\lim_{_{\substack{ r^{\ast }\rightarrow
0}}}\int_{0}^{\pi }\frac{r^{\ast }\mathrm{e}^{\mathrm{i}\varphi }}{\Psi
\left( \rho ^{\ast }\mathrm{e}^{\mathrm{i}\pi }+r^{\ast }\mathrm{e}^{\mathrm{%
i}\varphi }\right) }\mathrm{d}\varphi ,  \notag \\
&=&\mathrm{i}\pi \frac{1+a_{1}\left( \rho ^{\ast }\right) ^{\alpha }\mathrm{e%
}^{\mathrm{i}\alpha \pi }+a_{3}\left( \frac{a_{2}}{a_{3}}-\frac{b_{1}}{b_{2}}%
\right) \left( \rho ^{\ast }\right) ^{\beta }\mathrm{e}^{\mathrm{i}\beta \pi
}}{\alpha a_{1}\left( \rho ^{\ast }\right) ^{\alpha }\mathrm{e}^{\mathrm{i}%
\alpha \pi }+\beta a_{2}\left( \rho ^{\ast }\right) ^{\beta }\mathrm{e}^{%
\mathrm{i}\beta \pi }+\left( \beta +\eta \right) a_{3}\left( \rho ^{\ast
}\right) ^{\beta +\eta }\mathrm{e}^{\mathrm{i}\left( \beta +\eta \right) \pi
}}\rho ^{\ast }\mathrm{e}^{-\rho ^{\ast }t}.  \label{intgama^*}
\end{eqnarray}%
In calculating (\ref{intgama^*}), the expression (\ref{limint}), with $%
\gamma =\beta +\eta ,$ is used. The integral corresponding to contour $%
\Gamma _{\ast }$, parametrized by $s-\rho ^{\ast }\mathrm{e}^{-\mathrm{i}\pi
}=r^{\ast }\mathrm{e}^{\mathrm{i}\varphi },$ $\varphi \in (-\pi ,0),$ in the
limit when $r^{\ast }\rightarrow 0$ reads%
\begin{eqnarray}
\lim_{_{\substack{ r^{\ast }\rightarrow 0}}}I_{\Gamma _{\ast }} &=&\lim_{ 
_{\substack{ r^{\ast }\rightarrow 0}}}\int_{\Gamma _{\ast }}\tilde{f}%
_{sr}\left( s\right) \mathrm{e}^{st}\mathrm{d}s  \notag \\
&=&\mathrm{i}\left( 1+a_{1}\left( \rho ^{\ast }\right) ^{\alpha }\mathrm{e}%
^{-\mathrm{i}\alpha \pi }+a_{3}\left( \frac{a_{2}}{a_{3}}-\frac{b_{1}}{b_{2}}%
\right) \left( \rho ^{\ast }\right) ^{\beta }\mathrm{e}^{-\mathrm{i}\beta
\pi }\right) \mathrm{e}^{-\rho ^{\ast }t}\lim_{_{\substack{ r^{\ast
}\rightarrow 0}}}\int_{0}^{-\pi }\frac{r^{\ast }\mathrm{e}^{\mathrm{i}%
\varphi }}{\Psi \left( \rho ^{\ast }\mathrm{e}^{-\mathrm{i}\pi }+r^{\ast }%
\mathrm{e}^{\mathrm{i}\varphi }\right) }\mathrm{d}\varphi  \notag \\
&=&\mathrm{i}\pi \frac{1+a_{1}\left( \rho ^{\ast }\right) ^{\alpha }\mathrm{e%
}^{-\mathrm{i}\alpha \pi }+a_{3}\left( \frac{a_{2}}{a_{3}}-\frac{b_{1}}{b_{2}%
}\right) \left( \rho ^{\ast }\right) ^{\beta }\mathrm{e}^{-\mathrm{i}\beta
\pi }}{\alpha a_{1}\left( \rho ^{\ast }\right) ^{\alpha }\mathrm{e}^{-%
\mathrm{i}\alpha \pi }+\beta a_{2}\left( \rho ^{\ast }\right) ^{\beta }%
\mathrm{e}^{-\mathrm{i}\beta \pi }+\left( \beta +\eta \right) a_{3}\left(
\rho ^{\ast }\right) ^{\beta +\eta }\mathrm{e}^{-\mathrm{i}\left( \beta
+\eta \right) \pi }}\rho ^{\ast }\mathrm{e}^{-\rho ^{\ast }t},
\label{intgama_*}
\end{eqnarray}%
using the similar procedure as in calculating $\lim_{_{\substack{ r^{\ast
}\rightarrow 0}}}I_{\Gamma ^{\ast }}$.

In the limit when $R\rightarrow \infty ,$ $r\rightarrow 0,$ and $r^{\ast
}\rightarrow 0,$ the inverse Laplace transform $\mathcal{L}^{-1}\left[ 
\tilde{f}_{sr}\left( s\right) \right] $ (\ref{int-gama0}), i.e., the second
and the third term in the relaxation modulus (\ref{DrugoResenje-6-7}), is
obtained from the Cauchy integral theorem (\ref{kit}) as the sum of
integrals along contours $\Gamma _{3a}\cup \Gamma _{3b},$ $\Gamma _{5a}\cup
\Gamma _{5b},$ $\Gamma ^{\ast },$ and $\Gamma _{\ast },$ respectively given
by (\ref{Intgama3ab-6-7}), (\ref{Intgama5ab-6-7}), (\ref{intgama^*}), and (%
\ref{intgama_*}), since the integrals along $\Gamma _{1},$ $\Gamma _{2},$ $%
\Gamma _{4},$ $\Gamma _{6},$ $\Gamma _{7}$ tend to zero as $R\rightarrow
\infty $ and $r\rightarrow 0,$ as already proved in Section \ref{Nemanula}.
Therefore, one has%
\begin{equation*}
\mathcal{L}^{-1}\left[ \tilde{f}_{sr}\left( s\right) \right] \left( t\right)
=f_{sr}\left( t\right) -\rho ^{\ast }f_{sr}^{\ast }\left( \rho ^{\ast
}\right) \mathrm{e}^{-\rho ^{\ast }t},
\end{equation*}%
with $f_{sr}$ given by (\ref{f-sr}) and%
\begin{eqnarray*}
f_{sr}^{\ast }\left( \rho ^{\ast }\right) &=&\frac{1}{2}\left( \frac{%
1+a_{1}\left( \rho ^{\ast }\right) ^{\alpha }\mathrm{e}^{\mathrm{i}\alpha
\pi }+a_{3}\left( \frac{a_{2}}{a_{3}}-\frac{b_{1}}{b_{2}}\right) \left( \rho
^{\ast }\right) ^{\beta }\mathrm{e}^{\mathrm{i}\beta \pi }}{\alpha
a_{1}\left( \rho ^{\ast }\right) ^{\alpha }\mathrm{e}^{\mathrm{i}\alpha \pi
}+\beta a_{2}\left( \rho ^{\ast }\right) ^{\beta }\mathrm{e}^{\mathrm{i}%
\beta \pi }+\left( \beta +\eta \right) a_{3}\left( \rho ^{\ast }\right)
^{\beta +\eta }\mathrm{e}^{\mathrm{i}\left( \beta +\eta \right) \pi }}\right.
\\
&&+\left. \frac{1+a_{1}\left( \rho ^{\ast }\right) ^{\alpha }\mathrm{e}^{-%
\mathrm{i}\alpha \pi }+a_{3}\left( \frac{a_{2}}{a_{3}}-\frac{b_{1}}{b_{2}}%
\right) \left( \rho ^{\ast }\right) ^{\beta }\mathrm{e}^{-\mathrm{i}\beta
\pi }}{\alpha a_{1}\left( \rho ^{\ast }\right) ^{\alpha }\mathrm{e}^{-%
\mathrm{i}\alpha \pi }+\beta a_{2}\left( \rho ^{\ast }\right) ^{\beta }%
\mathrm{e}^{-\mathrm{i}\beta \pi }+\left( \beta +\eta \right) a_{3}\left(
\rho ^{\ast }\right) ^{\beta +\eta }\mathrm{e}^{-\mathrm{i}\left( \beta
+\eta \right) \pi }}\right) ,
\end{eqnarray*}%
yielding (\ref{f-sr-zvezda}).

\subsubsection{Case when function $\Psi $ has a pair of complex conjugated
zeros}

\textbf{Relaxation modulus in the case of models having zero glass
compliance. }The Cauchy residue theorem with the relaxation modulus in
complex domain (\ref{sr-lap}) as an integrand takes the form 
\begin{equation}
\oint_{\Gamma ^{\left( \mathrm{I}\right) }}\tilde{\sigma}_{sr}\left(
s\right) \mathrm{e}^{st}\mathrm{d}s=2\pi \mathrm{i}\left( \func{Res}\left( 
\tilde{\sigma}_{sr}\left( s\right) \mathrm{e}^{st},s_{0}\right) +\func{Res}%
\left( \tilde{\sigma}_{sr}\left( s\right) \mathrm{e}^{st},\bar{s}_{0}\right)
\right) ,  \label{Kit2}
\end{equation}%
due to the existence of complex conjugated zeroes $s_{0}$ and $\bar{s}_{0}$
having negative real part of function $\Psi ,$ given by (\ref{psi}), with
the contour $\Gamma ^{\left( \mathrm{I}\right) }=\Gamma _{0}\cup \Gamma
_{1}\cup \Gamma _{2}\cup \Gamma _{3}\cup \Gamma _{4}\cup \Gamma _{5}\cup
\Gamma _{6}\cup \Gamma _{7}$ chosen as in Figure \ref{fig-nema nula}.

The relaxation modulus is obtained in the form%
\begin{equation}
\sigma _{sr}\left( t\right) =\frac{1}{\pi }\int_{0}^{\infty }\frac{K\left(
\rho \right) }{\left\vert \Psi \left( \rho \mathrm{e}^{\mathrm{i}\pi
}\right) \right\vert ^{2}}\frac{\mathrm{e}^{-\rho t}}{\rho ^{1-\mu }}\mathrm{%
d}\rho +f_{sr}^{\left( r\right) }\left( t\right) ,  \label{Trece resenje}
\end{equation}%
with functions $K$ and $f_{sr}^{\left( r\right) }$ given by (\ref{funct-K})
and (\ref{f-sr-res-1-5}), using the Cauchy residue theorem (\ref{Kit2}).
Namely, the integration along the contour $\Gamma ^{\left( \mathrm{I}\right)
}$ yields the inverse Laplace transform (\ref{int-gama0-sr}) and the first
term in the relaxation modulus (\ref{Trece resenje}), that is already
obtained in Section \ref{Nemanula}, while the second term in (\ref{Trece
resenje}) consists of the residues of function $\tilde{\sigma}_{sr}\left(
s\right) \mathrm{e}^{st},$ since $s_{0}=\rho _{0}\mathrm{e}^{\mathrm{i}%
\varphi _{0}}$ and $\bar{s}_{0}=\rho _{0}\mathrm{e}^{-\mathrm{i}\varphi
_{0}},$ $\varphi _{0}\in \left( \frac{\pi }{2},\pi \right) ,$ are its poles
of the first order, i.e., the first order zeros of function $\Psi $ (\ref%
{psi}). Therefore, one has%
\begin{eqnarray*}
f_{sr}^{\left( r\right) }\left( t\right) &=&\left[ \frac{1}{s^{1-\mu }}\frac{%
b_{1}+b_{2}s^{\eta }}{\frac{\mathrm{d}}{\mathrm{d}s}\Psi \left( s\right) }%
\mathrm{e}^{st}\right] _{s=s_{0}}+\left[ \frac{1}{s^{1-\mu }}\frac{%
b_{1}+b_{2}s^{\eta }}{\frac{\mathrm{d}}{\mathrm{d}s}\Psi \left( s\right) }%
\mathrm{e}^{st}\right] _{s=\bar{s}_{0}} \\
&=&\left( \left. \frac{b_{1}+b_{2}s^{\eta }}{\alpha a_{1}s^{\alpha -1}+\beta
a_{2}s^{\beta -1}+\gamma a_{3}s^{\gamma -1}}\right\vert _{s=\rho _{0}\mathrm{%
e}^{\mathrm{i}\varphi _{0}}}\frac{\mathrm{e}^{-\mathrm{i}\left( 1-\mu
\right) \varphi _{0}}}{\rho _{0}^{1-\mu }}\mathrm{e}^{\mathrm{i}\rho
_{0}t\sin \varphi _{0}}\right. \\
&&+\left. \left. \frac{b_{1}+b_{2}s^{\eta }}{\alpha a_{1}s^{\alpha -1}+\beta
a_{2}s^{\beta -1}+\gamma a_{3}s^{\gamma -1}}\right\vert _{s=\rho _{0}\mathrm{%
e}^{-\mathrm{i}\varphi _{0}}}\frac{\mathrm{e}^{\mathrm{i}\left( 1-\mu
\right) \varphi _{0}}}{\rho _{0}^{1-\mu }}\mathrm{e}^{\mathrm{i}\rho
_{0}t\sin \varphi _{0}}\right) \mathrm{e}^{\rho _{0}t\cos \varphi _{0}},
\end{eqnarray*}%
yielding (\ref{f-sr-res-1-5}).

\textbf{Relaxation modulus in the case of models having non-zero glass
compliance. }Inverting the Laplace transform in the relaxation modulus in
complex domain (\ref{sr-lap-1}), one obtains%
\begin{eqnarray}
\sigma _{sr}\left( t\right) &=&\frac{b_{2}}{a_{3}}-\frac{b_{2}}{a_{3}}%
\int_{0}^{t}\mathcal{L}^{-1}\left[ \tilde{f}_{sr}\left( s\right) \right]
\left( \tau \right) \mathrm{d}\tau  \notag \\
&=&\frac{b_{2}}{a_{3}}-\frac{b_{2}}{a_{3}}\int_{0}^{t}f_{sr}\left( \tau
\right) \mathrm{d}\tau -\frac{b_{2}}{a_{3}}\int_{0}^{t}f_{sr}^{\left(
r\right) }\left( \tau \right) \mathrm{d}\tau ,  \label{Trece resenje-6-7}
\end{eqnarray}%
where functions $\tilde{f}_{sr},$ $f_{sr},$ and $f_{sr}^{\left( r\right) }$
are respectively given by (\ref{f-tilda-sr}), (\ref{f-sr}), and (\ref%
{f-sr-res}), due to the existence of complex conjugated zeroes $s_{0}$ and $%
\bar{s}_{0}$ of function $\Psi $ (\ref{psi}) having negative real part.

Using function $\tilde{f}_{sr}$ in the form (\ref{f-tilda-sr-1}) as an
integrand in the Cauchy residue theorem%
\begin{equation*}
\oint_{\Gamma ^{\left( \mathrm{I}\right) }}\tilde{f}_{sr}\left( s\right) 
\mathrm{e}^{st}\mathrm{d}s=2\pi \mathrm{i}\left( \func{Res}\left( \tilde{f}%
_{sr}\left( s\right) \mathrm{e}^{st},s_{0}\right) +\func{Res}\left( \tilde{f}%
_{sr}\left( s\right) \mathrm{e}^{st},\bar{s}_{0}\right) \right) ,
\end{equation*}%
with the contour $\Gamma ^{\left( \mathrm{I}\right) }=\Gamma _{0}\cup \Gamma
_{1}\cup \Gamma _{2}\cup \Gamma _{3}\cup \Gamma _{4}\cup \Gamma _{5}\cup
\Gamma _{6}\cup \Gamma _{7}$ chosen as in Figure \ref{fig-nema nula}, one
obtains the inverse Laplace transform $\mathcal{L}^{-1}\left[ \tilde{f}%
_{sr}\left( s\right) \right] $ (\ref{int-gama0}), i.e., the second and the
third term in (\ref{Trece resenje-6-7}), in the form 
\begin{equation}
\mathcal{L}^{-1}\left[ \tilde{f}_{sr}\left( s\right) \right] \left( t\right)
=f_{sr}\left( t\right) +f_{sr}^{\left( r\right) }\left( t\right) .
\label{inv-lt-f-tilde-sr}
\end{equation}%
The first term in (\ref{inv-lt-f-tilde-sr}), being a consequence of the
integration along the contour $\Gamma ^{\left( \mathrm{I}\right) },$ is
already obtained in Section \ref{Nemanula}, while the second one consists of
the residues of function $\tilde{f}_{sr}\left( s\right) \mathrm{e}^{st},$
since $s_{0}=\rho _{0}\mathrm{e}^{\mathrm{i}\varphi _{0}}$ and $\bar{s}%
_{0}=\rho _{0}\mathrm{e}^{-\mathrm{i}\varphi _{0}},$ $\varphi _{0}\in \left( 
\frac{\pi }{2},\pi \right) ,$ are its poles of the first order, i.e., the
first order zeros of function $\Psi ,$ as proved in Section \ref{CFpsi}.
Therefore in (\ref{inv-lt-f-tilde-sr}), one has $f_{sr}$ given by (\ref{f-sr}%
) and 
\begin{eqnarray*}
f_{sr}^{\left( r\right) }\left( t\right) &=&\left[ \frac{1+a_{1}s^{\alpha
}+a_{3}\left( \frac{a_{2}}{a_{3}}-\frac{b_{1}}{b_{2}}\right) s^{\beta }}{%
\frac{\mathrm{d}}{\mathrm{d}s}\Psi \left( s\right) }\mathrm{e}^{st}\right]
_{s=s_{0}}+\left[ \frac{1+a_{1}s^{\alpha }+a_{3}\left( \frac{a_{2}}{a_{3}}-%
\frac{b_{1}}{b_{2}}\right) s^{\beta }}{\frac{\mathrm{d}}{\mathrm{d}s}\Psi
\left( s\right) }\mathrm{e}^{st}\right] _{s=\bar{s}_{0}} \\
&=&\left( \left. \frac{1+a_{1}s^{\alpha }+a_{3}\left( \frac{a_{2}}{a_{3}}-%
\frac{b_{1}}{b_{2}}\right) s^{\beta }}{\alpha a_{1}s^{\alpha -1}+\beta
a_{2}s^{\beta -1}+\left( \beta +\eta \right) a_{3}s^{\beta +\eta -1}}%
\right\vert _{s=\rho _{0}\mathrm{e}^{\mathrm{i}\varphi _{0}}}\mathrm{e}^{%
\mathrm{i}\rho _{0}t\sin \varphi _{0}}\right. \\
&&+\left. \left. \frac{1+a_{1}s^{\alpha }+a_{3}\left( \frac{a_{2}}{a_{3}}-%
\frac{b_{1}}{b_{2}}\right) s^{\beta }}{\alpha a_{1}s^{\alpha -1}+\beta
a_{2}s^{\beta -1}+\left( \beta +\eta \right) a_{3}s^{\beta +\eta -1}}%
\right\vert _{s=\rho _{0}\mathrm{e}^{-\mathrm{i}\varphi _{0}}}\mathrm{e}^{-%
\mathrm{i}\rho _{0}t\sin \varphi _{0}}\right) \mathrm{e}^{\rho _{0}t\cos
\varphi _{0}},
\end{eqnarray*}%
yielding (\ref{f-sr-res}).

\section*{Acknowledgment}

This work is supported by the Serbian Ministry of Education, Science and
Technological Development under grant $174005$, as well as by the Provincial
Secretariat for Higher Education and Scientific Research under grant $%
142-451-2384/2018$.


\end{document}